\documentclass[letter]{aa}  

\usepackage{amsmath}
\usepackage{graphicx}
\usepackage{graphics}
\usepackage{epsfig}
\usepackage{pdfpages}
\usepackage{blindtext}
\usepackage{supertabular}
\usepackage{txfonts}
\usepackage[colorlinks=true,linkcolor=blue,citecolor=blue,urlcolor=blue]{hyperref}%
\begin{document}

\title{Unidentified quasars among stationary objects from \textit{Gaia} DR2}

\author{
K.~E.~Heintz\inst{1,2},
J.~P.~U.~Fynbo\inst{2},
E.~H\o g\inst{3},
P.~M\o ller\inst{4}, 
J.-K.~Krogager\inst{5},
S.~Geier\inst{6,7},
P.~Jakobsson\inst{1}
\and
L. Christensen\inst{8}
}

\institute{
Centre for Astrophysics and Cosmology, Science Institute, University of Iceland, Dunhagi 5, 107 Reykjav\'ik, Iceland
\and
The Cosmic Dawn Center, Niels Bohr Institute, University of Copenhagen, Juliane
Maries Vej 30, 2100 Copenhagen \O, Denmark  \\
\email{\href{mailto:keh14@hi.is}{keh14@hi.is}}
\and
Niels Bohr Institute, University of Copenhagen, Juliane Maries Vej 30, 2100 Copenhagen \O, Denmark 
\and
European Southern Observatory, Karl-Schwarzschildstrasse 2, D-85748 Garching bei M\"unchen, Germany
\and
Institut d'Astrophysique de Paris, CNRS-UPMC, UMR7095, 98bis bd Arago, 75014 Paris, France
\and
Gran Telescopio Canarias (GRANTECAN), Cuesta de San Jos\'e s/n, E-38712 , Bre\~na Baja, La Palma, Spain 
\and
Instituto de Astrof\'isica de Canarias, V\'ia L\'actea s/n, E38200, La Laguna, Tenerife, Spain
\and
Dark Cosmology Centre, Niels Bohr Institute, University of Copenhagen, Juliane
Maries Vej 30, 2100 Copenhagen \O, Denmark 
}

\date{Received 2018; accepted 2018}

\authorrunning{Heintz et al.}

 
  \abstract
{
We here apply a novel technique selecting quasar candidates purely as sources with zero proper motions in the \textit{Gaia} data release 2 (DR2). We demonstrate that
this approach is highly efficient toward high Galactic latitudes with $\lesssim
25\%$ contamination from stellar sources. Such a selection technique offers a
very pure sample completeness, since all cosmological point sources are
selected regardless of their intrinsic spectral properties within the limiting
magnitude of \textit{Gaia}. We carry out a pilot-study by defining a sample compiled 
by including all \textit{Gaia}-DR2 sources within one degree of the North Galactic Pole (NGP) selected to have proper motions consistent with zero within $2\sigma$ uncertainty. By cross-matching the sample to the optical Sloan Digital Sky Survey (SDSS) and the mid-infrared AllWISE photometric catalogues we investigate the colours of each of our sources. Together with already spectroscopically confirmed quasars we are therefore able to determine the efficiency of our
selection. The majority of the zero proper motion sources have optical to
mid-infrared colours consistent with known quasars. The remaining population
may be contaminating stellar sources, but some may also be quasars with colours
similar to stars. Spectroscopic follow-up of the zero proper motion sources is needed to unveil such a hitherto hidden quasar population. This approach has the potential to allow substantial progress on many important questions concerning quasars such as determining the fraction of dust-obscured quasars, the fraction of broad absorption line (BAL) quasars,  and the metallicity distribution of damped Lyman-$\alpha$ absorbers. The technique could also potentially reveal new types of quasars or even new classes of cosmological point sources.
}


\maketitle
%

\section{Introduction}

There is great interest in building unbiased catalogues of quasars
for a range of important astrophysical questions. These include 
understanding the quasar phenomenon itself and the growth and 
occurrence of supermassive black holes through cosmic time, 
the use of quasars as probes of intervening material and their
role in in re-ionisation of both hydrogen and helium, and  for the UV background 
levels throughout the universe
\citep{HP1994,Weymann}. Most current quasar surveys, however, rely on their specific intrinsic properties such as strong ultraviolet emission \citep{Schneider10}, distinct near/mid-infrared colours \citep{Maddox12,Secrest15}, X-ray output \citep{Brusa10} or prominent radio emission \citep{Ivezic02}.

In this {\it Letter} we apply an astrometric approach of identifying quasars as apparently
stationary sources on the sky, based purely on the astrometric measurement from
the \textit{Gaia} mission \citep{Heintz15}, and present the first pilot study of
such an approach. 
Our goal here is to quantify the efficiency and completeness of this
selection technique.
Identifying quasars based only on their zero proper motions
has the potential to open a novel route of selecting quasars in an unbiased
way, and might even lead to the discovery of new types of quasars or other
types of extragalactic point sources.


\section{Astrometric selection of quasars}

The \textit{Gaia} data release 2 \citep[DR2;][]{GaiaDR2} catalogue consists of
more than $1.3\times 10^9$ sources down to $G \approx 21$ mag, for which the
five-parameter astrometric solution (positions, parallaxes and proper motions)
has been determined \citep{Lindegren18}. The Gaia $G$ filter is very broad covering 
the spectral range from 400 to 1000 nm and hence quasars covering a wide range
of redshifts should be included in the catalogue. 

\begin{figure} [!t]
	\centering
	\epsfig{file=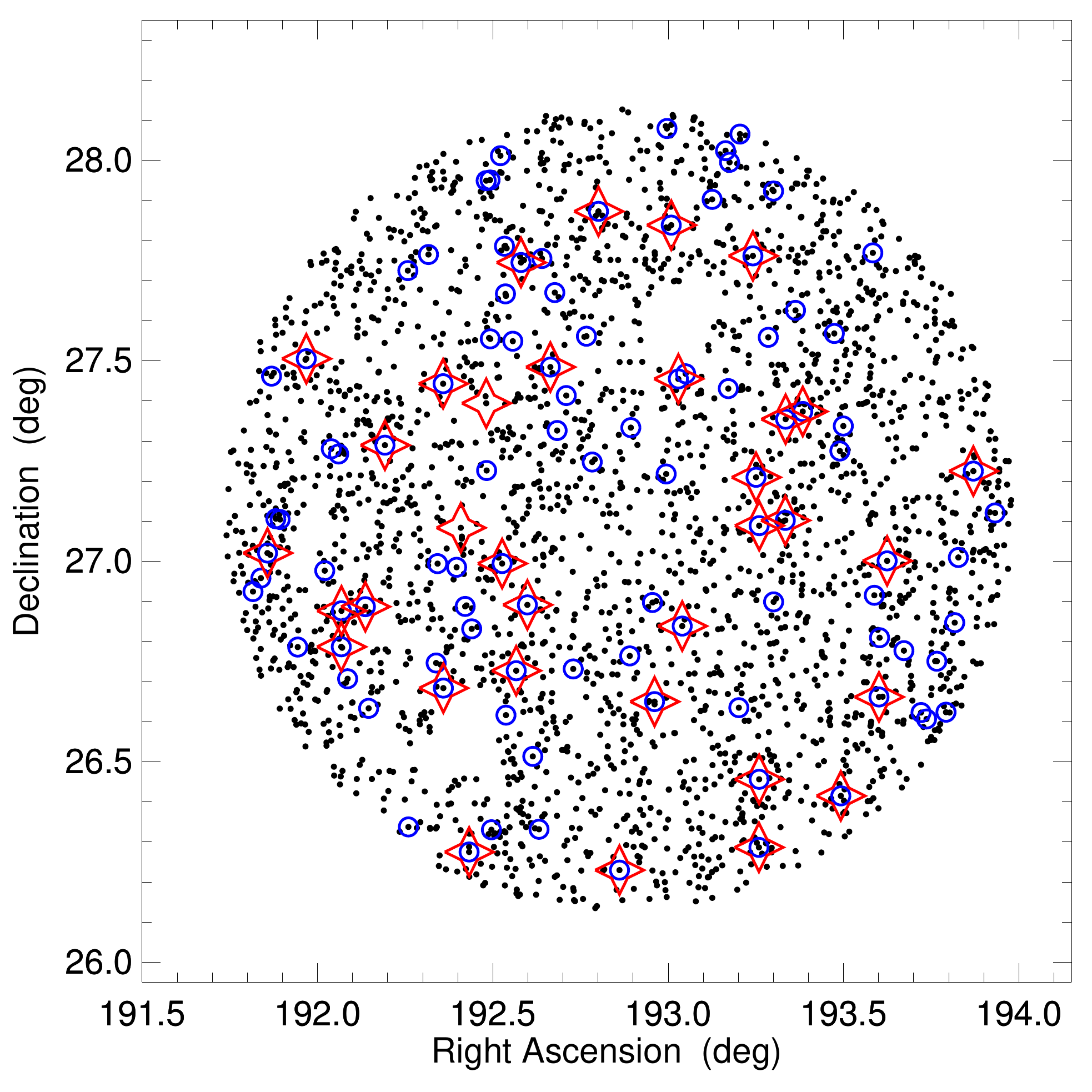,width=\columnwidth}
	\caption{Location on the sky of all point-like \textit{Gaia} sources with proper motions and $18 < G < 20$ mag (black dots) within one degree of the NGP. The subset of these with proper motions consistent with zero within $2\sigma$ are shown by the blue circles and those that are already spectroscopically confirmed quasars are shown by the red star symbols.}
	\label{fig:radec}
\end{figure}

We extract all sources within a radius of one degree from the North Galactic
Pole (NGP) centred on $(\alpha,\delta) =
12^{\mathrm{h}}\,51^{\mathrm{m}}\,26^{\mathrm{s}}.0~+27^{\mathrm{o}}\,07^{\mathrm{m}}\,42^{\mathrm{s}}.0$
from the \textit{Gaia} DR2 catalogue (see Fig.~\ref{fig:radec}). We then limit
our search to sources with $18 < G < 20$ mag, for which the associated
uncertainty is up to 1.2 mas\,yr$^{-1}$ in the respective proper motion
components. This is motivated by our pre-study \citep{Heintz15} in which we (based on pre-launch simulations of the Gaia-data)
found that the expected contamination of apparently stationary stars is lower
than $\approx 20\%$ at the Galactic poles, but increases significantly when
observing closer to the Galactic plane or at magnitudes brighter than $G < 18$
mag. We then select all point
sources with total proper motions, $\mu =\sqrt{\mu_{\mathrm{RA}}^2 +
\mu_{\mathrm{Dec}}^2}$, consistent with zero at the $2\sigma$ confidence level
(i.e. S/N$_{\mu} = \mu / \mu_{\mathrm{err}} < 2$). Finally, we identify the
counterpart to each source in the Sloan Digital Sky Survey (SDSS) and require
that all \textit{Gaia} sources have morphologies consistent with being point
sources (\texttt{class} = 6 in the SDSS) to limit our search to quasars only
(i.e. excluding Seyferts and potential contaminating extended galaxies). This
results in about 2\% of the sample being removed due to extended morphology.
Matching the \textit{Gaia} sample to the SDSS with a matching radius of less than 1 arcsec also allows us to investigate the properties of our sample in optical
colour-colour space.

In total, we find that there are 2,634 spatially unresolved \textit{Gaia} sources with
proper motions and $18 < G < 20$ mag within one degree of the NGP, of which 100
sources ($\approx 4\%$) have proper motions consistent with zero (within
$2\sigma$). These are shown as the blue circles in Fig.~\ref{fig:radec}.
Cross-matching our extracted catalogue with the SDSS data release 14 quasar
sample \citep[SDSS-DR14Q,][]{Paris17} and the NASA/IPAC Extragalactic Database (NED) we find that 34 quasars are already spectroscopically confirmed
within the same magnitude limit and region on the sky, of which 32 ($\approx
95\%$) also have S/N$_{\mu} = \mu / \mu_{\mathrm{err}} < 2$. 
For the remaining two the measured proper motions are $2.76\pm 1.09$ and
$1.22\pm 0.58$ mas yr$^{-1}$. We also discover two spectroscopically confirmed stars, observed as part of the SDSS-APOGEE survey \citep{Alam15}. An extract of the full sample of \textit{Gaia}
sources with zero proper motions is presented in Table~\ref{tab:1}.
We also examine the additional
requirement that the sources have parallaxes consistent with zero within
$3\sigma$, but only five sources (GQs\,1255+2707, 1248+2658, 1247+2655,
1247+2706, and 1251+2804) were outside this criterion so we chose to include
them for completeness.

\begin{longtab}
\begin{longtable}{lccccccccc}
\caption{
Point sources within one degree of the NGP with
proper motions consistent with zero (within 2$\sigma$) and $18 < G < 20$ mag. \label{tab:1}}  \\
\hline\hline
Source & R.A. & Decl. & $u$ & $g$ & $r$ & $i$ & $z$ & $W1 - W2$ & $z_{\mathrm{QSO}}$  \\
\hline
\hline
\endfirsthead
\caption{continued.}\\
\hline\hline
Source & R.A. & Decl. & $u$ & $g$ & $r$ & $i$ & $z$ & $W1-W2$ &  $z_{\mathrm{QSO}}$  \\
\hline
\hline
\endhead
GQ125420+274609 & 12:54:20.0 & +27:46:09.6 & 23.2 & 21.4 & 20.0 & 19.1 & 18.7 & - & - \\
GQ125302+261711 & 12:53:02.1 & +26:17:11.2 & 20.0 & 19.4 & 19.6 & 19.5 & 19.2 & 0.86 & 2.32 \\
GQ125358+262453 & 12:53:58.1 & +26:24:53.7 & 20.9 & 20.1 & 19.8 & 19.9 & 19.8 & 0.91 & 3.10 \\
GQ125302+262722 & 12:53:02.2 & +26:27:22.5 & 19.7 & 19.5 & 19.1 & 19.0 & 19.1 & 1.21 & 1.26 \\
GQ125456+263623 & 12:54:56.8 & +26:36:23.8 & 23.2 & 20.8 & 19.6 & 19.0 & 18.8 & -0.09 & - \\
GQ125510+263724 & 12:55:10.1 & +26:37:25.0 & 21.0 & 20.2 & 19.9 & 19.9 & 19.8 & - & - \\
GQ125424+263941 & 12:54:24.3 & +26:39:41.8 & 19.3 & 19.1 & 19.2 & 19.0 & 19.1 & 1.59 & 1.60 \\
GQ125453+263721 & 12:54:53.0 & +26:37:21.8 & 20.9 & 19.8 & 19.5 & 19.4 & 19.2 & - & - \\
GQ125503+264502 & 12:55:03.8 & +26:45:02.7 & 21.3 & 20.2 & 19.6 & 19.4 & 19.2 & - & - \\
GQ125441+264637 & 12:54:41.3 & +26:46:37.8 & 20.9 & 19.7 & 19.3 & 19.0 & 19.0 & - & - \\
GQ125424+264833 & 12:54:24.6 & +26:48:33.8 & 20.1 & 19.9 & 19.7 & 19.6 & 19.7 & 1.22 & 1.47P \\
GQ125516+265049 & 12:55:16.1 & +26:50:49.0 & 25.0 & 21.1 & 19.8 & 19.2 & 18.9 & - & - \\
GQ125518+270034 & 12:55:18.6 & +27:00:34.3 & 23.9 & 21.2 & 19.7 & 18.2 & 17.5 & 0.16 & - \\
GQ125543+270714 & 12:55:43.7 & +27:07:14.1 & 23.8 & 21.0 & 19.7 & 18.2 & 17.5 & 0.14 & star \\
GQ125126+261346 & 12:51:26.6 & +26:13:46.9 & 19.6 & 19.7 & 19.4 & 19.3 & 19.4 & 1.11 & 1.43 \\
GQ124943+261629 & 12:49:43.8 & +26:16:29.7 & 18.9 & 18.9 & 18.7 & 18.5 & 18.5 & 0.34 & 1.84 \\
GQ124959+261949 & 12:49:59.0 & +26:19:49.3 & 21.1 & 20.2 & 19.9 & 19.7 & 19.7 & - & - \\
GQ125031+261953 & 12:50:31.6 & +26:19:53.7 & 19.6 & 19.4 & 19.3 & 19.4 & 19.4 & 1.00 & 0.88P \\
GQ125027+263048 & 12:50:27.3 & +26:30:48.1 & 20.7 & 19.5 & 19.0 & 18.8 & 18.7 & - & - \\
GQ125248+263805 & 12:52:48.3 & +26:38:05.6 & 24.8 & 21.6 & 20.1 & 19.0 & 18.3 & 0.08 & - \\
GQ125150+263900 & 12:51:50.7 & +26:39:00.1 & 17.7 & 17.6 & 17.4 & 17.2 & 17.1 & 0.90 & 1.91 \\
GQ125133+264550 & 12:51:33.7 & +26:45:50.4 & 20.0 & 19.9 & 19.4 & 19.4 & 19.5 & 1.42 & 1.28P \\
GQ125209+265018 & 12:52:09.6 & +26:50:18.5 & 21.9 & 20.4 & 19.7 & 19.5 & 19.3 & 0.41 & 3.44 \\
GQ124902+262014 & 12:49:02.2 & +26:20:14.7 & 22.6 & 20.8 & 19.7 & 19.1 & 18.8 & - & - \\
GQ125008+263658 & 12:50:09.0 & +26:36:58.6 & 19.7 & 19.3 & 19.3 & 19.2 & 19.2 & 1.05 & 0.68P \\
GQ124835+263759 & 12:48:35.0 & +26:37:59.8 & 20.1 & 20.1 & 19.9 & 19.8 & 19.9 & 1.28 & 1.15P \\
GQ124820+264225 & 12:48:20.6 & +26:42:25.1 & 20.4 & 19.4 & 19.4 & 19.4 & 19.6 & - & - \\
GQ124926+264101 & 12:49:26.1 & +26:41:01.4 & 19.7 & 19.3 & 19.2 & 19.2 & 19.2 & 1.10 & 0.67 \\
GQ124921+264446 & 12:49:21.2 & +26:44:46.8 & 19.9 & 19.9 & 19.6 & 19.6 & 18.9 & 0.99 & 0.31P \\
GQ125054+264353 & 12:50:55.0 & +26:43:53.6 & 20.1 & 19.8 & 19.6 & 19.2 & 19.2 & 1.22 & 1.59P \\
GQ125015+264337 & 12:50:16.0 & +26:43:37.6 & 18.5 & 18.4 & 18.4 & 18.2 & 18.3 & 1.38 & 1.79 \\
GQ125149+265349 & 12:51:49.2 & +26:53:50.0 & 23.1 & 21.0 & 19.8 & 19.3 & 19.1 & 0.50 & star \\
GQ124945+264953 & 12:49:45.5 & +26:49:53.3 & 25.0 & 21.2 & 19.8 & 19.1 & 18.8 & 0.25 & - \\
GQ124941+265314 & 12:49:41.0 & +26:53:14.8 & 20.5 & 19.9 & 19.7 & 19.7 & 19.6 & - & 2.46P \\
GQ125023+265328 & 12:50:23.6 & +26:53:28.7 & 19.8 & 19.4 & 19.0 & 18.9 & 18.9 & 1.08 & 1.49 \\
GQ124935+265906 & 12:49:35.2 & +26:59:06.4 & 25.0 & 21.4 & 19.9 & 19.1 & 18.7 & 0.15 & - \\
GQ125006+265939 & 12:50:06.3 & +26:59:39.9 & 19.9 & 19.8 & 19.5 & 19.7 & 19.7 & 1.20 & 0.97 \\
GQ125421+265454 & 12:54:21.0 & +26:54:54.7 & 20.2 & 20.0 & 19.6 & 19.6 & 19.6 & 1.16 & 1.30P \\
GQ125429+270003 & 12:54:29.8 & +27:00:03.8 & 19.7 & 19.5 & 19.4 & 19.1 & 19.2 & 1.10 & 1.64 \\
GQ125312+265355 & 12:53:12.1 & +26:53:55.7 & 24.6 & 21.5 & 20.0 & 18.9 & 18.4 & - & - \\
GQ125302+270519 & 12:53:02.3 & +27:05:19.9 & 22.6 & 20.4 & 20.1 & 20.0 & 20.0 & 1.02 & 2.99 \\
GQ125320+270607 & 12:53:20.1 & +27:06:07.4 & 20.1 & 19.7 & 19.1 & 18.8 & 18.9 & 1.32 & 1.15 \\
GQ125528+271330 & 12:55:28.8 & +27:13:30.1 & 20.0 & 19.8 & 19.5 & 19.7 & 19.7 & 1.30 & 1.03 \\
GQ125357+271630 & 12:53:57.5 & +27:16:30.6 & 22.2 & 19.9 & 18.9 & 18.5 & 18.3 & - & - \\
GQ125359+272014 & 12:53:59.8 & +27:20:14.8 & 20.6 & 20.3 & 20.3 & 19.9 & 20.4 & 1.58 & 0.08P \\
GQ125158+271304 & 12:51:58.6 & +27:13:04.1 & 21.6 & 20.2 & 19.7 & 19.5 & 19.6 & - & - \\
GQ125300+271234 & 12:53:00.1 & +27:12:34.2 & 19.5 & 19.2 & 19.1 & 19.0 & 19.0 & 1.42 & 1.48 \\
GQ125241+272550 & 12:52:41.1 & +27:25:50.8 & 21.4 & 21.2 & 20.7 & 20.1 & 19.9 & 1.22 & 0.22P \\
GQ125134+272000 & 12:51:34.4 & +27:20:00.2 & 20.9 & 19.9 & 19.7 & 19.7 & 19.7 & - & - \\
GQ125107+271451 & 12:51:08.0 & +27:14:51.5 & 20.2 & 19.8 & 19.4 & 19.2 & 18.8 & 0.79 & 0.38P \\
GQ125206+272717 & 12:52:07.0 & +27:27:17.5 & 18.7 & 18.6 & 18.6 & 18.4 & 18.5 & 1.22 & 1.68 \\
GQ125211+272803 & 12:52:11.9 & +27:28:03.8 & 20.7 & 20.6 & 20.4 & 20.0 & 20.1 & 1.41 & 1.87P \\
GQ125332+272225 & 12:53:32.1 & +27:22:25.1 & 19.6 & 19.3 & 19.2 & 18.9 & 18.8 & 1.39 & 1.63 \\
GQ125320+272116 & 12:53:20.3 & +27:21:16.0 & 18.0 & 17.8 & 18.0 & 17.9 & 17.8 & 1.04 & 0.51 \\
GQ125353+273405 & 12:53:53.7 & +27:34:05.9 & 21.4 & 20.3 & 19.8 & 19.5 & 19.5 & 0.39 & - \\
GQ125308+273331 & 12:53:08.5 & +27:33:31.2 & 23.8 & 21.6 & 20.4 & 19.2 & 18.5 & -0.09 & - \\
GQ125327+273733 & 12:53:27.1 & +27:37:33.1 & 19.9 & 19.9 & 19.8 & 19.6 & 19.4 & 1.08 & 1.96P \\
GQ125257+274542 & 12:52:58.0 & +27:45:42.5 & 19.0 & 18.7 & 18.6 & 18.5 & 18.4 & 1.32 & 2.00 \\
GQ124816+264712 & 12:48:16.3 & +26:47:12.3 & 18.1 & 18.1 & 18.2 & 17.9 & 17.8 & 1.35 & 1.86 \\
GQ124746+264709 & 12:47:46.5 & +26:47:09.8 & 21.2 & 20.2 & 20.0 & 19.8 & 19.7 & - & - \\
GQ124816+265235 & 12:48:16.4 & +26:52:35.3 & 20.1 & 19.5 & 19.4 & 19.4 & 19.3 & - & 2.51 \\
GQ124832+265312 & 12:48:32.8 & +26:53:12.7 & 18.2 & 17.8 & 17.9 & 17.7 & 17.8 & 1.11 & 0.59 \\
GQ124804+265836 & 12:48:05.0 & +26:58:36.1 & 24.8 & 22.2 & 20.7 & 19.1 & 18.3 & 0.15 & - \\
GQ124715+265528 & 12:47:15.9 & +26:55:28.1 & 22.4 & 20.1 & 18.9 & 18.6 & 18.3 & - & - \\
GQ124721+265728 & 12:47:21.3 & +26:57:28.9 & 21.3 & 20.2 & 19.7 & 19.5 & 19.4 & - & - \\
GQ124725+270114 & 12:47:25.8 & +27:01:14.1 & 19.8 & 19.3 & 19.2 & 19.2 & 19.0 & 1.07 & 0.80 \\
GQ124731+270622 & 12:47:31.5 & +27:06:22.2 & 20.9 & 20.0 & 19.8 & 19.7 & 19.8 & - & - \\
GQ124734+270615 & 12:47:34.4 & +27:06:15.2 & 23.7 & 21.8 & 20.3 & 19.0 & 18.3 & 0.05 & - \\
GQ124922+265938 & 12:49:22.2 & +26:59:38.9 & 21.3 & 20.1 & 19.7 & 19.5 & 19.5 & - & - \\
GQ124955+271335 & 12:49:55.7 & +27:13:35.3 & 19.5 & 19.4 & 19.3 & 19.2 & 19.2 & 1.37 & 1.47P \\
GQ124846+271722 & 12:48:46.1 & +27:17:22.7 & 19.8 & 19.4 & 19.1 & 18.8 & 18.7 & 1.29 & 1.51 \\
GQ124814+271603 & 12:48:14.5 & +27:16:03.6 & 22.8 & 20.4 & 19.2 & 18.6 & 18.3 & -0.15 & - \\
GQ124809+271650 & 12:48:09.4 & +27:16:50.1 & 20.2 & 20.0 & 19.7 & 19.8 & 19.7 & 1.16 & 0.95P \\
GQ124925+272634 & 12:49:26.0 & +27:26:34.5 & 18.7 & 18.5 & 18.3 & 18.4 & 18.5 & 1.37 & 1.16 \\
GQ124728+272742 & 12:47:28.6 & +27:27:42.5 & 21.2 & 20.7 & 20.3 & 19.9 & 19.9 & - & 0.46P \\
GQ124752+273018 & 12:47:52.3 & +27:30:18.9 & 20.4 & 20.0 & 19.9 & 19.9 & 19.8 & 0.91 & 0.91 \\
GQ125043+271934 & 12:50:44.0 & +27:19:34.3 & 20.3 & 20.2 & 20.1 & 20.2 & 20.0 & 0.83 & 0.93P \\
GQ125050+272448 & 12:50:50.2 & +27:24:48.4 & 24.6 & 21.5 & 20.1 & 18.7 & 17.9 & 0.06 & - \\
GQ125039+272904 & 12:50:39.4 & +27:29:04.4 & 20.7 & 20.1 & 19.6 & 19.1 & 18.9 & 1.15 & 1.89 \\
GQ125104+273341 & 12:51:04.1 & +27:33:41.2 & 20.2 & 20.2 & 19.8 & 19.8 & 20.1 & 1.24 & 1.15P \\
GQ124958+273317 & 12:49:58.2 & +27:33:17.5 & 20.0 & 19.8 & 19.8 & 19.7 & 19.6 & 1.26 & 1.72P \\
GQ125013+273256 & 12:50:13.7 & +27:32:56.4 & 19.9 & 19.8 & 19.6 & 19.5 & 19.7 & 1.14 & 1.15P \\
GQ125008+274001 & 12:50:08.6 & +27:40:01.5 & 21.1 & 20.1 & 19.9 & 19.7 & 19.8 & 0.32 & - \\
GQ125042+274012 & 12:50:42.4 & +27:40:13.0 & 20.2 & 20.0 & 19.7 & 19.6 & 19.6 & 1.41 & 1.48P \\
GQ125033+274519 & 12:50:33.6 & +27:45:19.1 & 20.0 & 19.8 & 19.7 & 19.6 & 19.7 & 1.41 & 1.48P \\
GQ125019+274443 & 12:50:19.2 & +27:44:43.1 & 20.5 & 19.7 & 19.4 & 19.1 & 18.7 & 1.02 & 2.67 \\
GQ125007+274709 & 12:50:07.9 & +27:47:09.4 & 23.6 & 21.4 & 19.8 & 18.7 & 18.0 & 0.28 & - \\
GQ125202+275018 & 12:52:02.1 & +27:50:18.6 & 19.3 & 18.8 & 18.7 & 18.7 & 18.5 & 1.17 & 2.48 \\
GQ125312+275524 & 12:53:12.1 & +27:55:24.6 & 19.7 & 19.5 & 19.3 & 19.3 & 19.2 & 1.18 & 1.00P \\
GQ125230+275410 & 12:52:30.1 & +27:54:10.0 & 19.4 & 19.3 & 19.0 & 19.1 & 19.1 & 1.03 & 1.04P \\
GQ125241+275942 & 12:52:42.0 & +27:59:42.7 & 24.0 & 21.8 & 20.6 & 19.1 & 18.3 & 0.10 & - \\
GQ125239+280127 & 12:52:39.2 & +28:01:27.2 & 19.6 & 19.6 & 19.5 & 19.4 & 19.2 & 1.29 & 2.04P \\
GQ125249+280356 & 12:52:49.1 & +28:03:56.4 & 19.4 & 19.3 & 19.2 & 19.4 & 19.2 & 1.01 & 0.93P \\
GQ125112+275222 & 12:51:12.3 & +27:52:22.4 & 19.0 & 18.8 & 18.5 & 18.5 & 18.6 & 1.24 & 1.04 \\
GQ125159+280446 & 12:51:59.1 & +28:04:46.3 & 23.1 & 21.4 & 19.9 & 18.8 & 18.2 & 0.01 & - \\
GQ124901+274330 & 12:49:01.8 & +27:43:30.2 & 21.1 & 20.2 & 19.8 & 19.7 & 19.5 & - & - \\
GQ124915+274554 & 12:49:15.9 & +27:45:54.8 & 19.6 & 19.3 & 19.3 & 19.2 & 18.9 & 1.43 & 2.16P \\
GQ124955+275657 & 12:49:55.4 & +27:56:57.7 & 21.2 & 19.9 & 19.3 & 19.0 & 18.9 & 0.49 & - \\
GQ124958+275703 & 12:49:58.1 & +27:57:03.7 & 19.8 & 19.6 & 19.4 & 19.1 & 19.0 & 1.34 & 1.91P \\
GQ125005+280041 & 12:50:05.2 & +28:00:41.9 & 21.0 & 20.2 & 20.0 & 19.8 & 19.7 & - & - \\

\hline\noalign{\smallskip}

\end{longtable}
\centering
\begin{minipage}{5.3in}
\tablefoot{
Right ascension and declination are from the \textit{Gaia} DR2
catalogue. Optical magnitudes are in the AB system and are from the SDSS
photometric data. The mid-infrared $W1-W2$ colours are from the AllWISE catalogue
and are in the Vega system. 
Spectroscopic redshifts are based on the already identified
quasars in the SDSS-DR14 quasar survey \citep{Paris17} or the NED database 
\citep[specifically from][]{Crampton87}. Photometric redshifts
(marked by a "P") are from \cite{Richards09}.
}

\end{minipage}
\end{longtab}

\section{Selection efficiency and completeness} \label{sec:res}

\begin{figure} [!t]
	\centering
	\epsfig{file=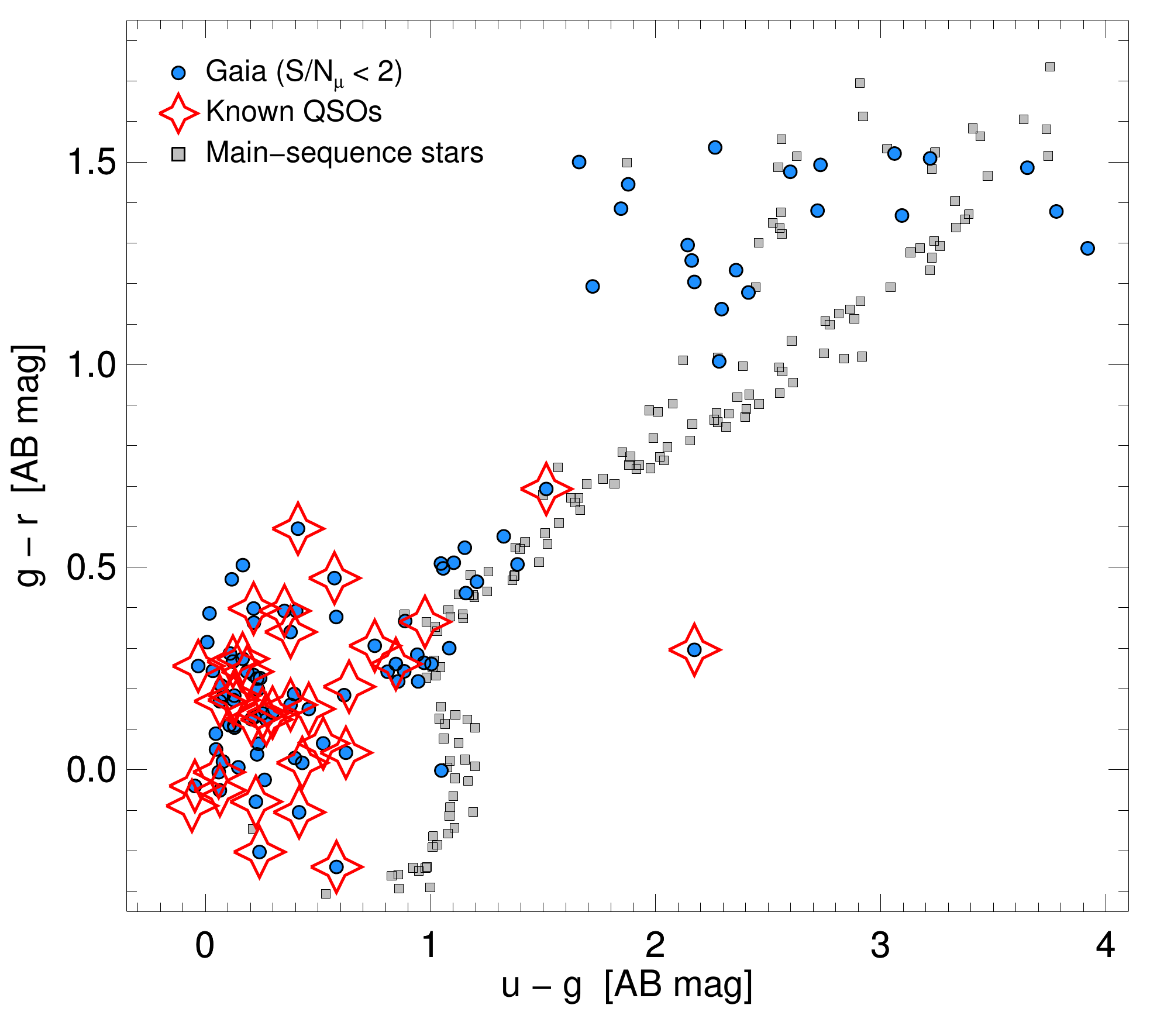,width=\columnwidth}
	\caption{Optical colour-colour plots of the \textit{WISE}-detected Gaia point sources with proper motions and $18 < G < 20$ mag (black dots) within one degree of the NGP. \textit{Gaia} point sources with zero proper motions are represented by the blue dots and the spectroscopically confirmed quasars are shown by the red star symbols. Typical stellar colours are shown as grey dots.}
	\label{fig:sdsscol}
\end{figure}

We now investigate the location of the \textit{Gaia} sources with zero proper
motions in optical colour-colour space. By doing so, we can examine whether
these candidate quasars have, e.g., ultraviolet excess typical of unobscured,
low-$z$ quasars \citep[e.g.][]{Sandage65,Schmidt83}. About 70\% of the zero
proper motion sources have blue ($u-g < 1$) colours (see Fig.~\ref{fig:sdsscol}). 
For quasars at $z \gtrsim
2.2$, the Lyman-$\alpha$ emission line will move out of the $u$-band, such that
the quasars appear redder in $u-g$ colour space. At red $g-r$ colours ($g-r >
1$) the zero proper motion sources have optical colours consistent with M or G
dwarf stars. While some of these are likely to be stellar contaminations,
removing these candidates will also exclude dust-reddened quasars
and broad absorption line (BAL) quasars from the sample, which are
found to have very red optical colours and to be systematically missing in
most existing quasar samples \citep{Fynbo13,Krogager15,Ross2015,Krawczyk2015,Krogager16}.

\begin{figure} [!t]
	\centering
	\epsfig{file=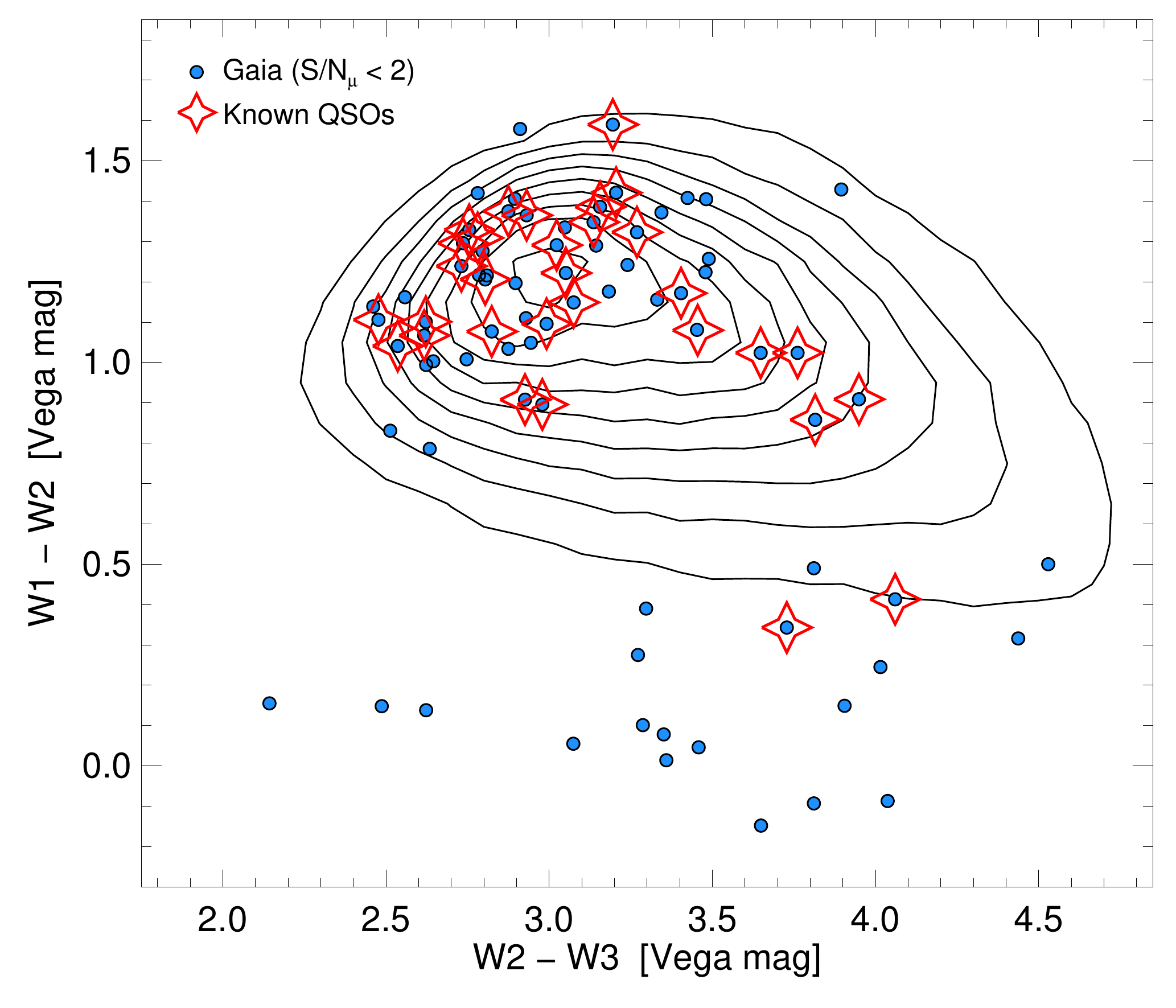,width=\columnwidth}
	\caption{\textit{WISE} colour-colour plot of \textit{Gaia} point sources with zero proper motions (blue dots) and SDSS DR14 quasars (red star symbols) within one degree of the NGP. Overplotted are contours of the full SDSS-DR14 quasar sample with mid-infrared counterparts in the AllWISE catalogue.}
	\label{fig:wisecol}
\end{figure}

To assess the efficiency of our selection we cross-match our sample of
\textit{Gaia} sources with zero proper motions to the all-sky mid-infrared
survey based on the \textit{WISE} satellite \citep[AllWISE;][]{Wright10}.
Mid-infrared selection of quasars is efficient at separating stars and
galaxies from quasars and is not affected by dust extinction while also being
sensitive to high-redshift quasars. Of the 100 \textit{Gaia} point sources with
zero proper motions, we identify 76 of the counterparts in the AllWISE catalogue
within 1 arcsec. This cross-match might introduce a bias
excluding quasars with weak infrared emission. Stellar contaminants will also
have weak infrared emission, however, and we find that of the 24 sources
excluded in this approach, roughly half have a significant ultraviolet excess
whereas the other half have optical colours consistent with the main-sequence
stellar track. In Fig.~\ref{fig:wisecol} we show the zero proper motion
\textit{Gaia} sources in mid-infrared colour-colour space. Overplotted are
contours of the SDSS-DR14Q sample for which \textit{WISE} photometry exists. A
simple color criterion of $W1-W2 > 0.8$ has been found to be robust in
identifying quasars at most redshifts \citep{Stern12}. In our sample of zero
proper motion sources with \textit{WISE} photometry, 55 (70\%) have $W1-W2 >
0.8$ (of which 29 are already identified quasars). We consider the remaining 26
sources as high-likelihood quasars. All these have also been photometrically identifed as quasars by \citep{Richards09}, and we list their estimated photometric redshifts in Table~\ref{tab:1} as well, marked by a "P". We note, however, that at $W1-W2 < 0.8$,
two spectroscopically confirmed quasars have also been observed, one being a high-$z$ quasar with optical colours consistent with known quasars in this redshift range and the other being a typical UV-excess quasar. We therefore
consider the sources with zero proper motions and $W1-W2 < 0.8$ as possible contaminants
(excluding the two already known quasars). We then infer a conservative selection efficiency
of $N_{\mathrm{QSO}}/N_{\mathrm{star}} \gtrsim 75\%$. This is a
lower limit due to the population of quasars with blue $W1-W2$
colours that also populates our sample.

\begin{figure} [!t]
	\centering
	\epsfig{file=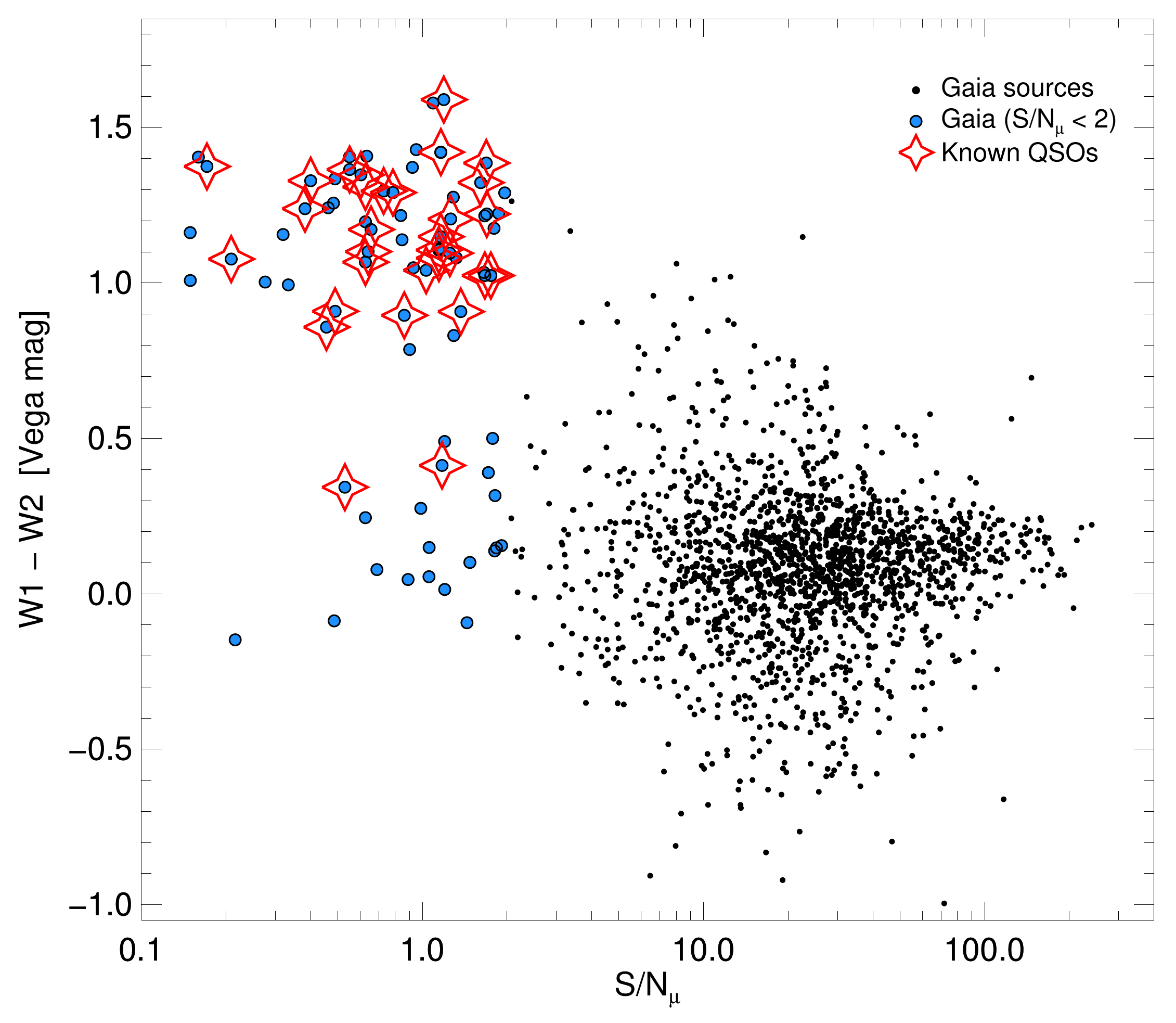,width=8cm}
	\caption{W1-W2 colour as a function of S/N$_{\mu}$ of the \textit{WISE}-detected Gaia point sources with proper motions and $18 < G < 20$ mag (black dots) within one degree of the NGP. \textit{Gaia} point sources with zero proper motions are represented by blue dots and spectroscopically confirmed quasars are shown with red star symbols.}
	\label{fig:wisepm}
\end{figure}

We present our main result in Fig.~\ref{fig:wisepm} where we show the full sample of \textit{Gaia} sources with $18 < G < 20$ mag and within one degree of the NGP for which a counterpart in the AllWISE catalogue could be identified. It is clear from the figure that the majority of point sources selected on the basis of zero proper motions occupy a distinct region in S/N$_\mu$ -- \textit{WISE} colour parameter space. This demonstrates that selecting quasars as stationary sources on the sky is definitely feasible and has a high efficiency of $\gtrsim 75\%$. The completeness is close to 100\% within the defined magnitude limit, since all cosmological objects are selected without any prior assumptions on the spectral energy distributions.


\section{Discussion and conclusions}

We have here demonstrated the possibility to select quasars as
stationary objects in the {\it Gaia} DR2 data set. When observing fields well
away from the Galactic stellar disk (here the NGP) the contamination from stars is very modest (below 25\%) when targeting the most relevant magnitudes (here $18 < G < 20$). Hence, astrometric selection offers both a complete and clean 
selection of quasars.

This technique offers the possibility to take major steps ahead on some very
interesting problems relating to the quasar phenomenon. We will mention a
few examples here. First, getting a more 
complete picture of dust obscuration in quasar hosts will be possible with
a sample of quasars selected from proper motion. Second, the 
redshift dependence of the frequency of BAL quasars 
can be determined. Third, using a purely astrometrically selected sample
of quasars we can get an independent gauge of the metallicity distribution 
of intervening galaxies, in particular the damped Lyman-$\alpha$ absorbers. Fourth, the identification of quasars via zero proper motion also provides unbiased measures of number densities of various absorbers, such as C\,\textsc{iv}, Mg\,\textsc{ii}, or H\,\textsc{i}.
Such a sample will still be subject to a flux limit, but this is easier to
model than the combined effect of a flux limit and the effect of dust reddening
on the quasar selection efficiency in optical quasar surveys. 
We also note that the {\it Gaia} DR2 data have been applied to find new 
gravitationally lensed quasars \citep{Krone-Martins2018}.

An interesting case is the confirmed quasar SDSS J125209.59+265018.4
(GQ125209+265018 in Table~\ref{tab:1}). In Fig.~\ref{fig:sdsscol} this is
located as the object on the stellar track at $u-g = 1.5$. In
Fig.~\ref{fig:wisecol} it is one of the two sources with blue WISE colours at
$W1-W2 < 0.8$. This illustrates well the potential 
of selection of quasars from astrometry in finding quasars that are otherwise
difficult to photometrically identify.

When the full \textit{Gaia} data is released, the errors on the proper motions will
decrease and it will thus be easier to disentangle objects that are truly
stationary (quasars) and stars with low proper motions. This will also make it possible to search for stationary sources at even fainter magnitudes. Also, since Gaia
astrometry exists for most of the sky, this proper motion criteria could help
reduce the contamination in other quasar surveys. Since Gaia covers the full
sky, the selection can also be carried out for a large sample of sources --
however, with the caveat that the contamination from apparently stationary
stars increases significantly closer to the Galactic plane.  
We can also estimate the expected contamination of e.g. the WISE $W1-W2$ color
selection, where it can be seen from Fig.~\ref{fig:wisepm} that 15\% of the
sources with $W1-W2 > 0.8$ have significant proper motions at more than
5$\sigma$. 


\begin{acknowledgements}
KEH and PJ acknowledge support by a Project Grant (162948--051) from The Icelandic Research Fund. The Cosmic Dawn Center is funded by the DNRF. LC is supported by DFF -- 4090-00079.
\end{acknowledgements}


\bibliographystyle{aa}
\bibliography{ref}

\begin{appendix}
\onecolumn
\section{Thumbnail images of all sources}
Thumbnails of all point sources within one degree of the NGP with
proper motions consistent with zero (within 2$\sigma$) and $18 < G < 20$ mag.

\begin{figure*} [!b]
\centering
\epsfig{file=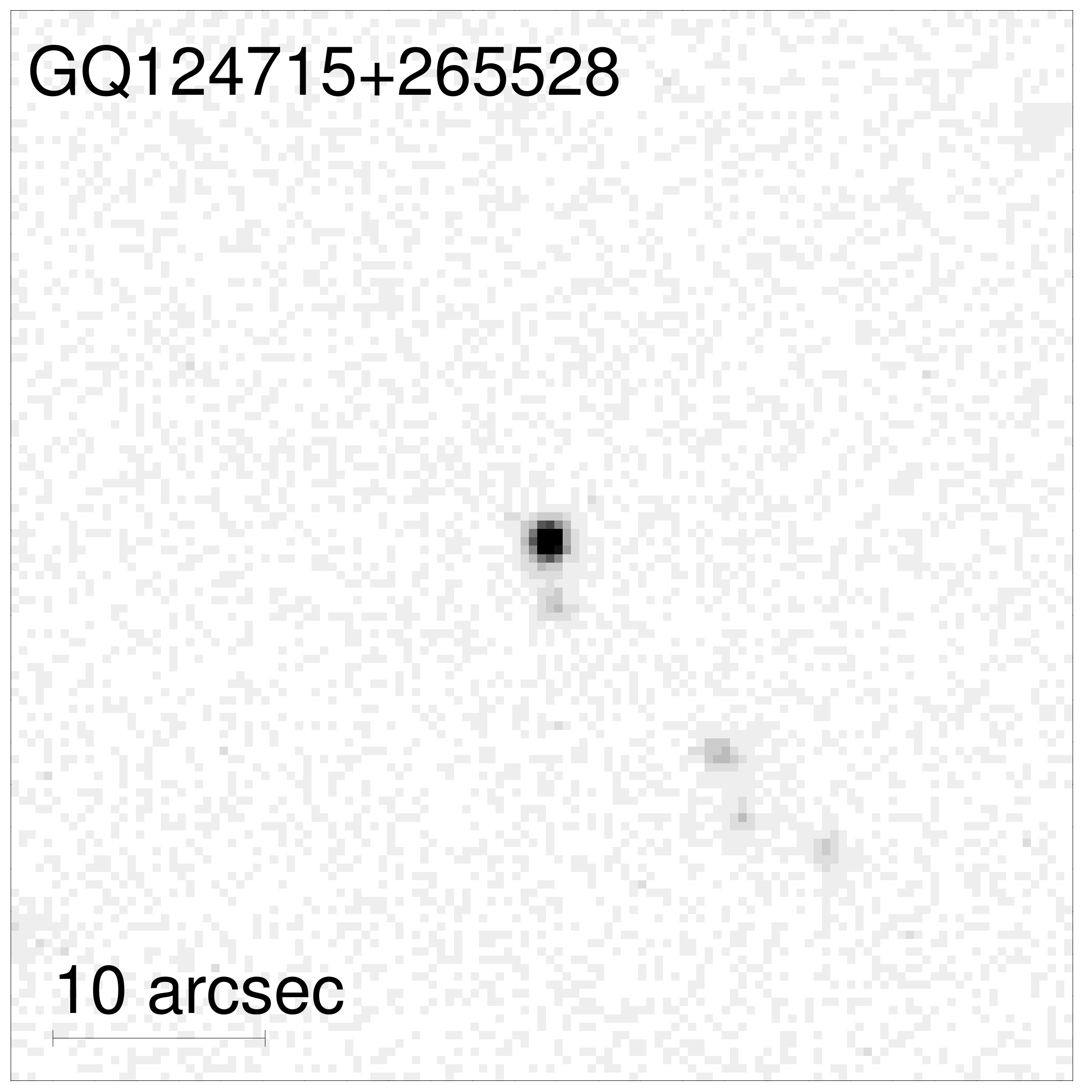,width=3.5cm} 
\epsfig{file=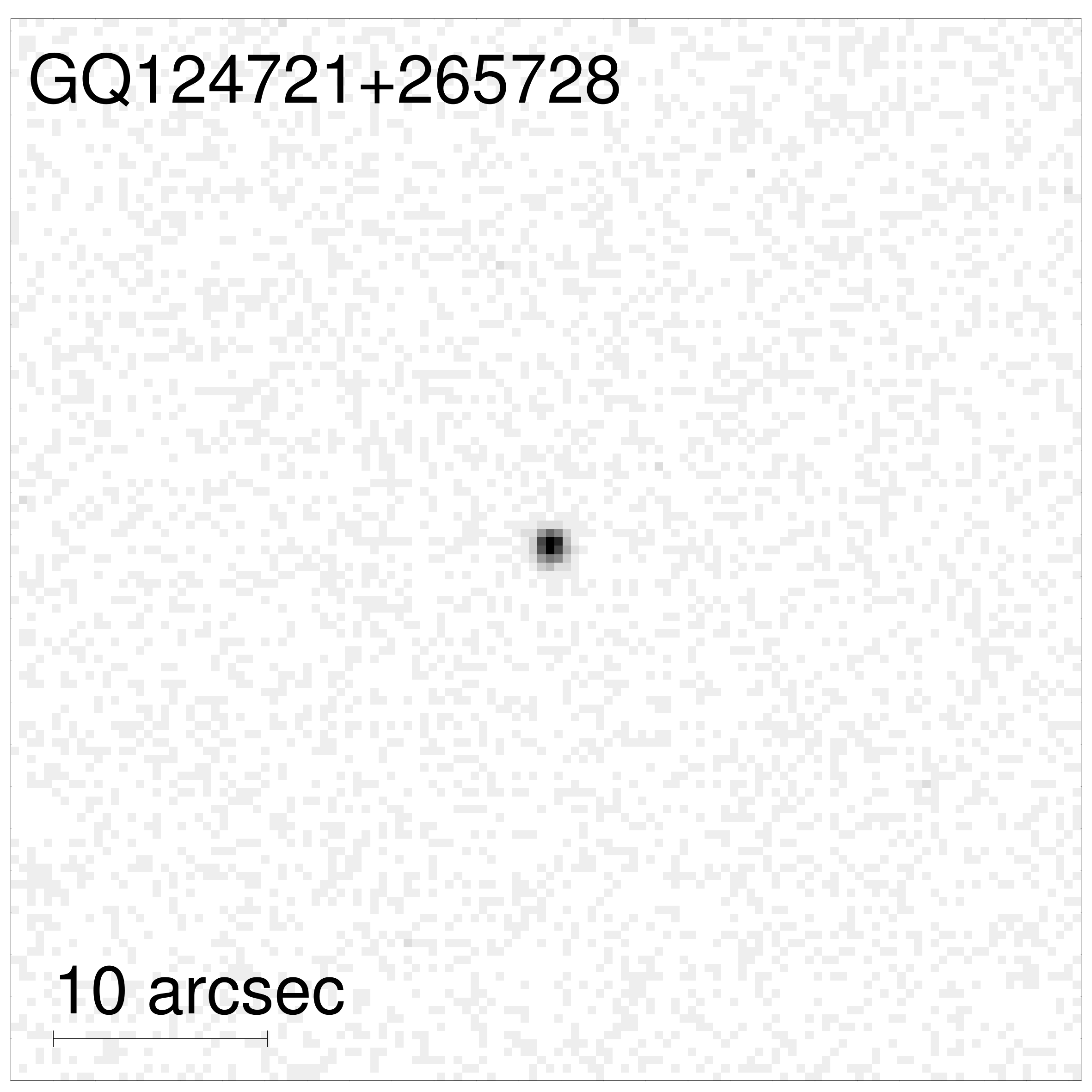,width=3.5cm} 
\epsfig{file=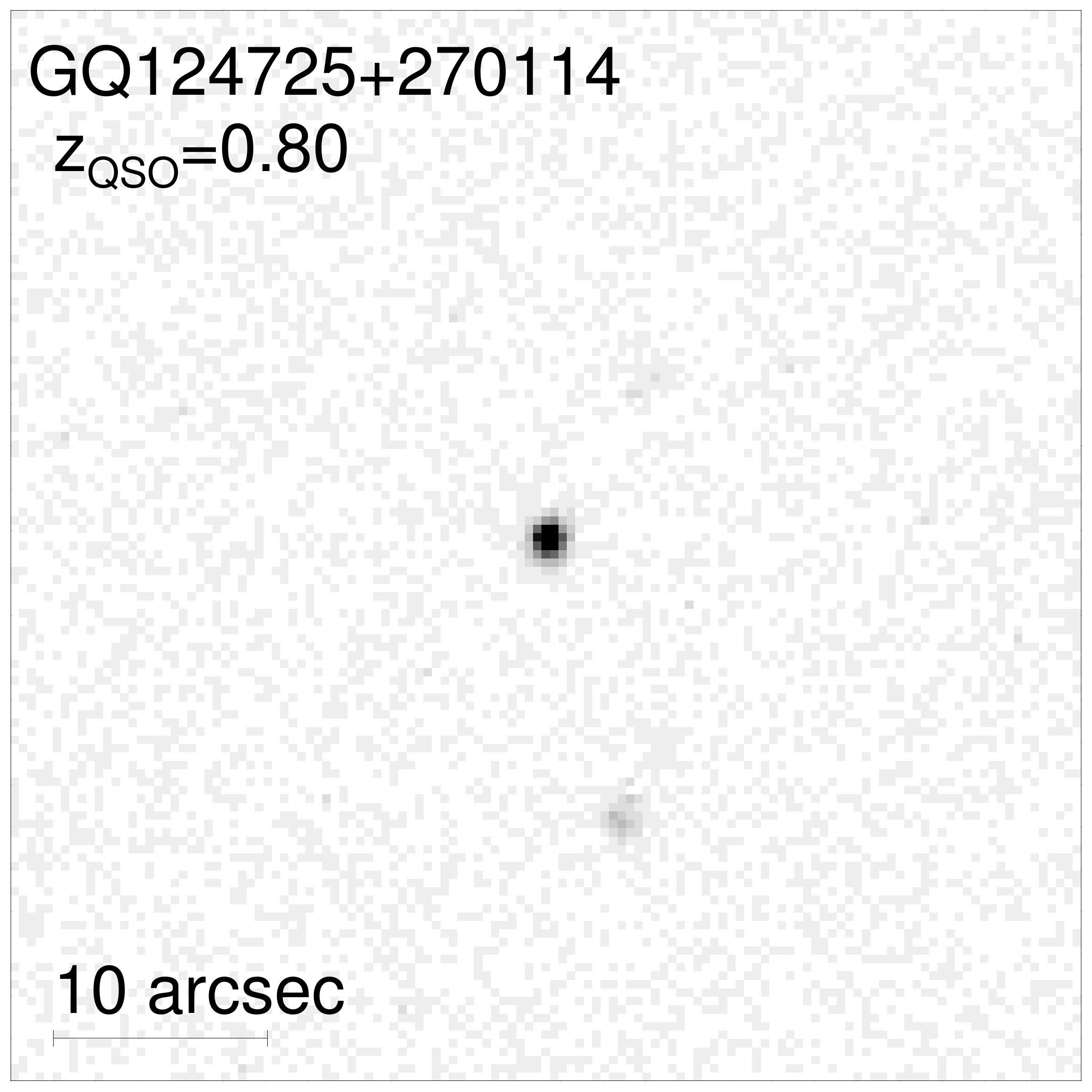,width=3.5cm} 
\epsfig{file=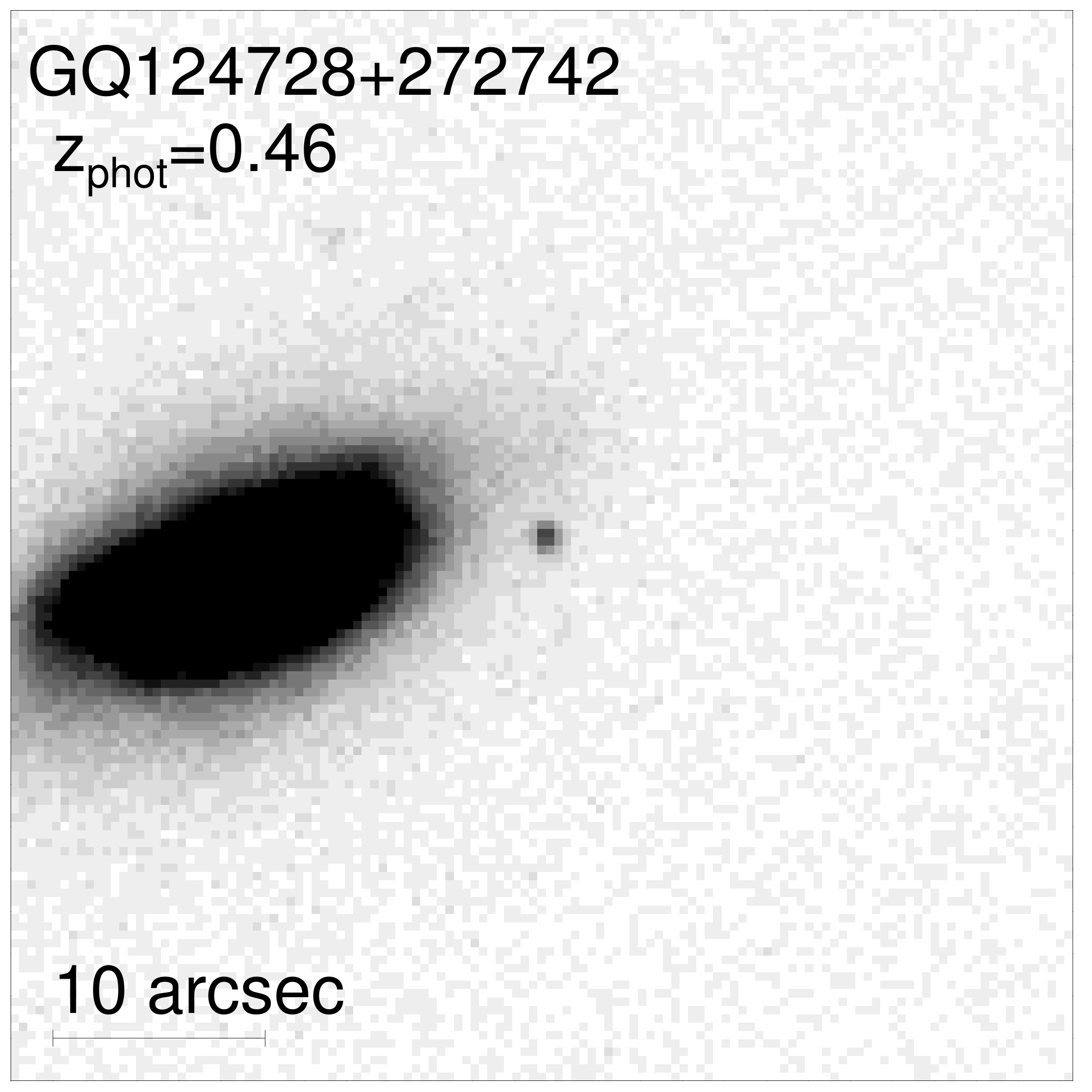,width=3.5cm} 
\epsfig{file=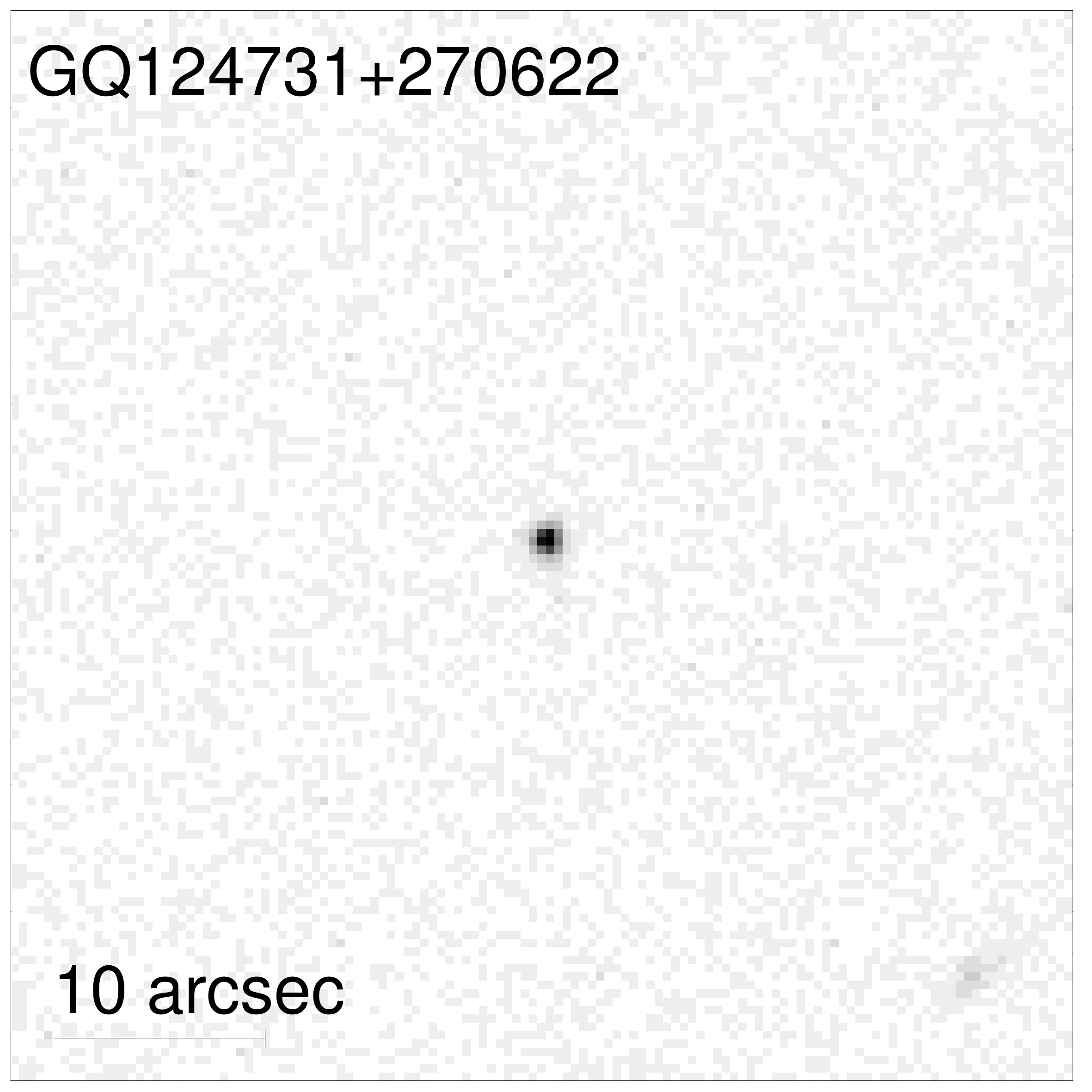,width=3.5cm} 
\epsfig{file=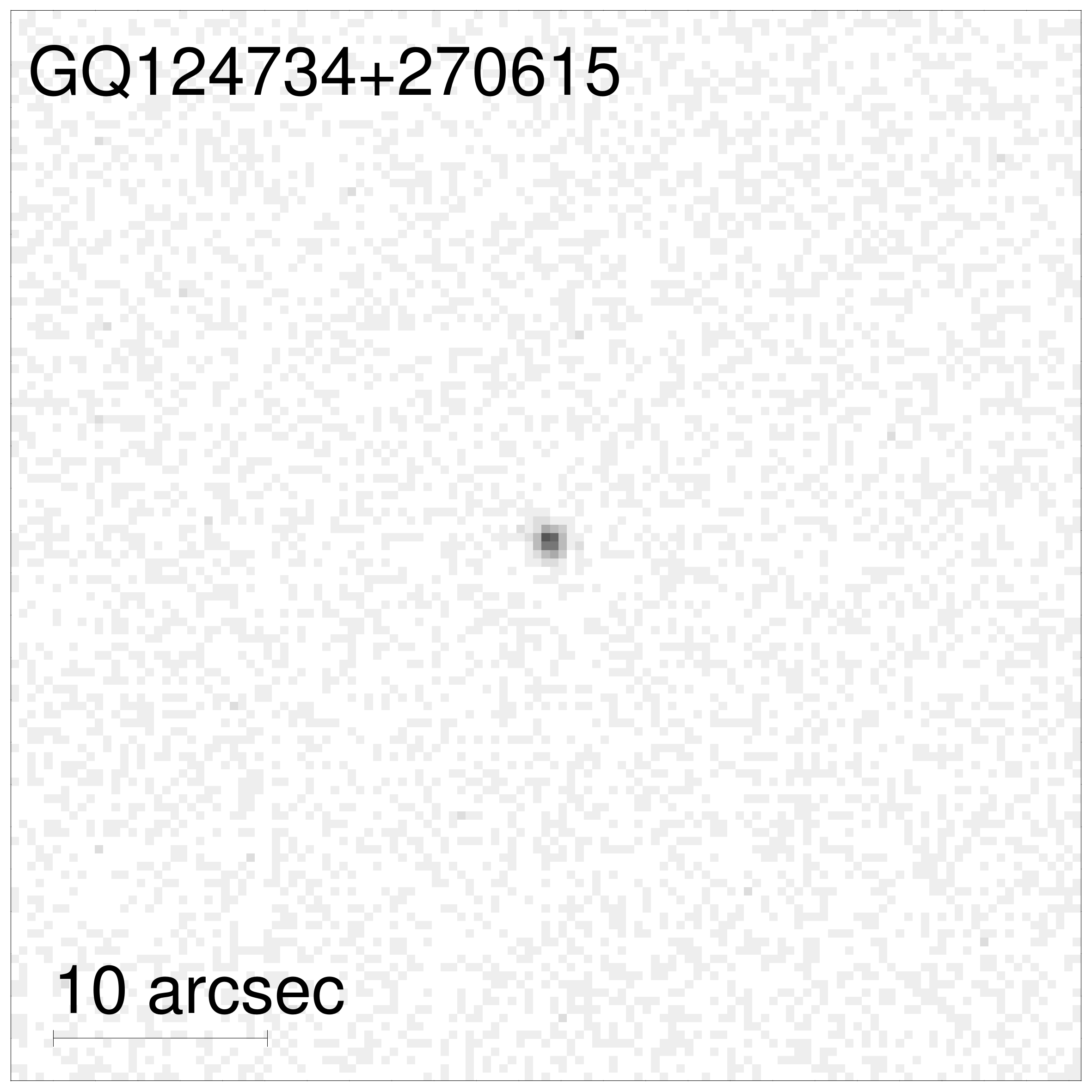,width=3.5cm} 
\epsfig{file=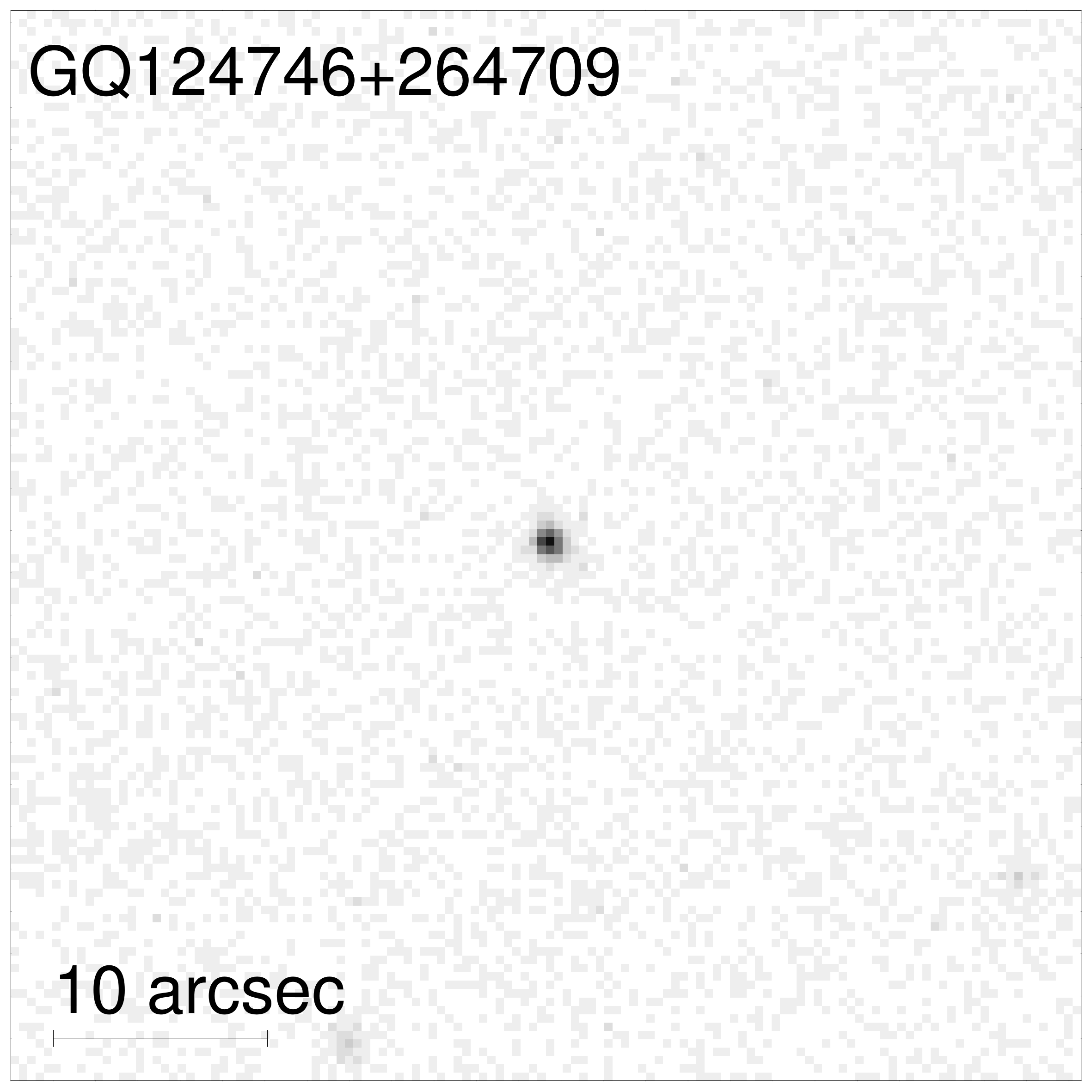,width=3.5cm} 
\epsfig{file=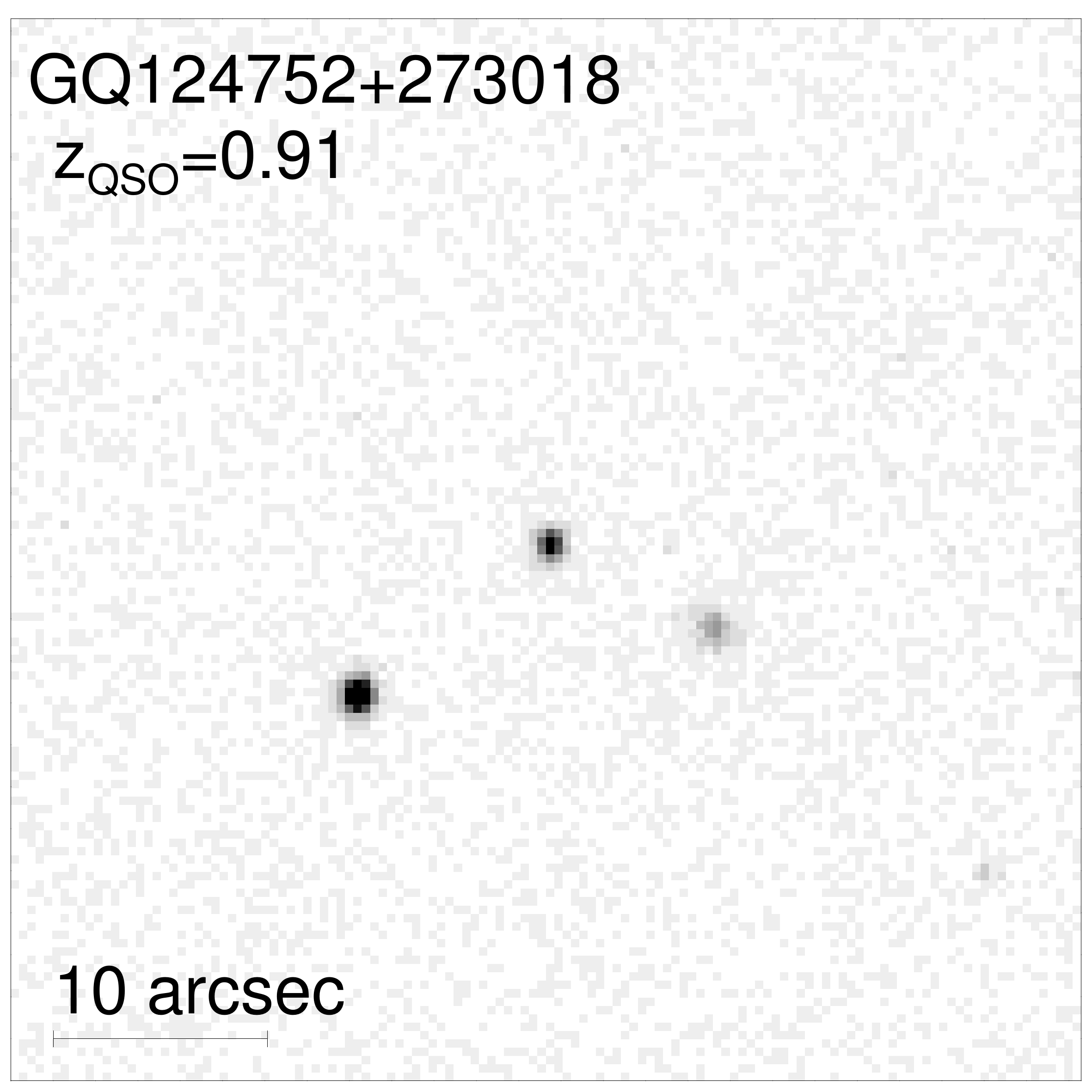,width=3.5cm} 
\epsfig{file=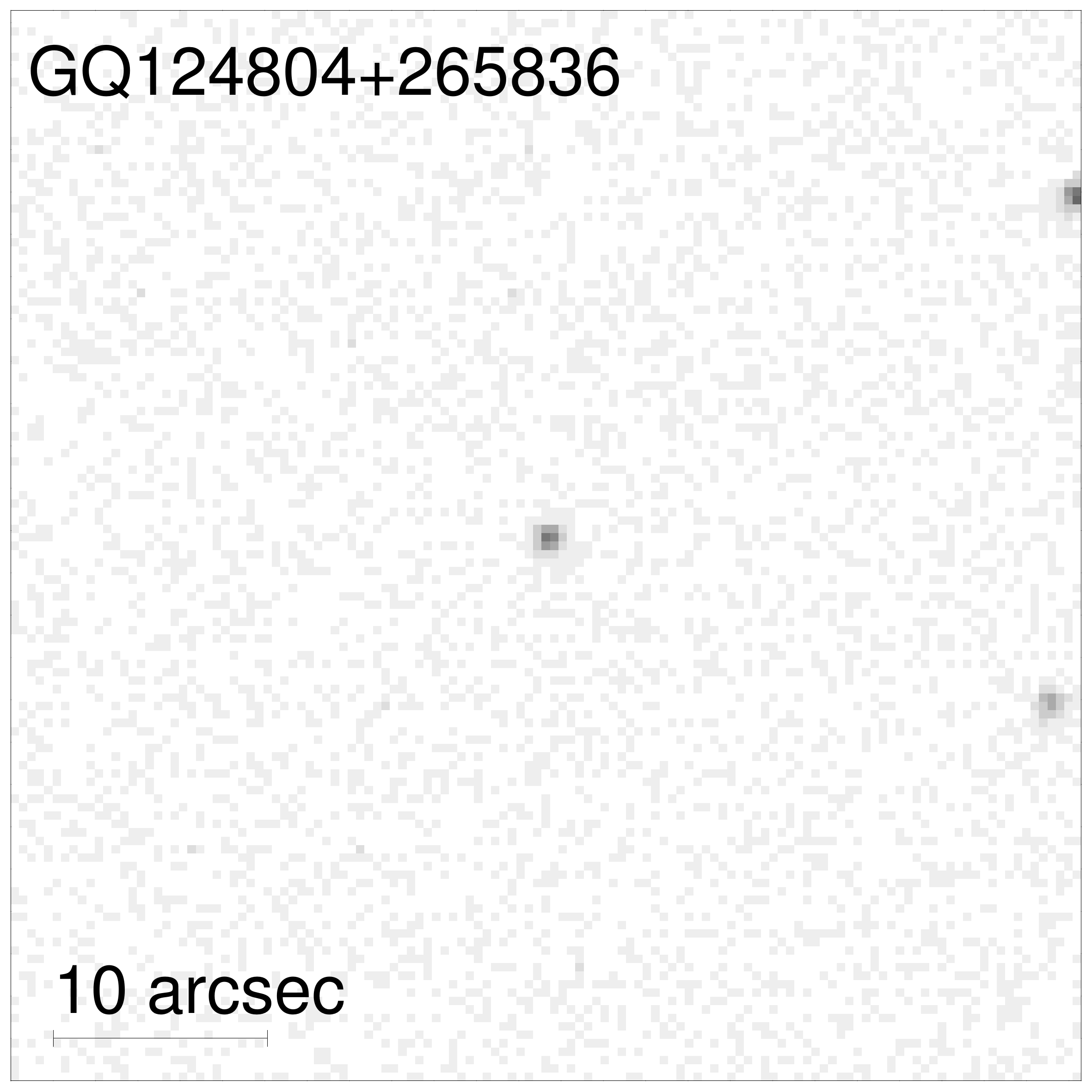,width=3.5cm} 
\epsfig{file=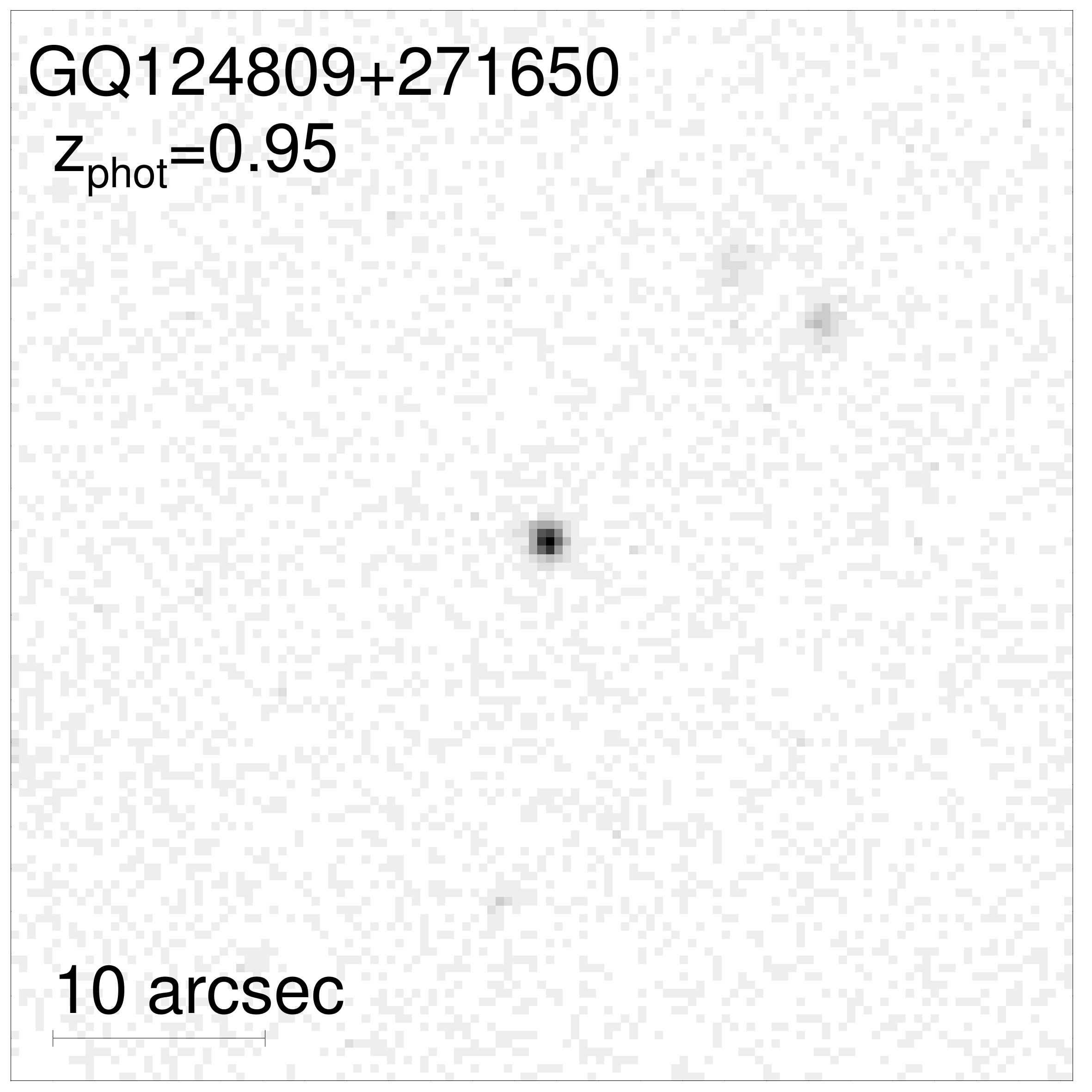,width=3.5cm} 
\epsfig{file=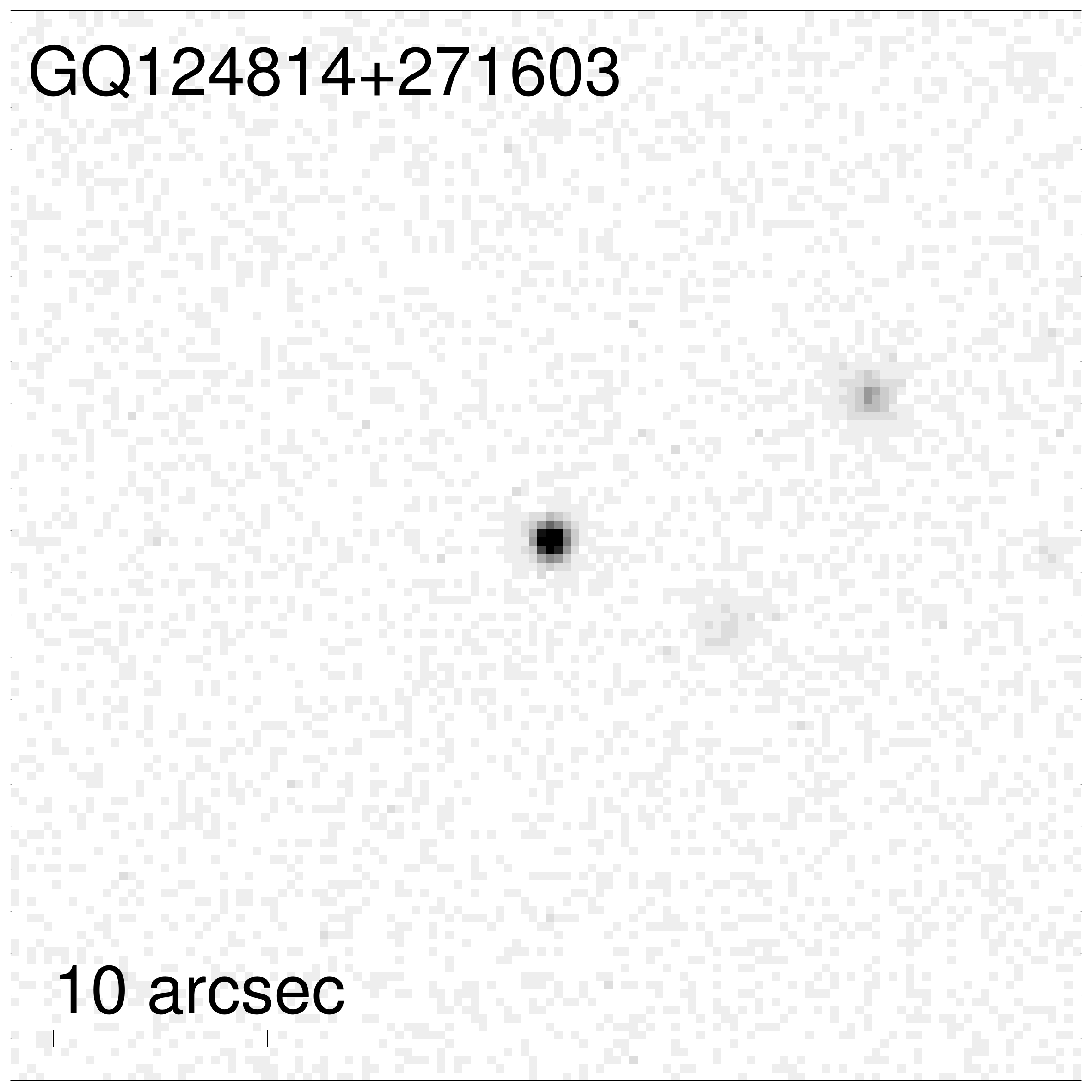,width=3.5cm} 
\epsfig{file=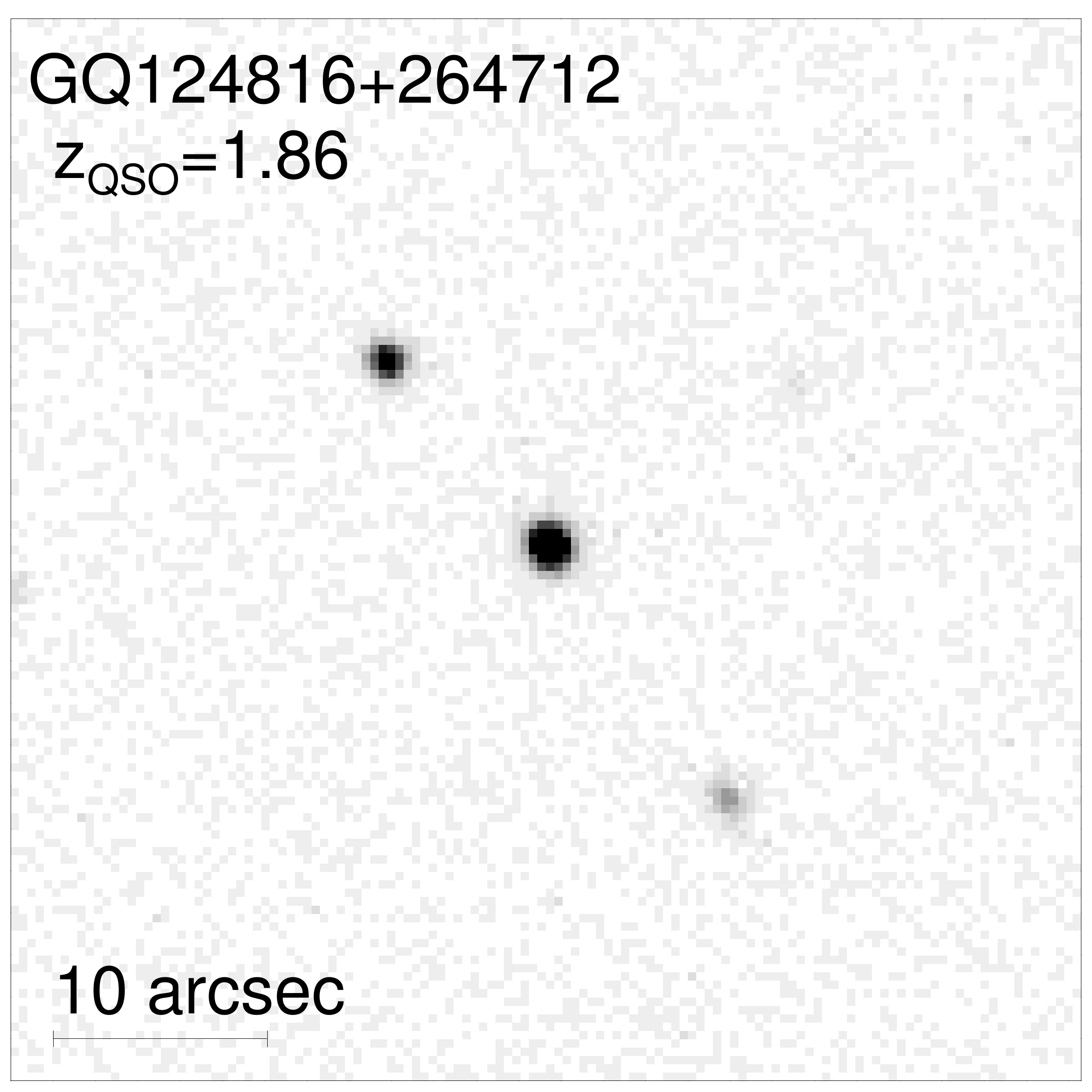,width=3.5cm} 
\epsfig{file=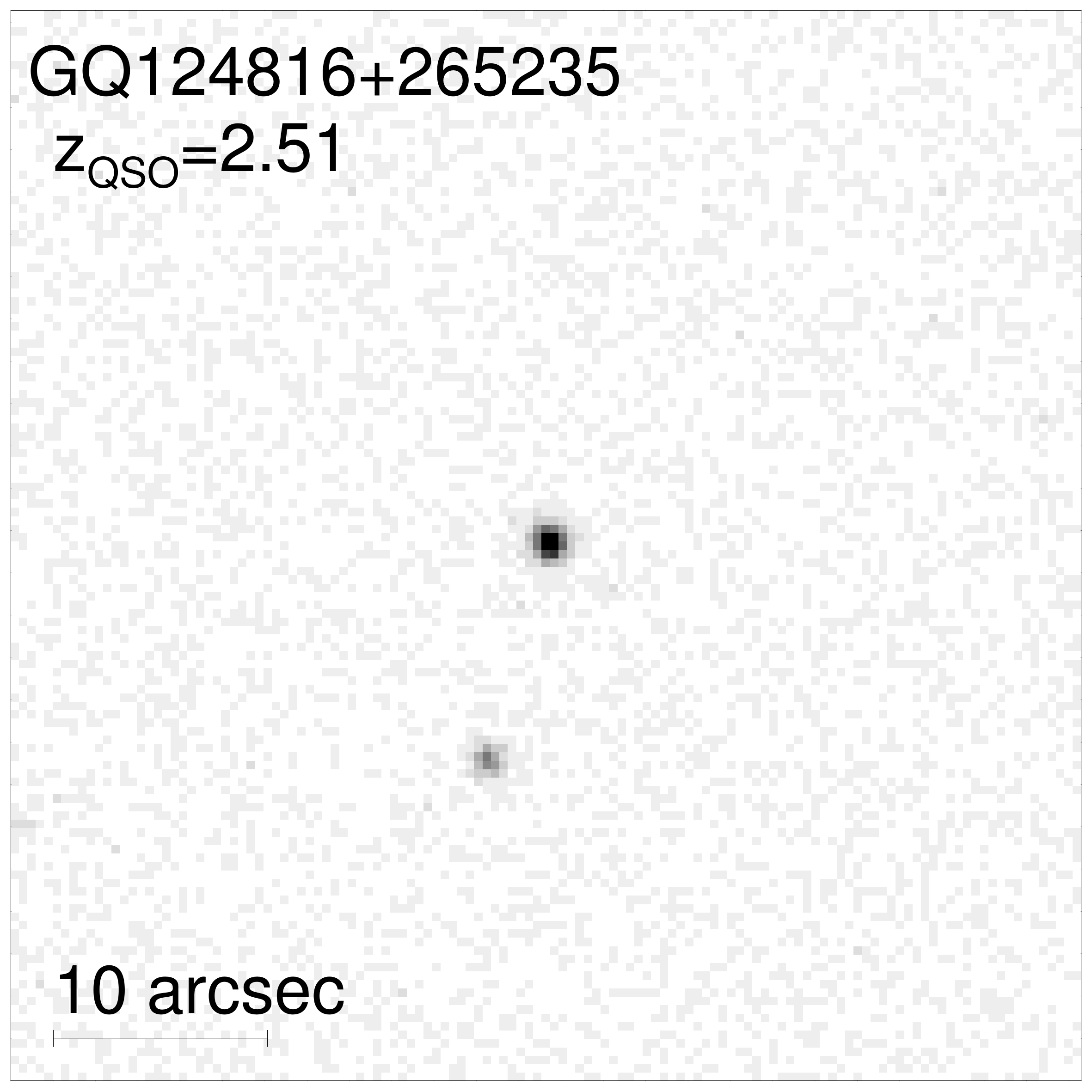,width=3.5cm} 
\epsfig{file=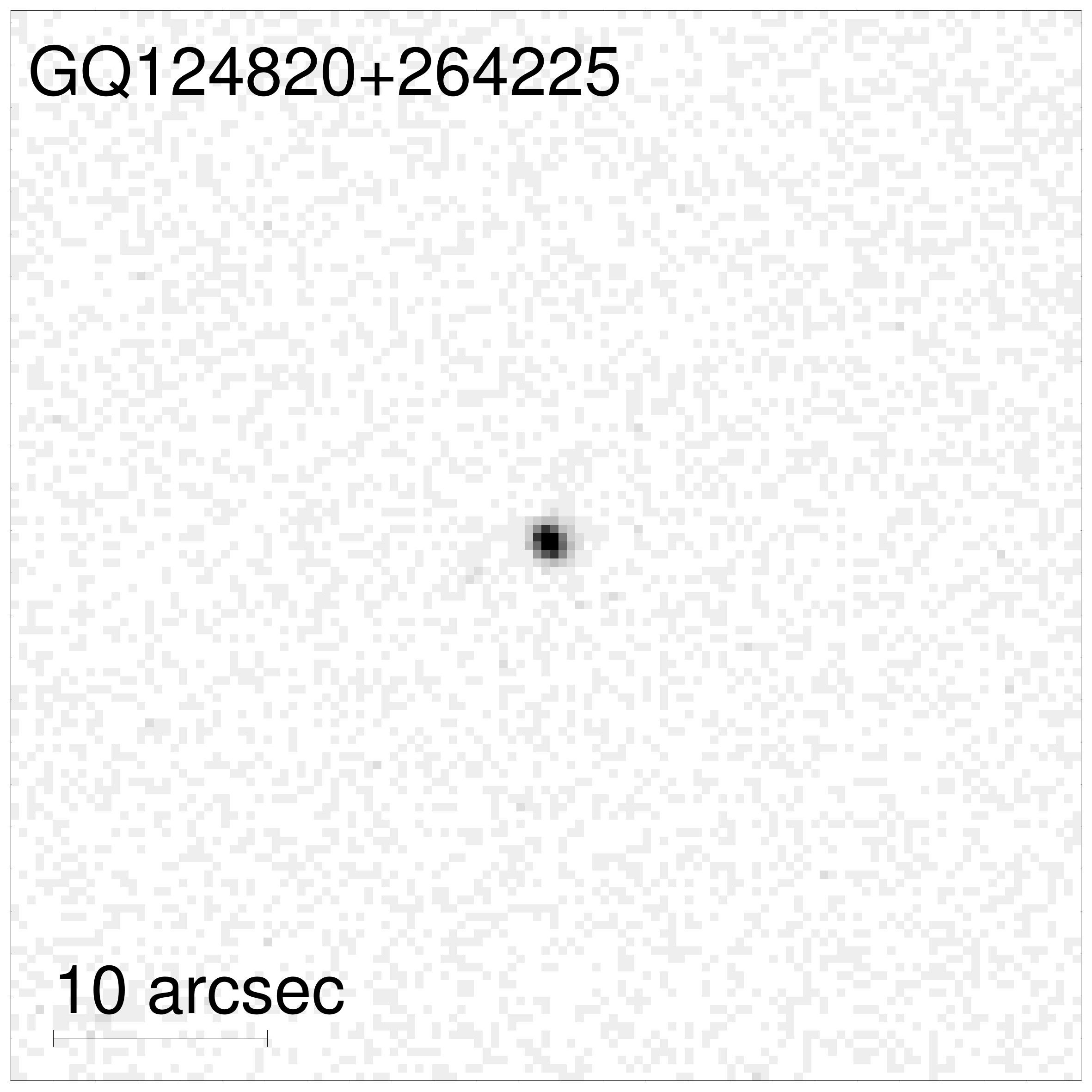,width=3.5cm} 
\epsfig{file=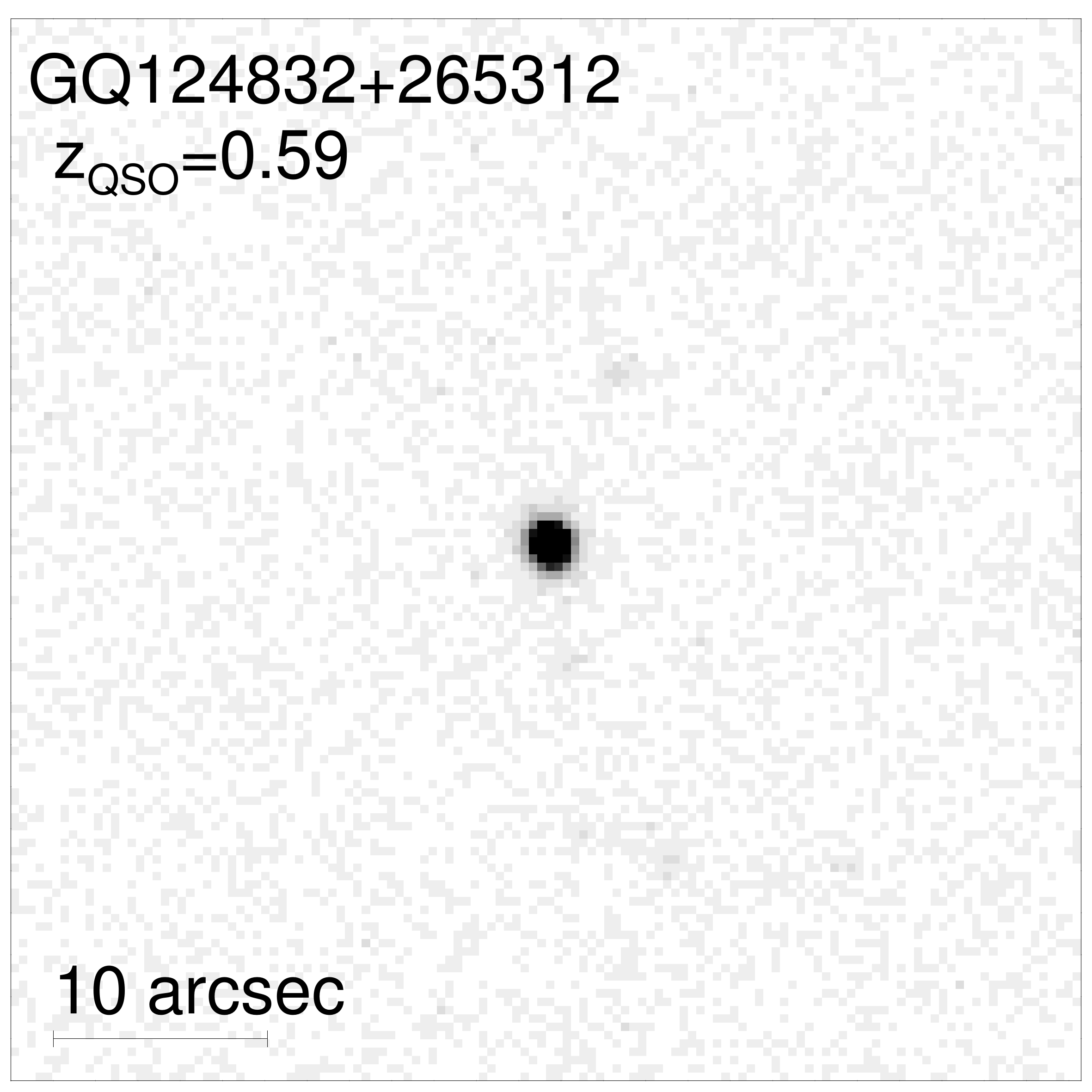,width=3.5cm} 
\epsfig{file=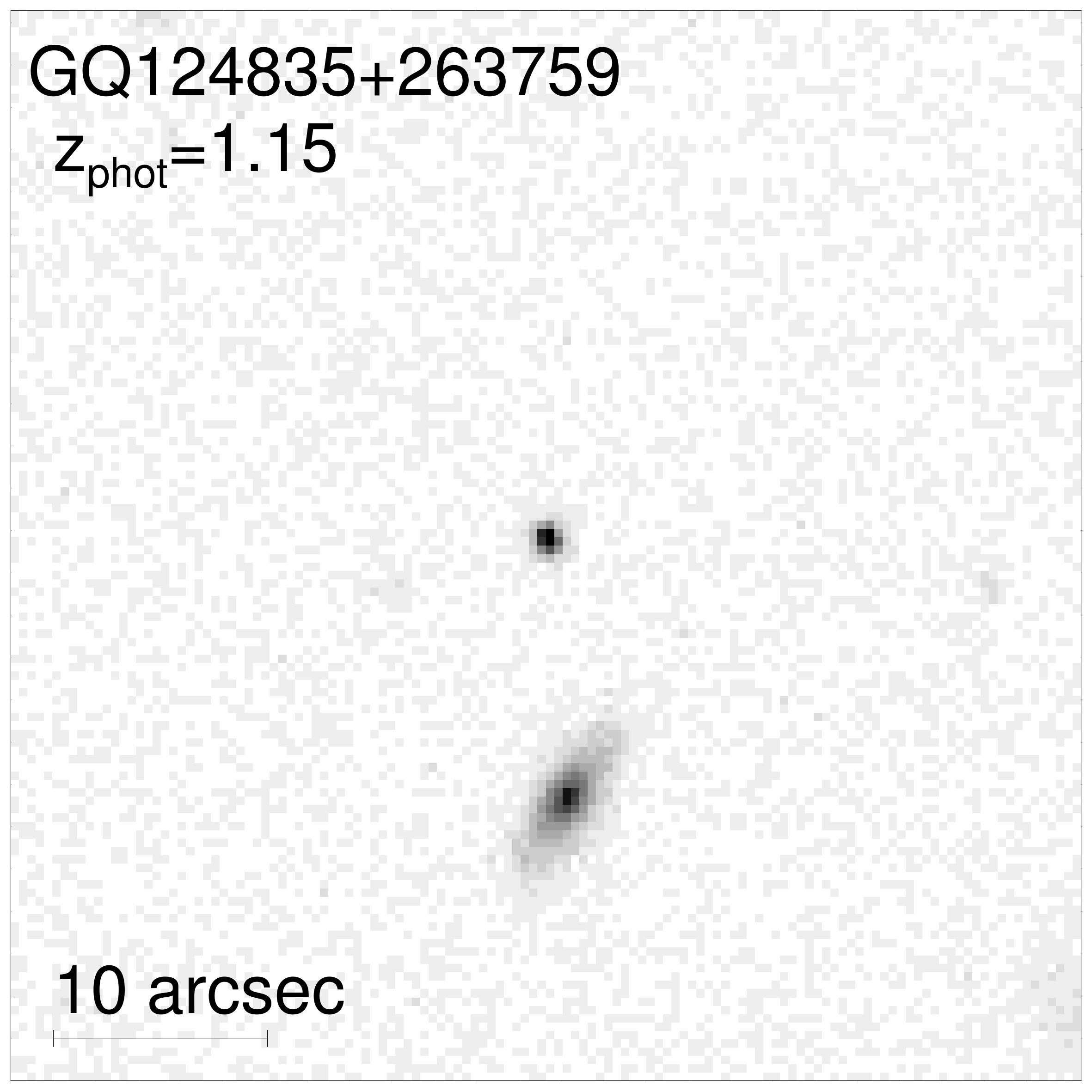,width=3.5cm} 
\epsfig{file=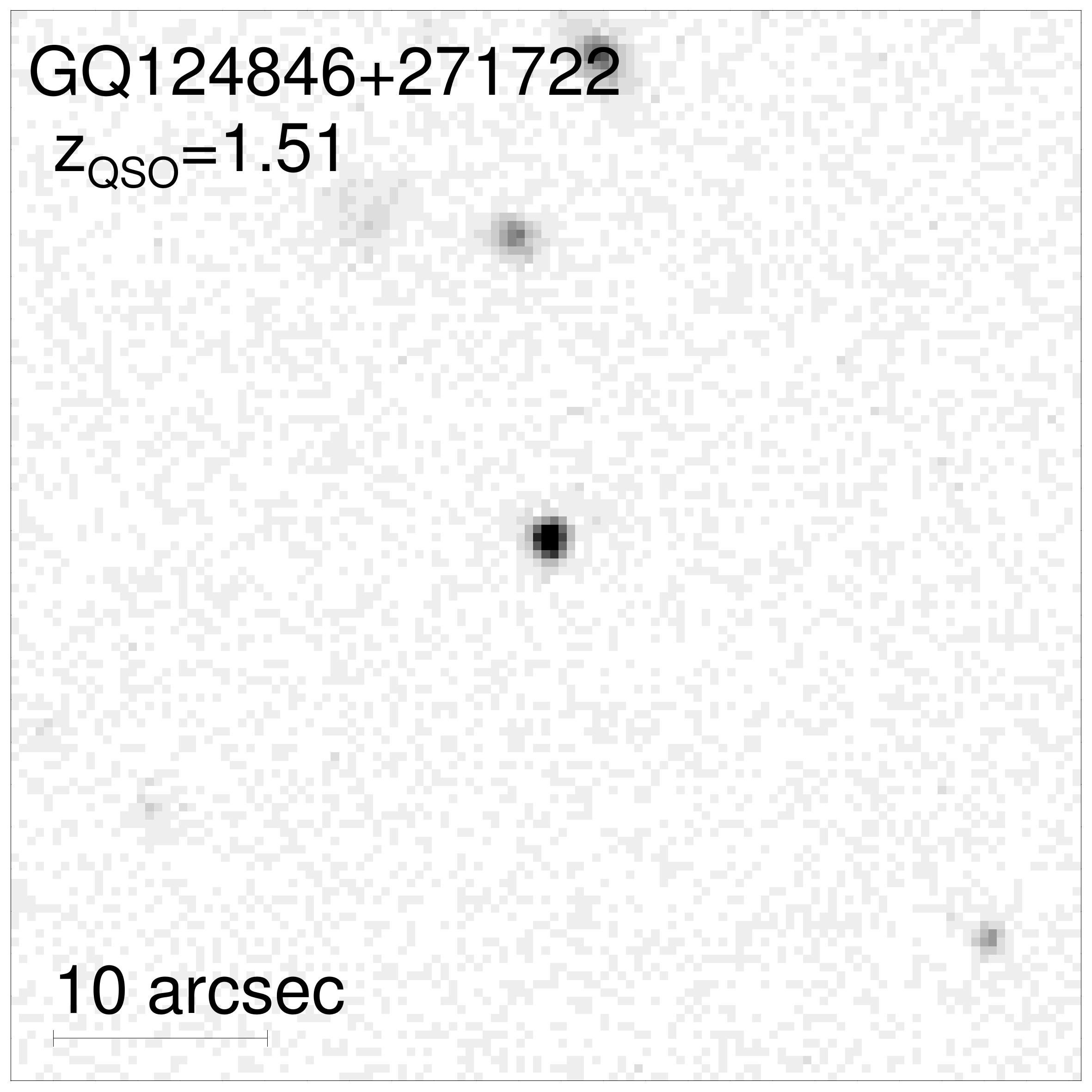,width=3.5cm} 
\epsfig{file=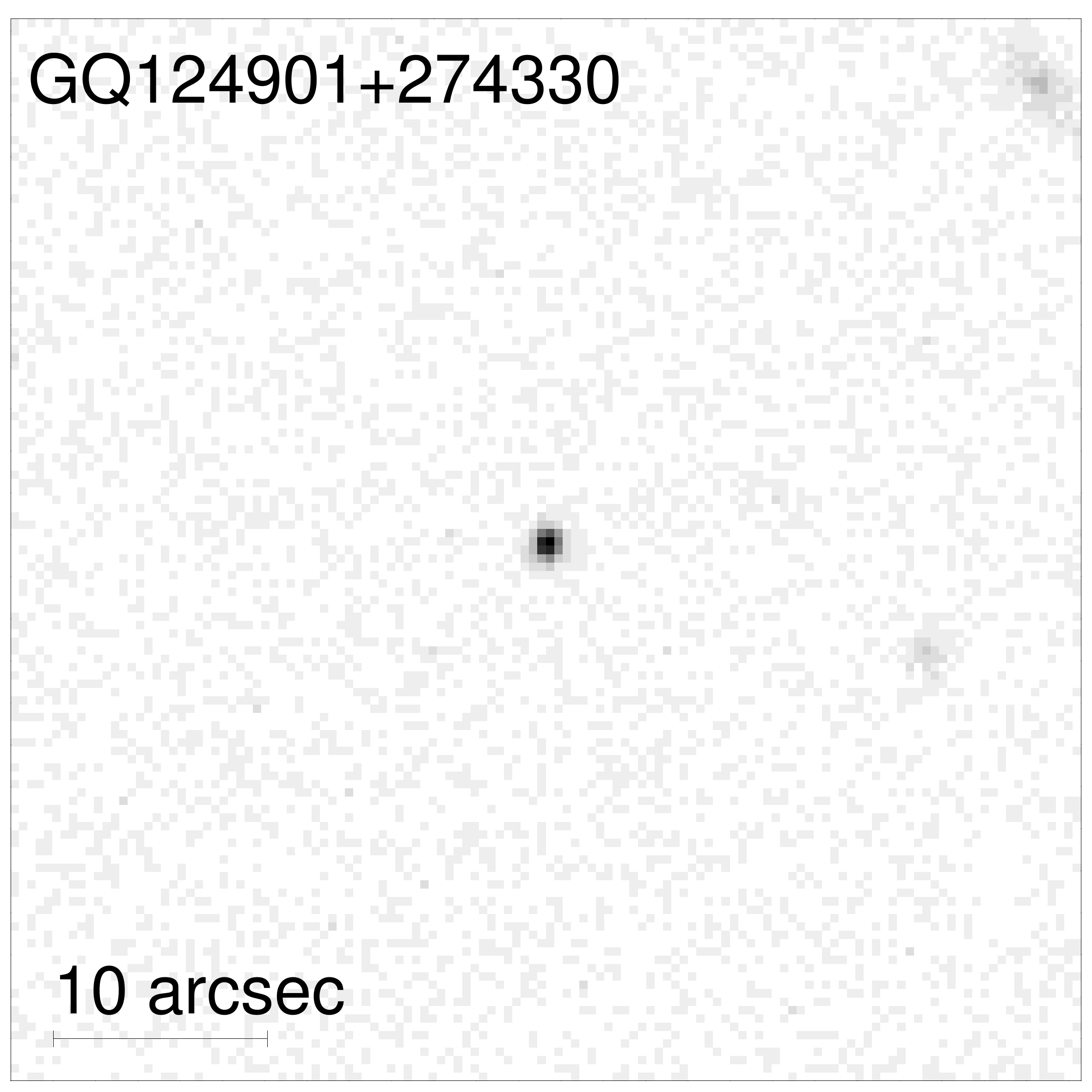,width=3.5cm} 
\epsfig{file=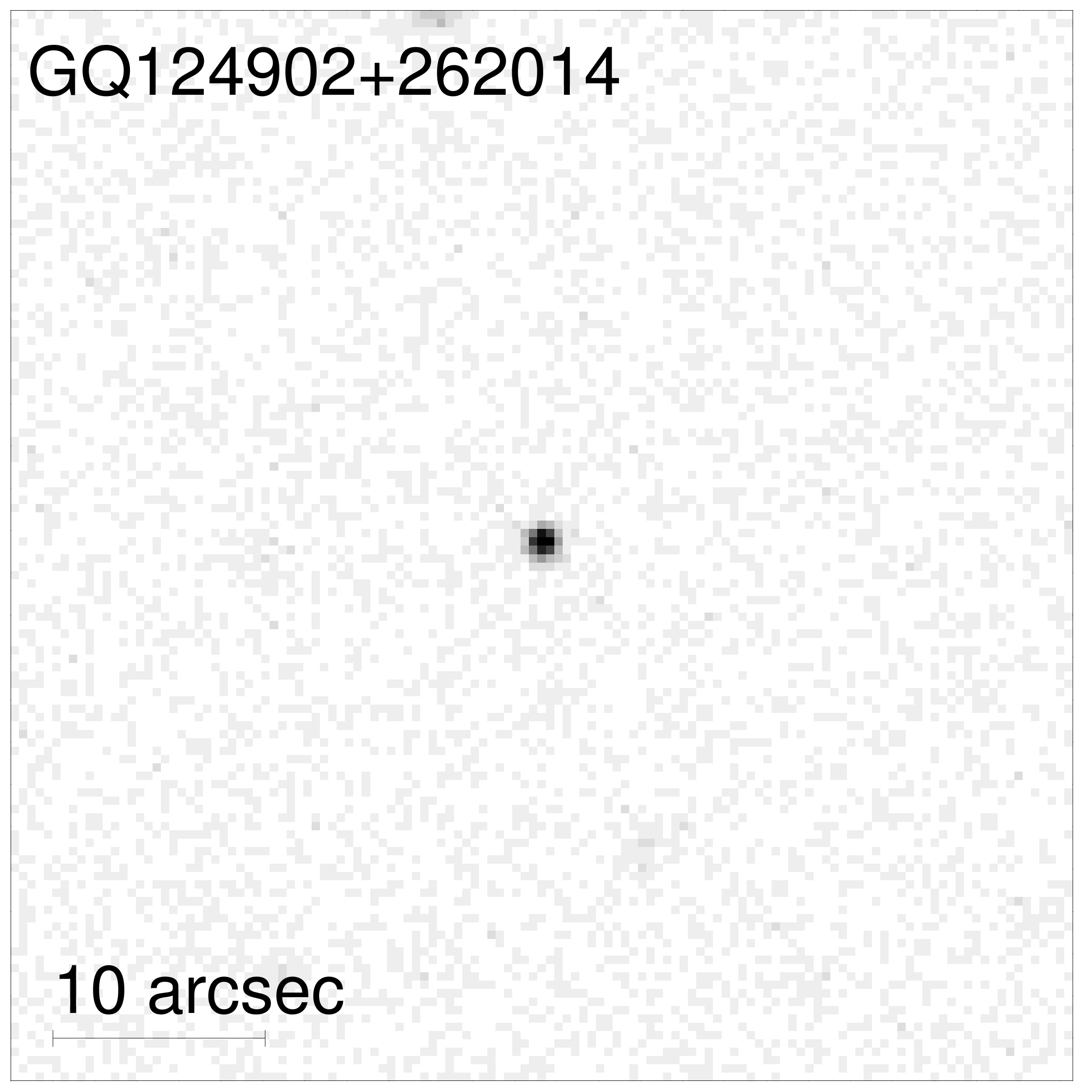,width=3.5cm} 
\epsfig{file=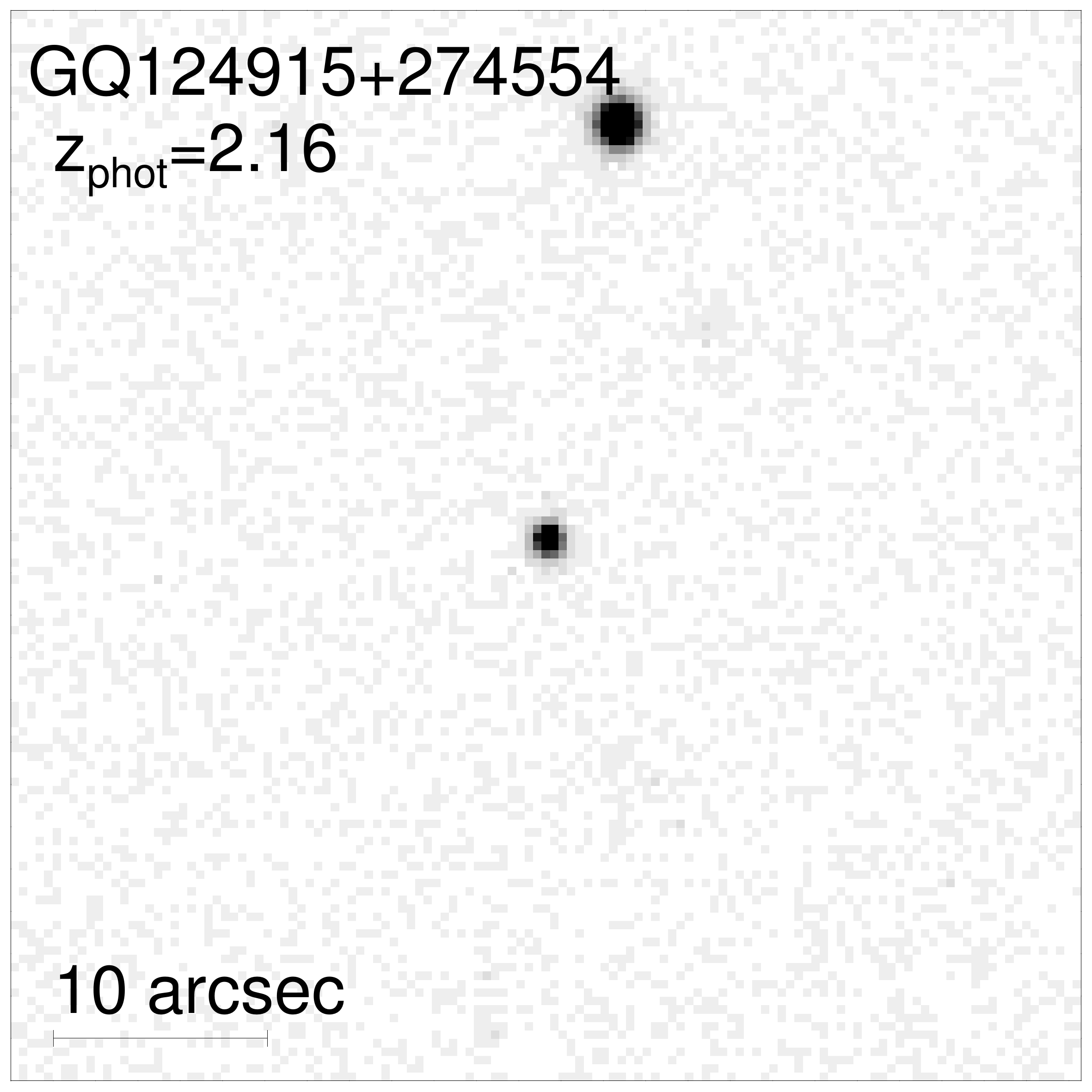,width=3.5cm} 
\epsfig{file=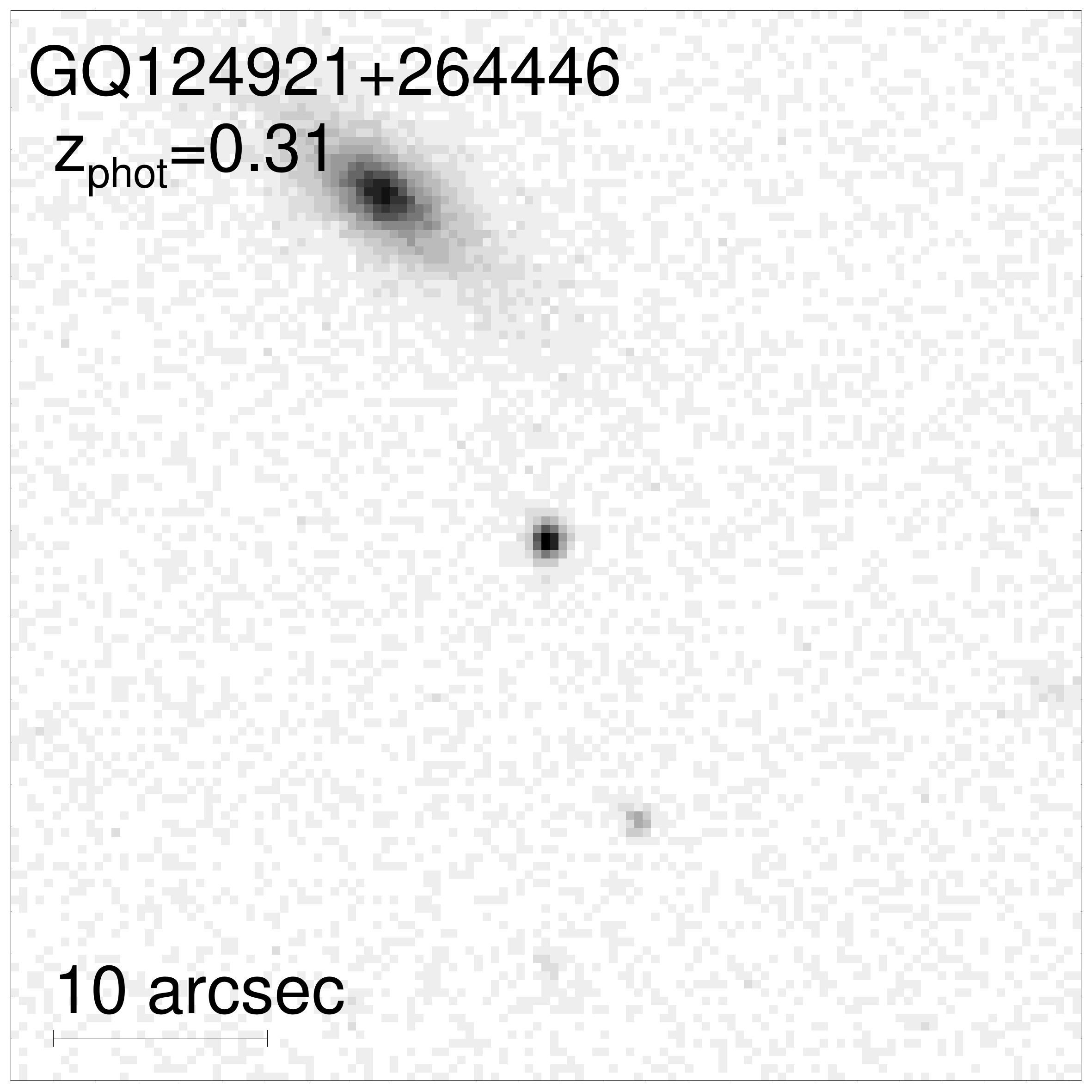,width=3.5cm} 
\epsfig{file=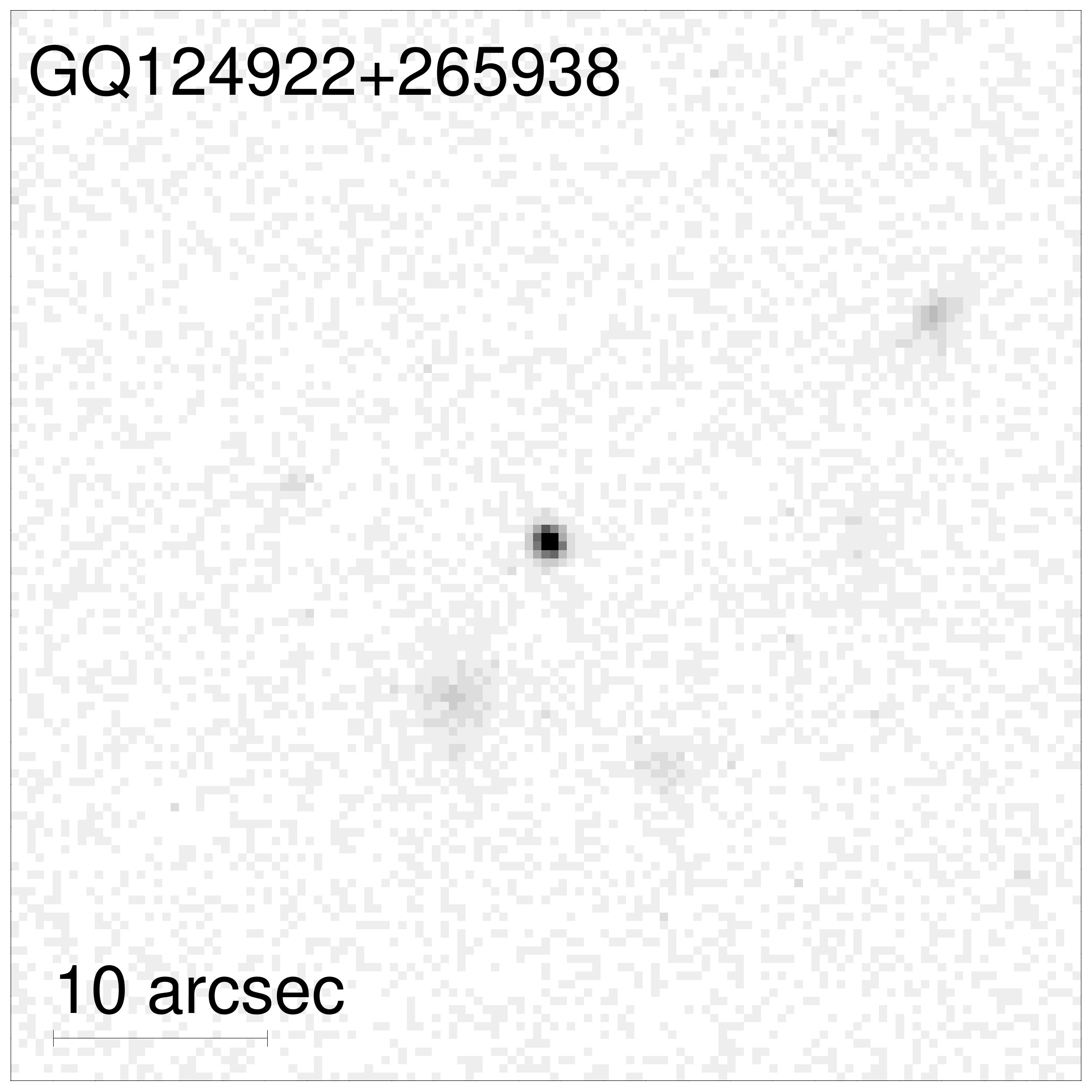,width=3.5cm} 
\epsfig{file=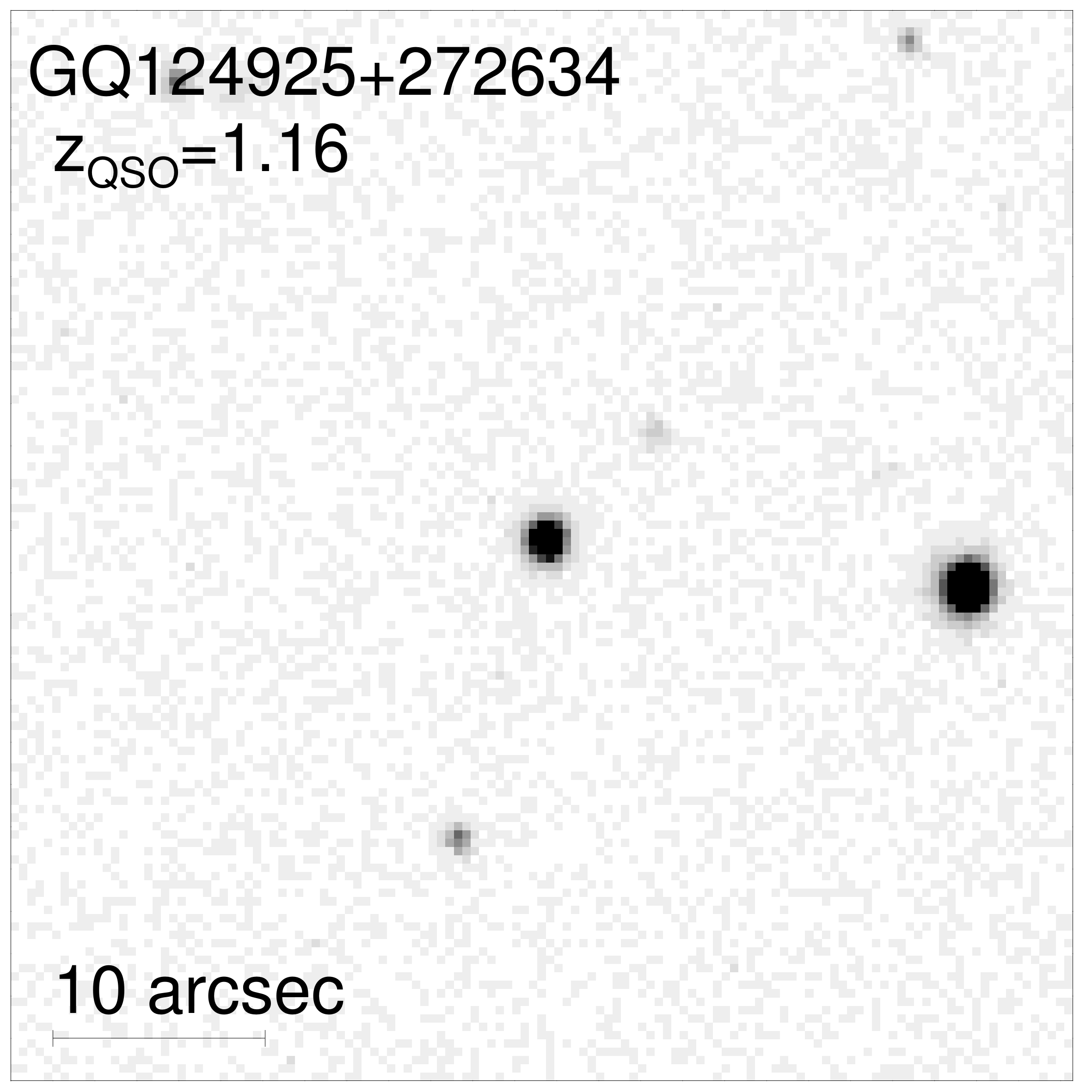,width=3.5cm} 
\epsfig{file=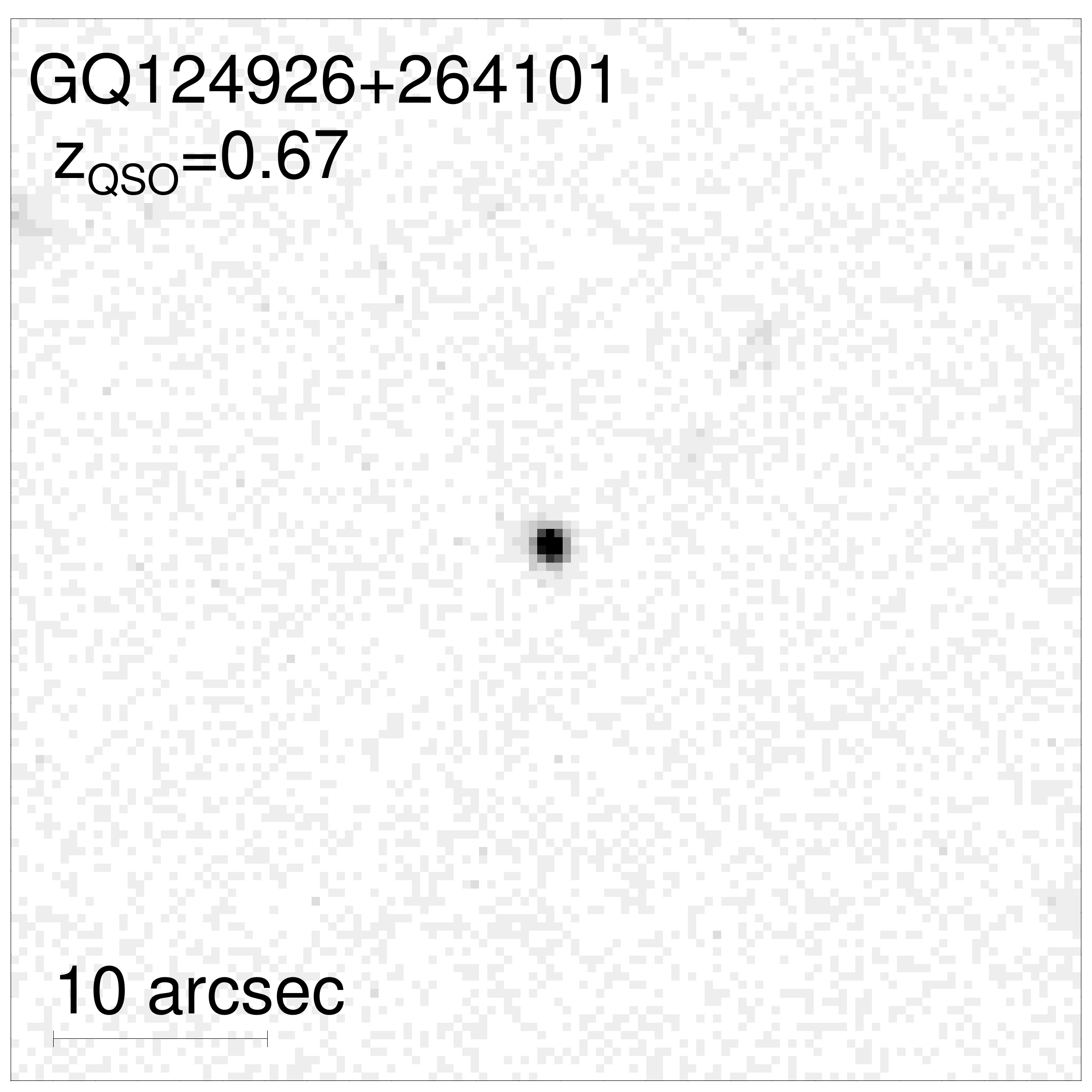,width=3.5cm} 
\epsfig{file=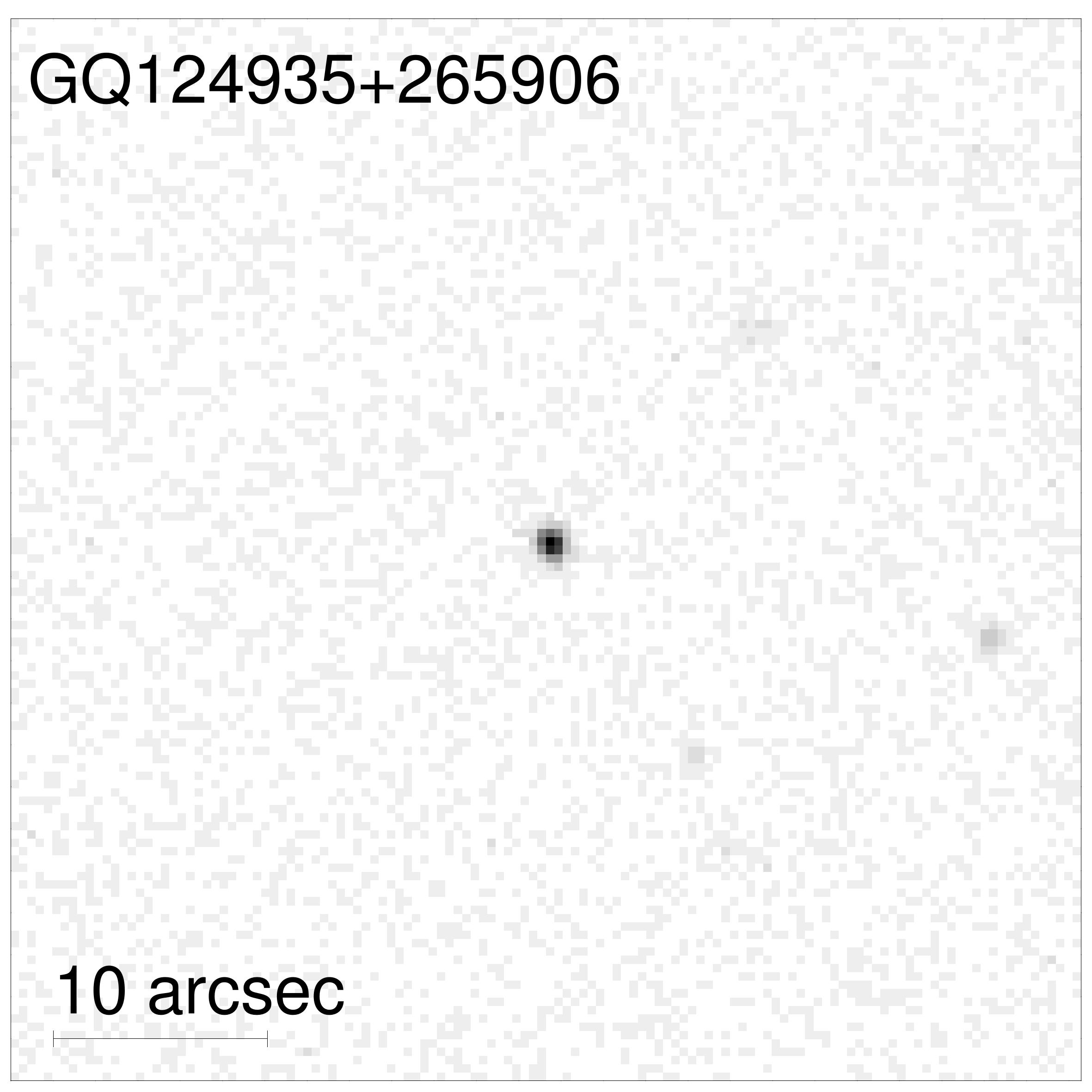,width=3.5cm} 
\epsfig{file=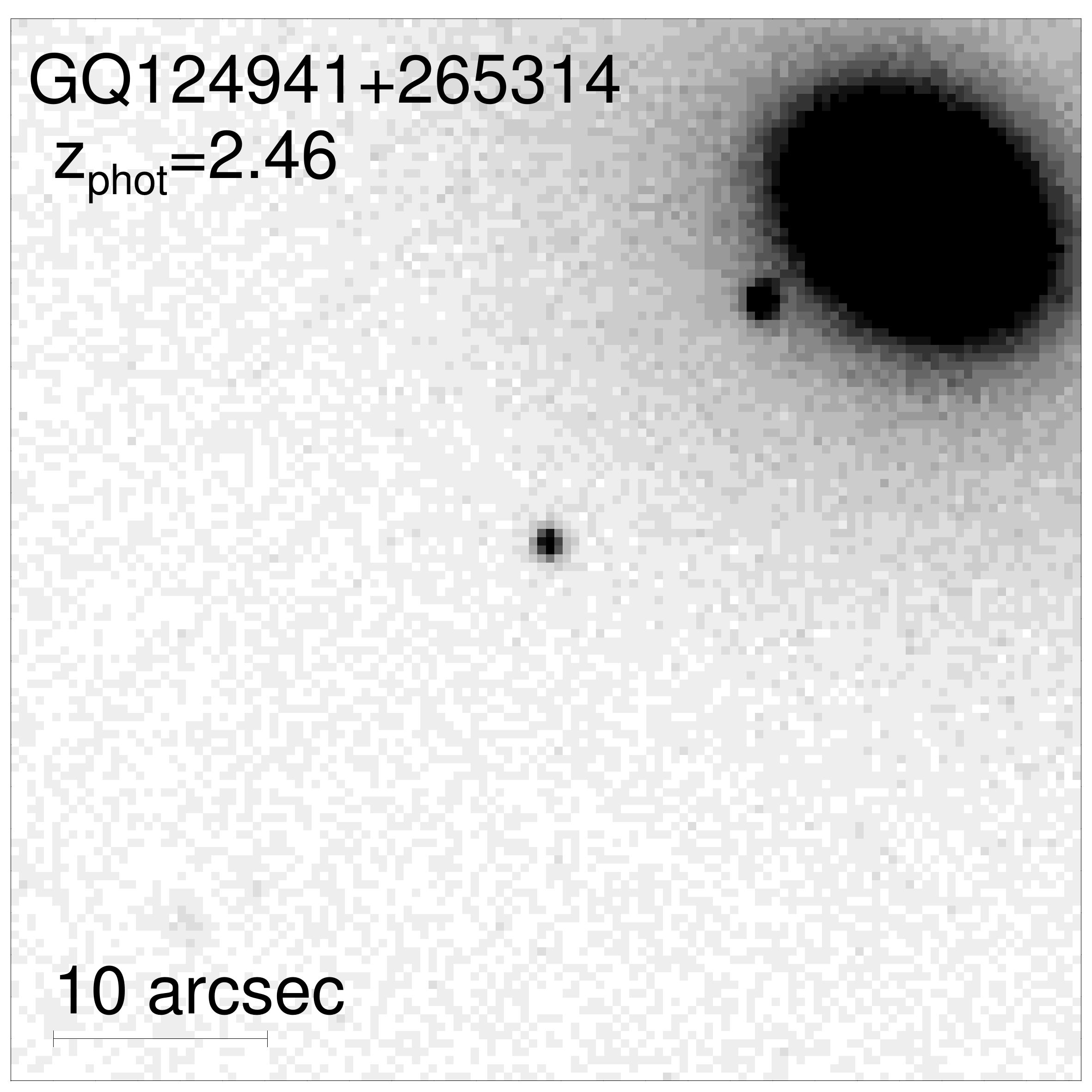,width=3.5cm} 
\epsfig{file=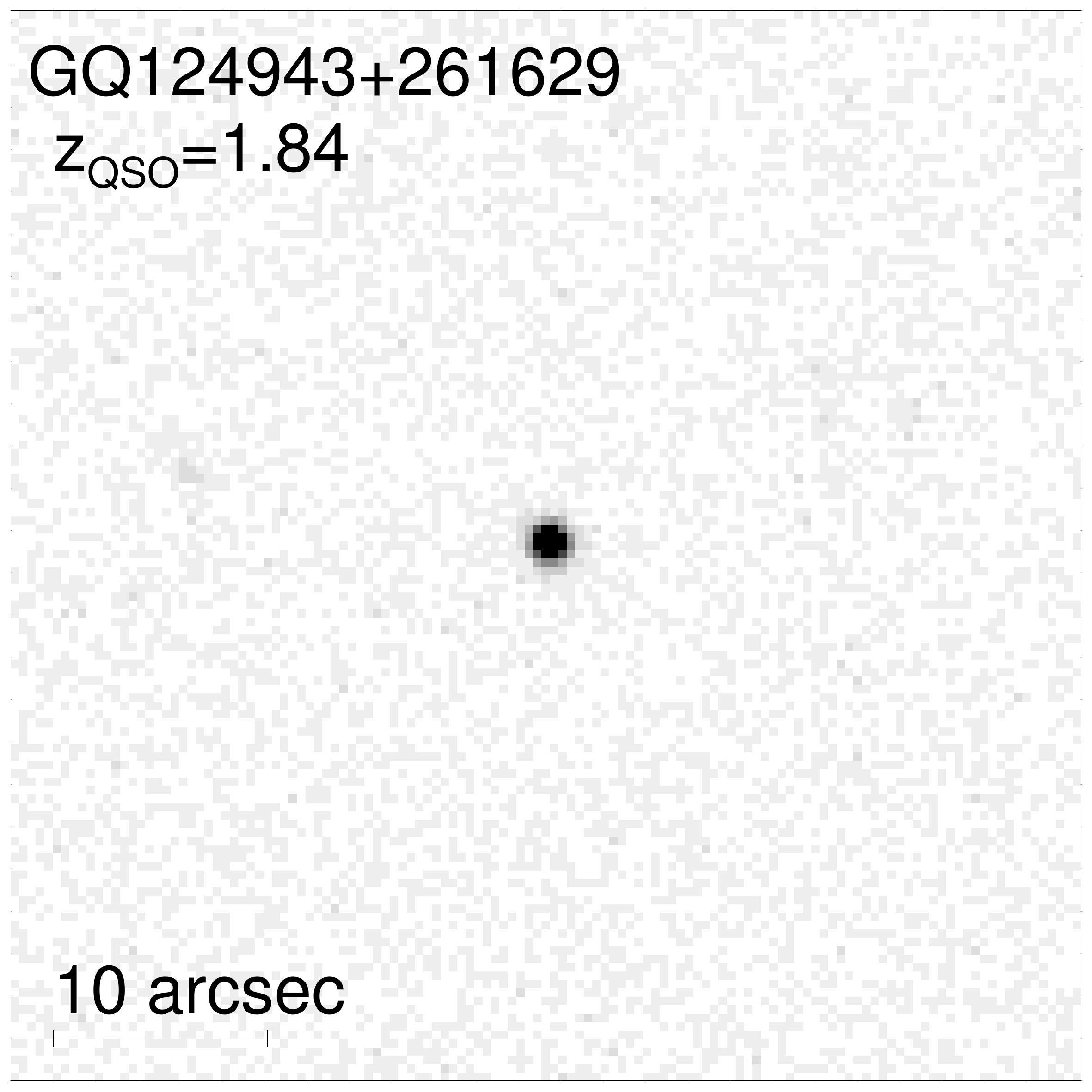,width=3.5cm} 
\epsfig{file=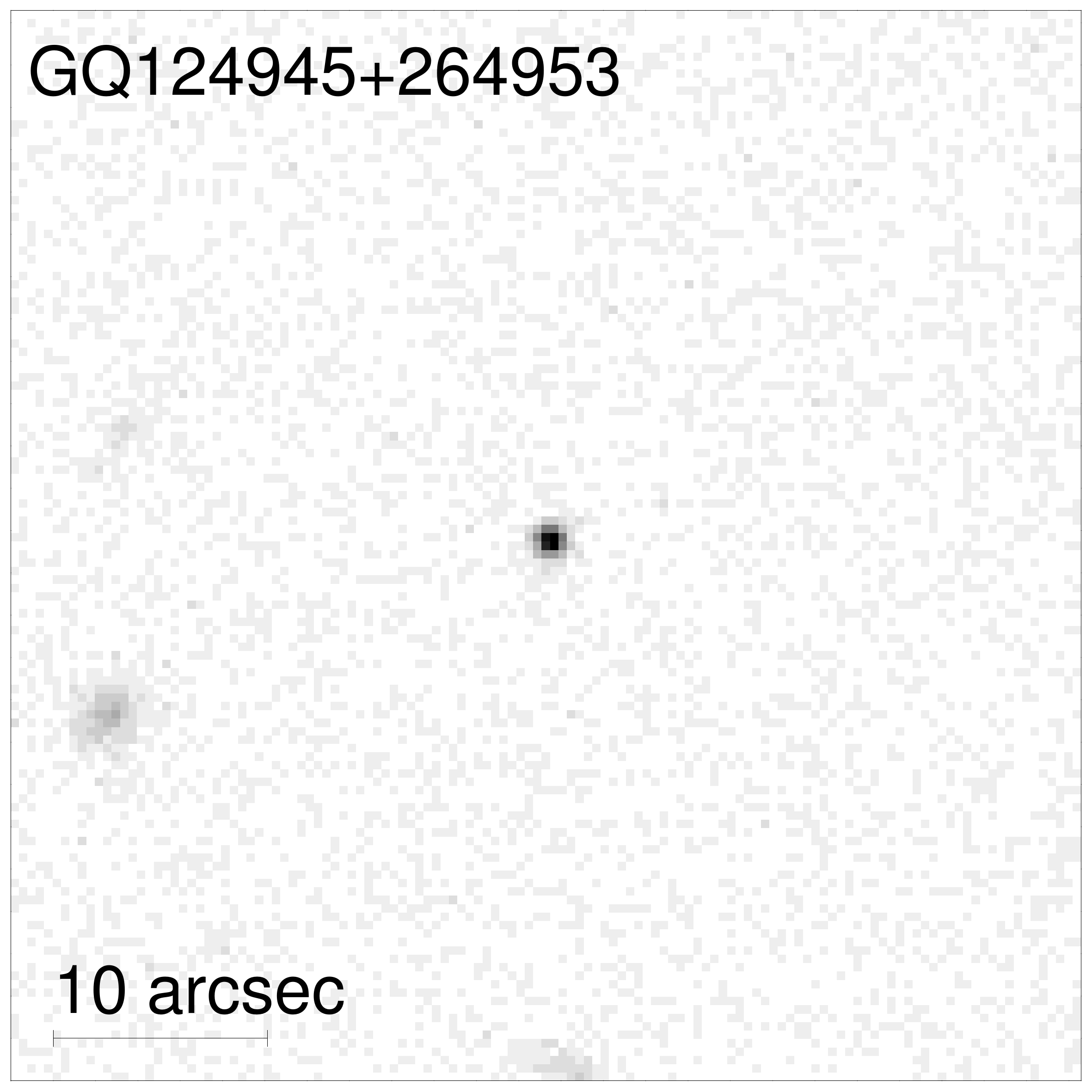,width=3.5cm} 
\epsfig{file=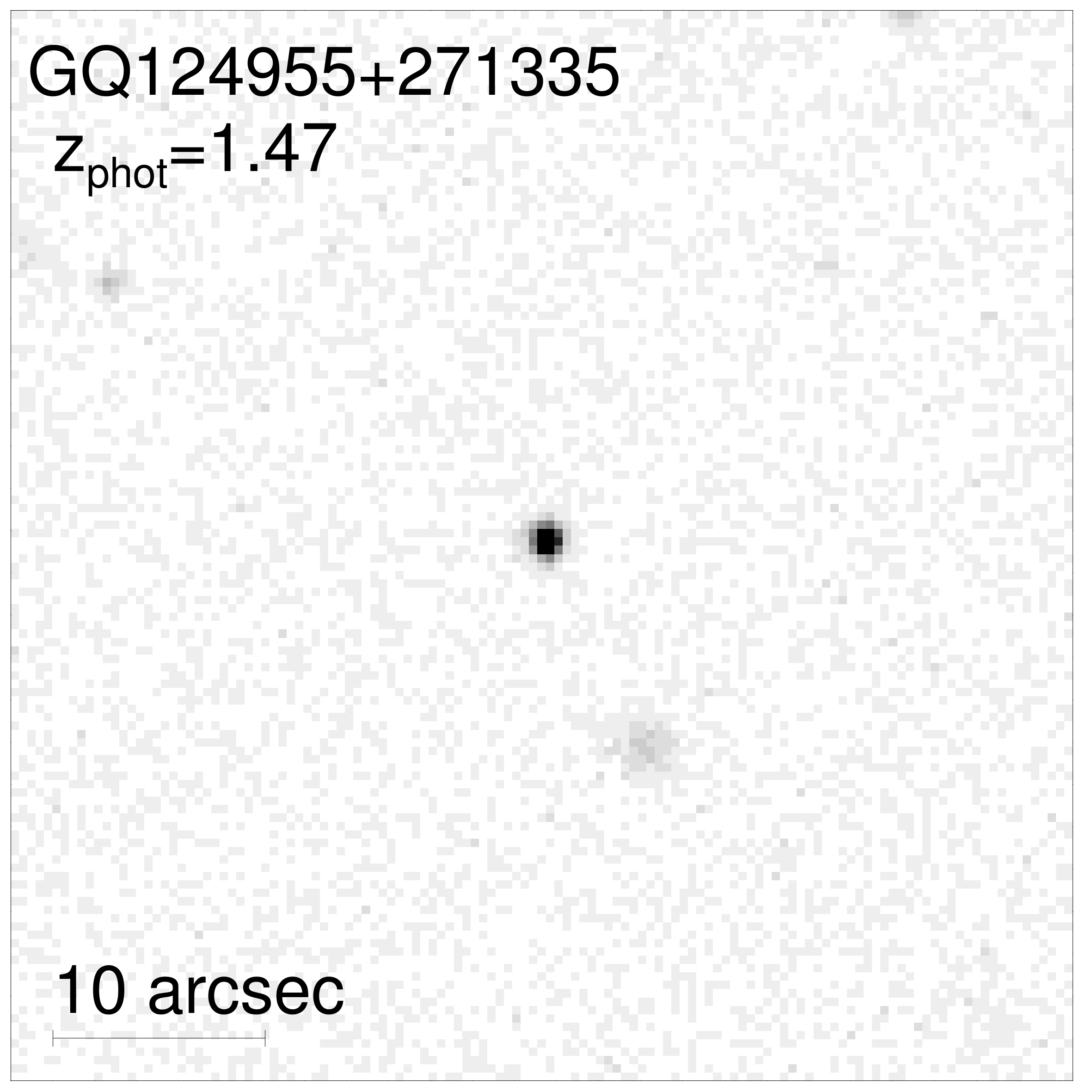,width=3.5cm} 
\epsfig{file=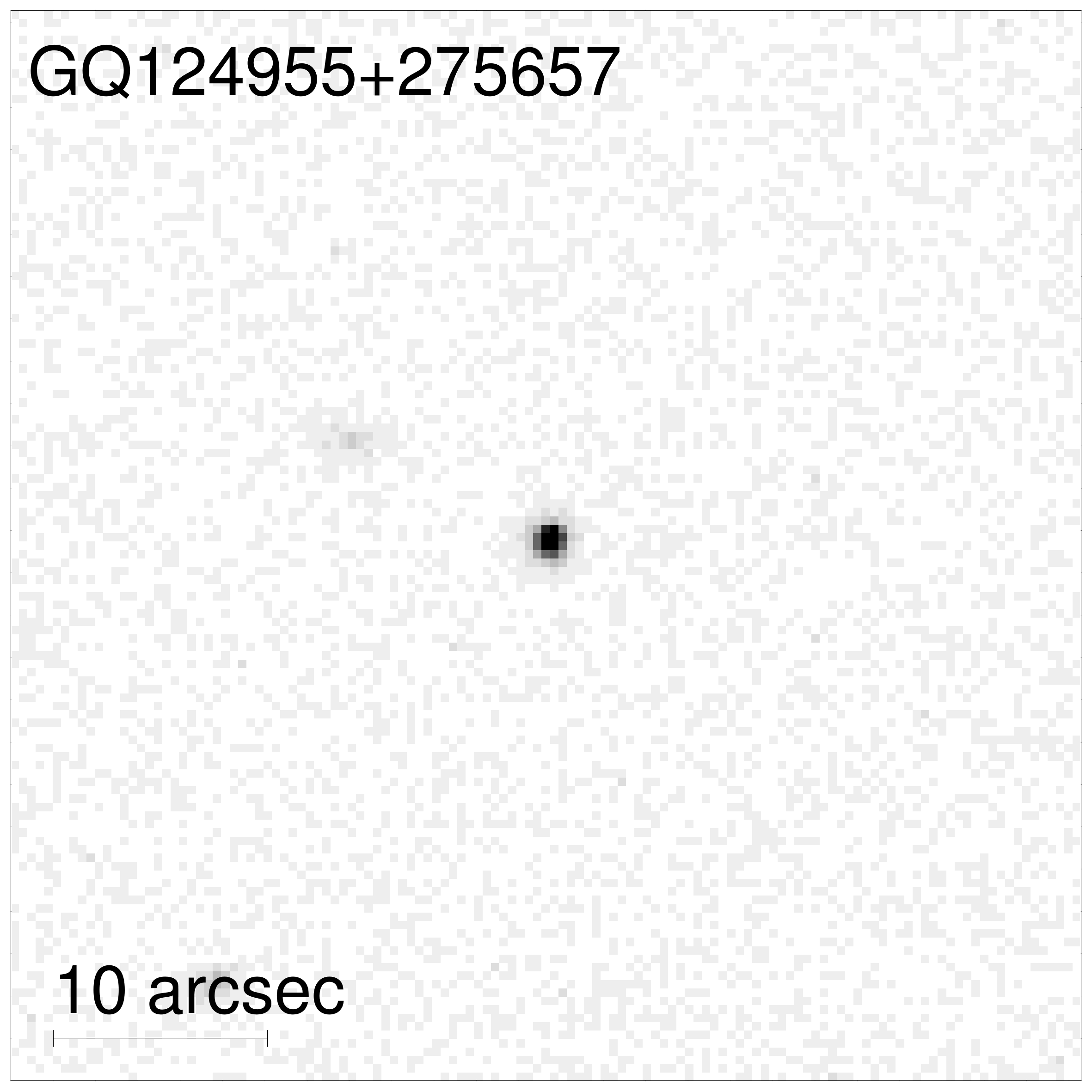,width=3.5cm} 
    \caption{50$\times$50 arcsec$^2$ thumbnails around each stationary source.}
	\label{fig:mosaic1}
\end{figure*}

\begin{figure*} [!t]
\centering
\epsfig{file=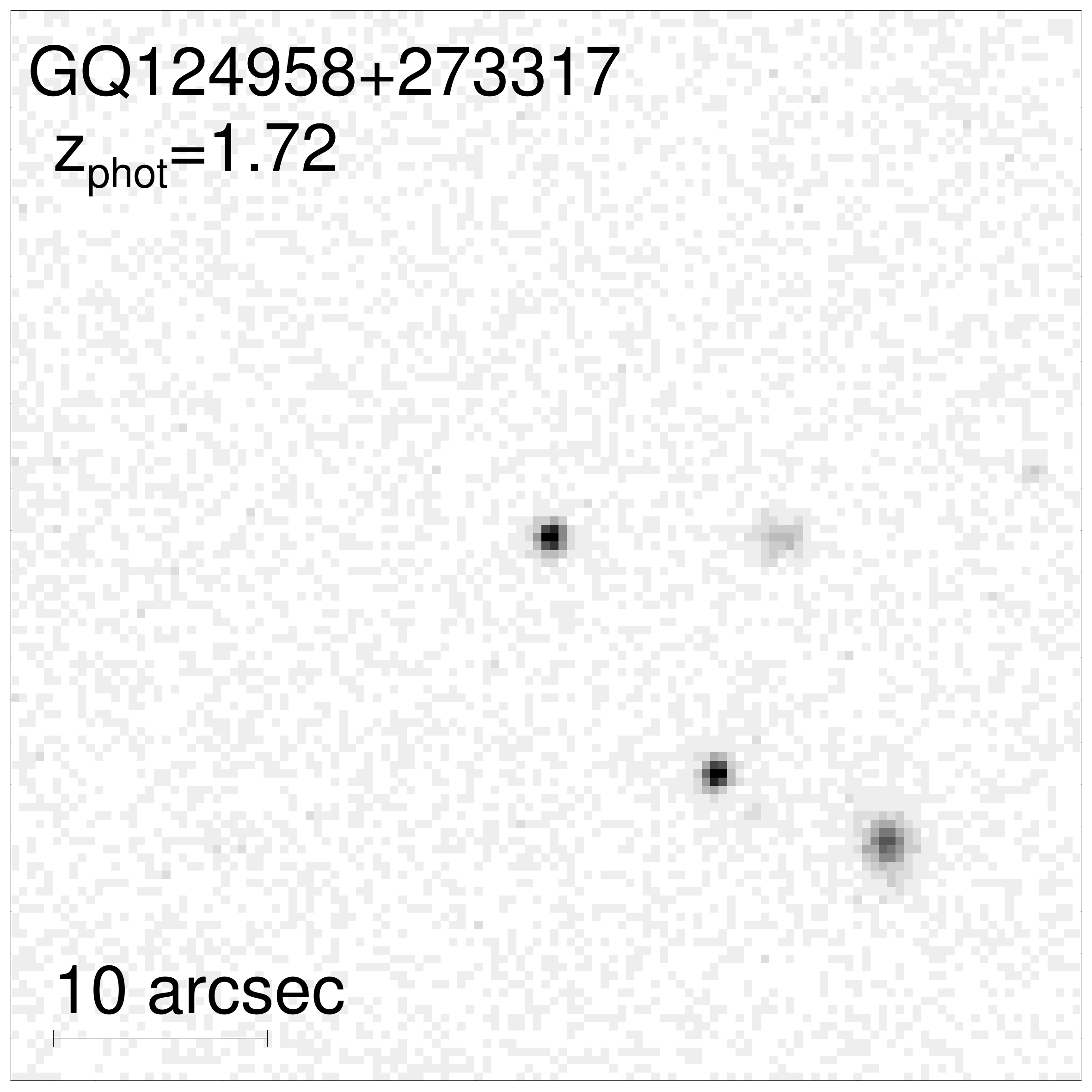,width=3.5cm} 
\epsfig{file=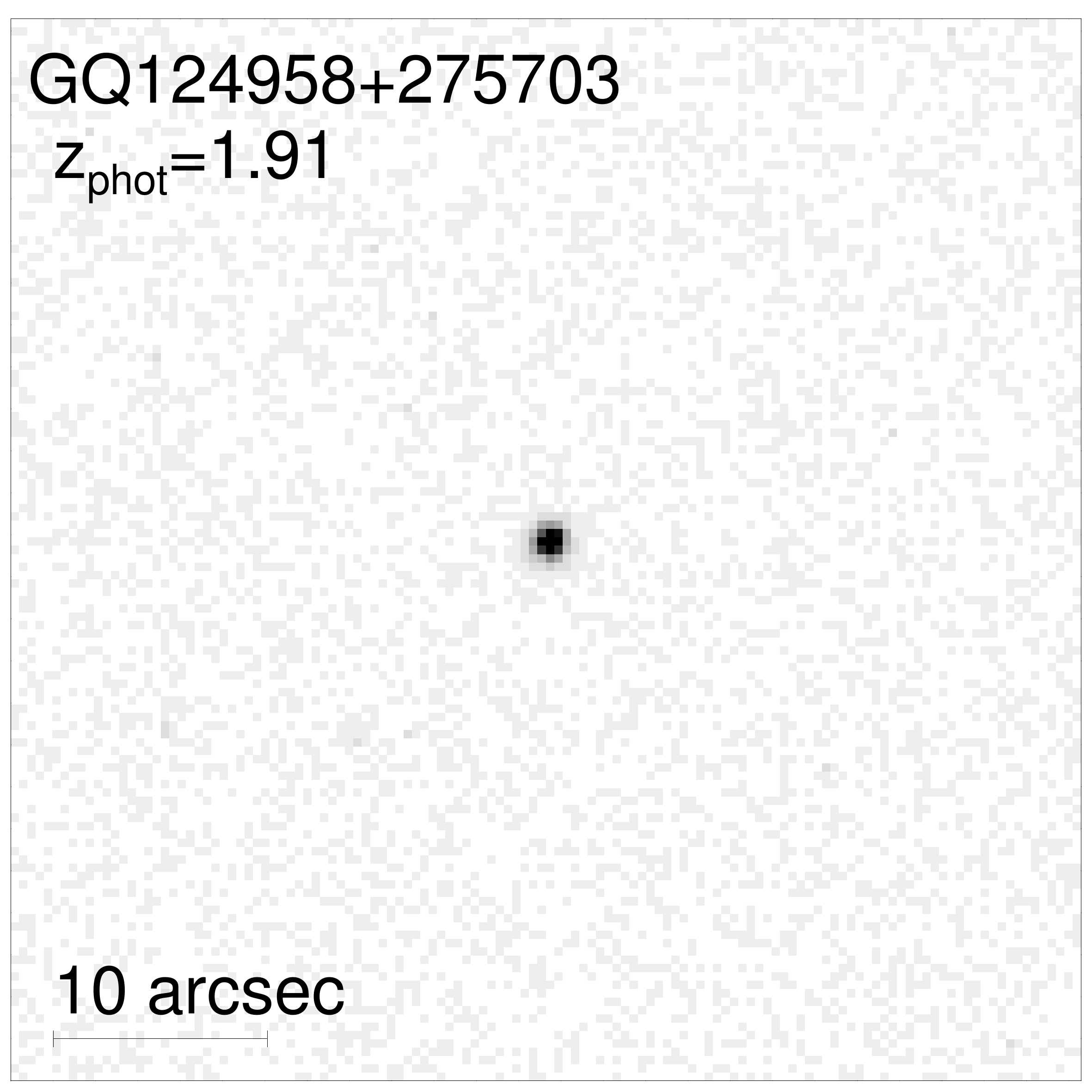,width=3.5cm} 
\epsfig{file=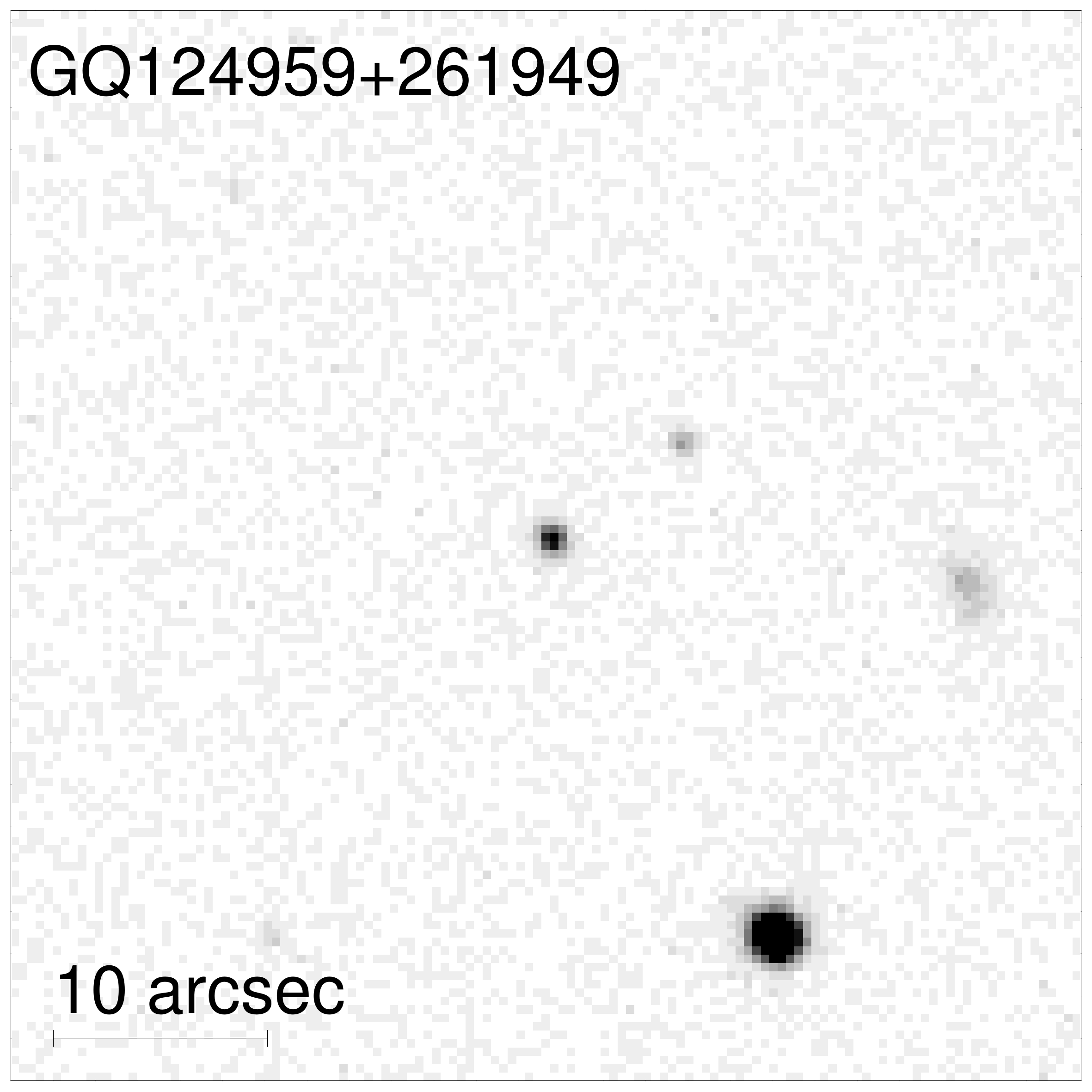,width=3.5cm} 
\epsfig{file=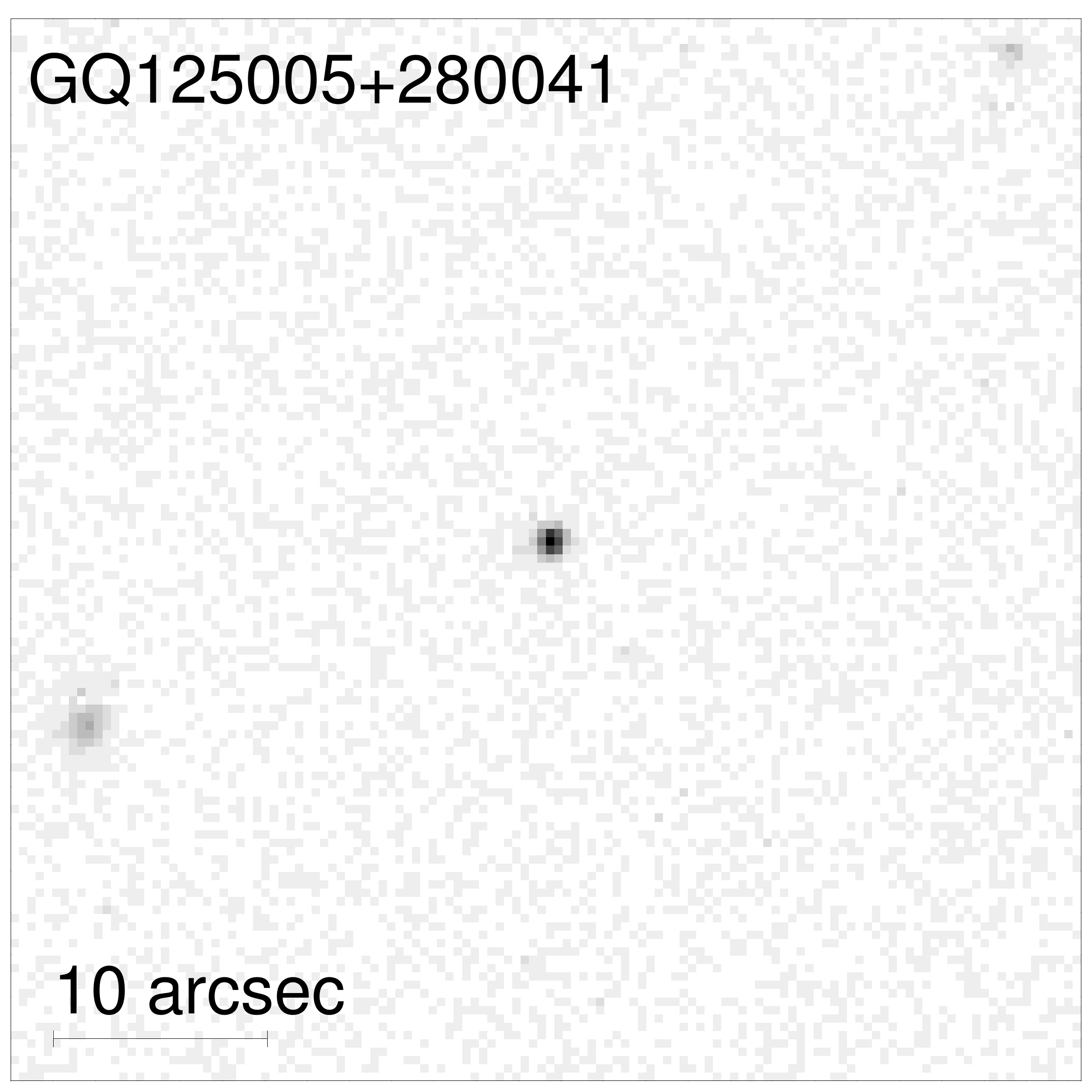,width=3.5cm} 
\epsfig{file=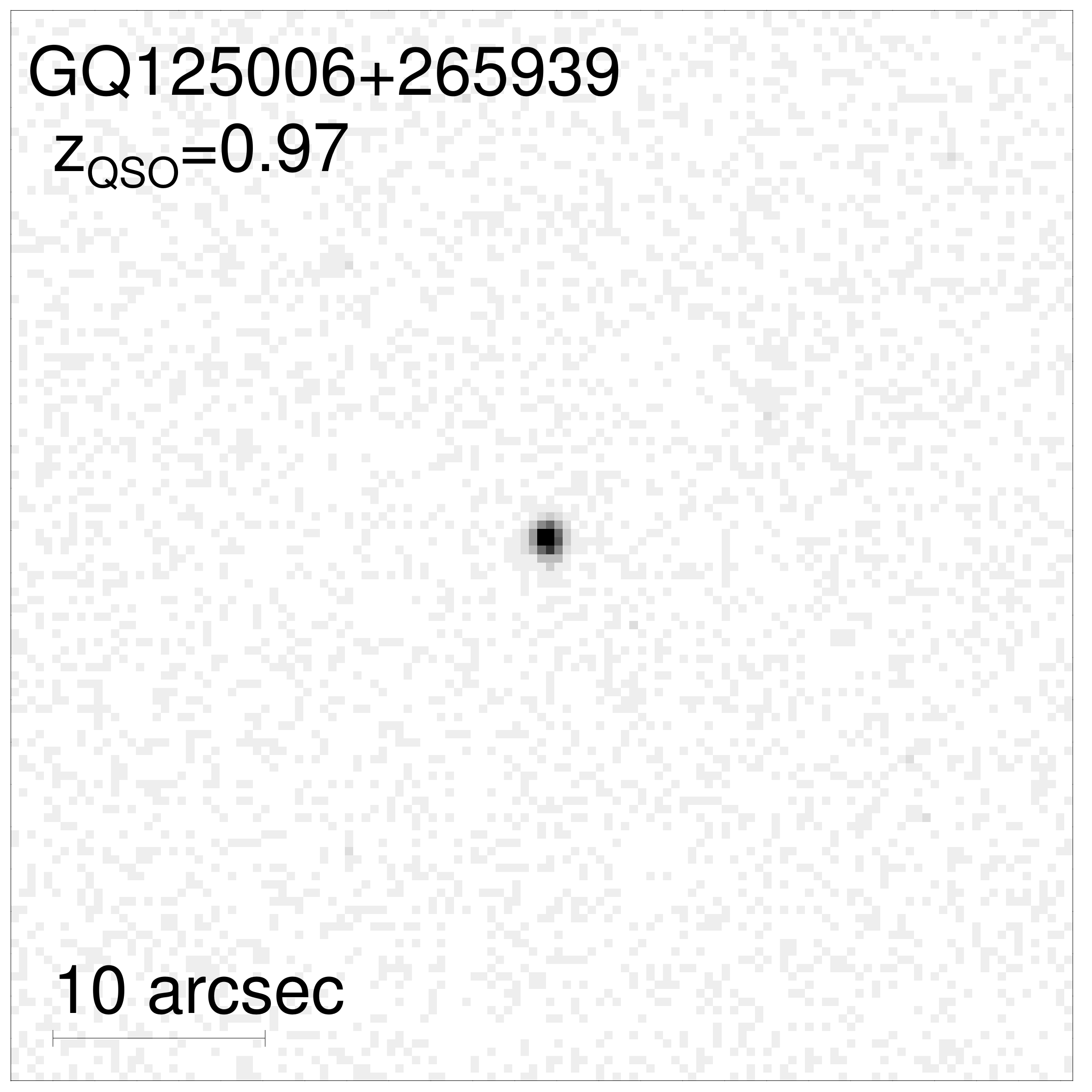,width=3.5cm} 
\epsfig{file=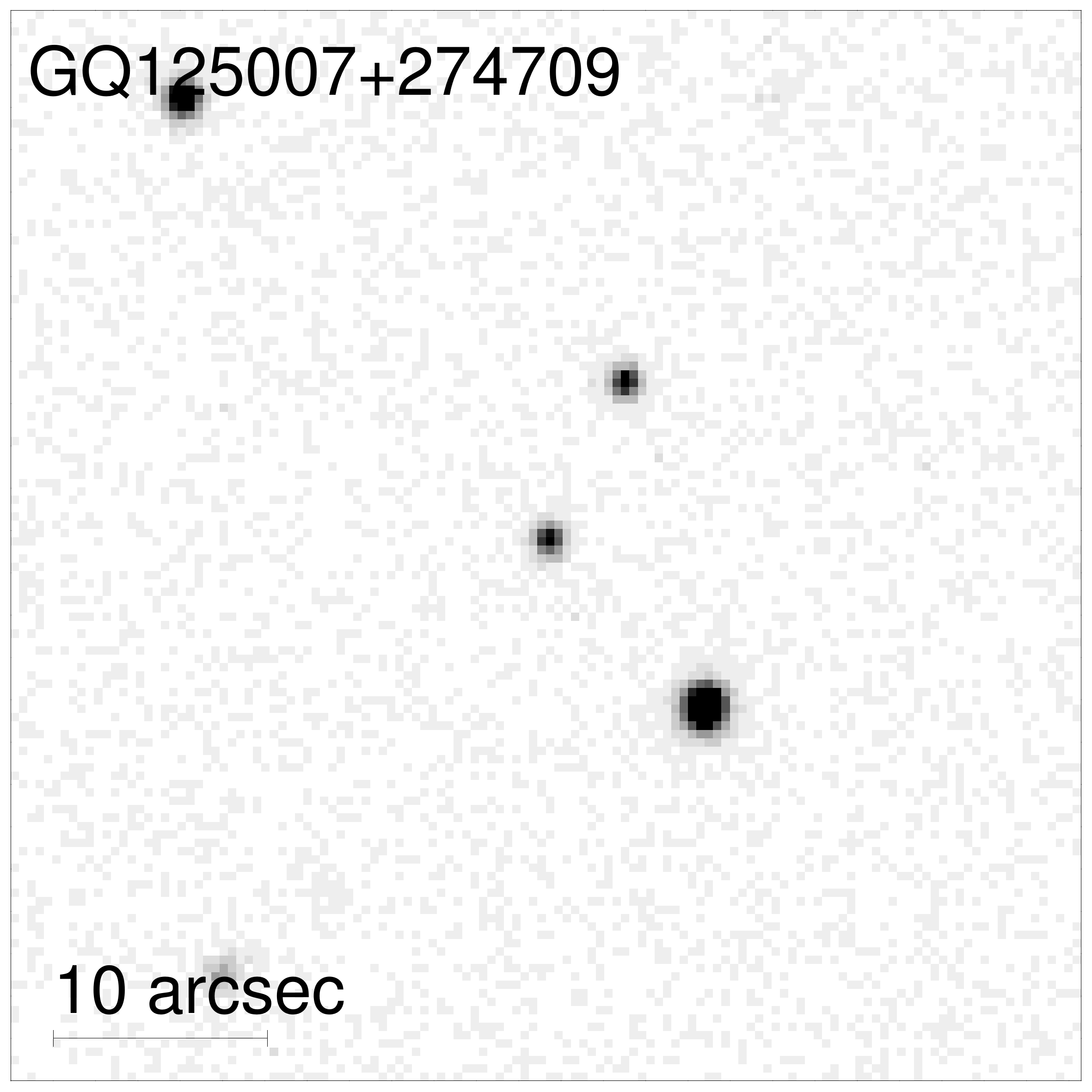,width=3.5cm} 
\epsfig{file=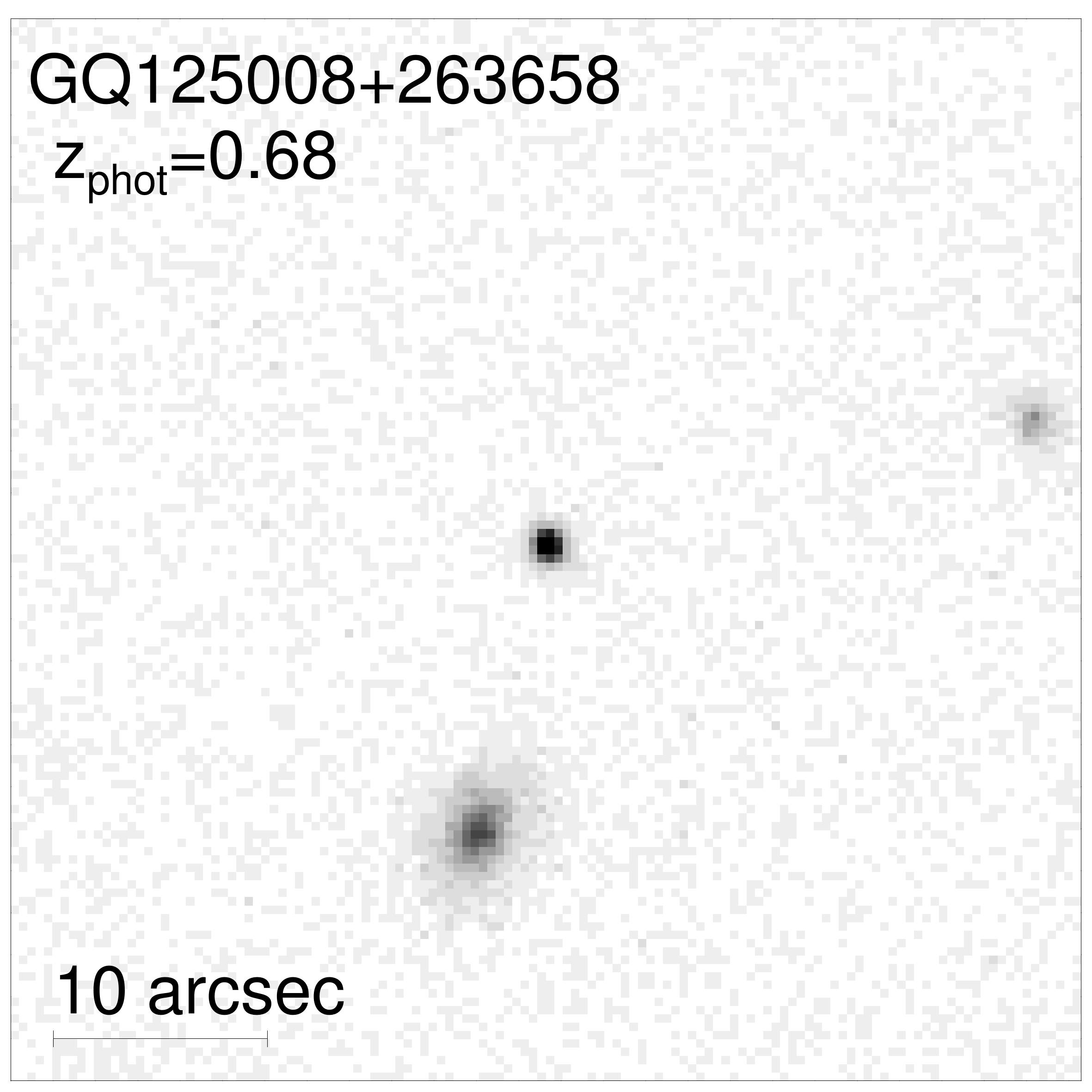,width=3.5cm} 
\epsfig{file=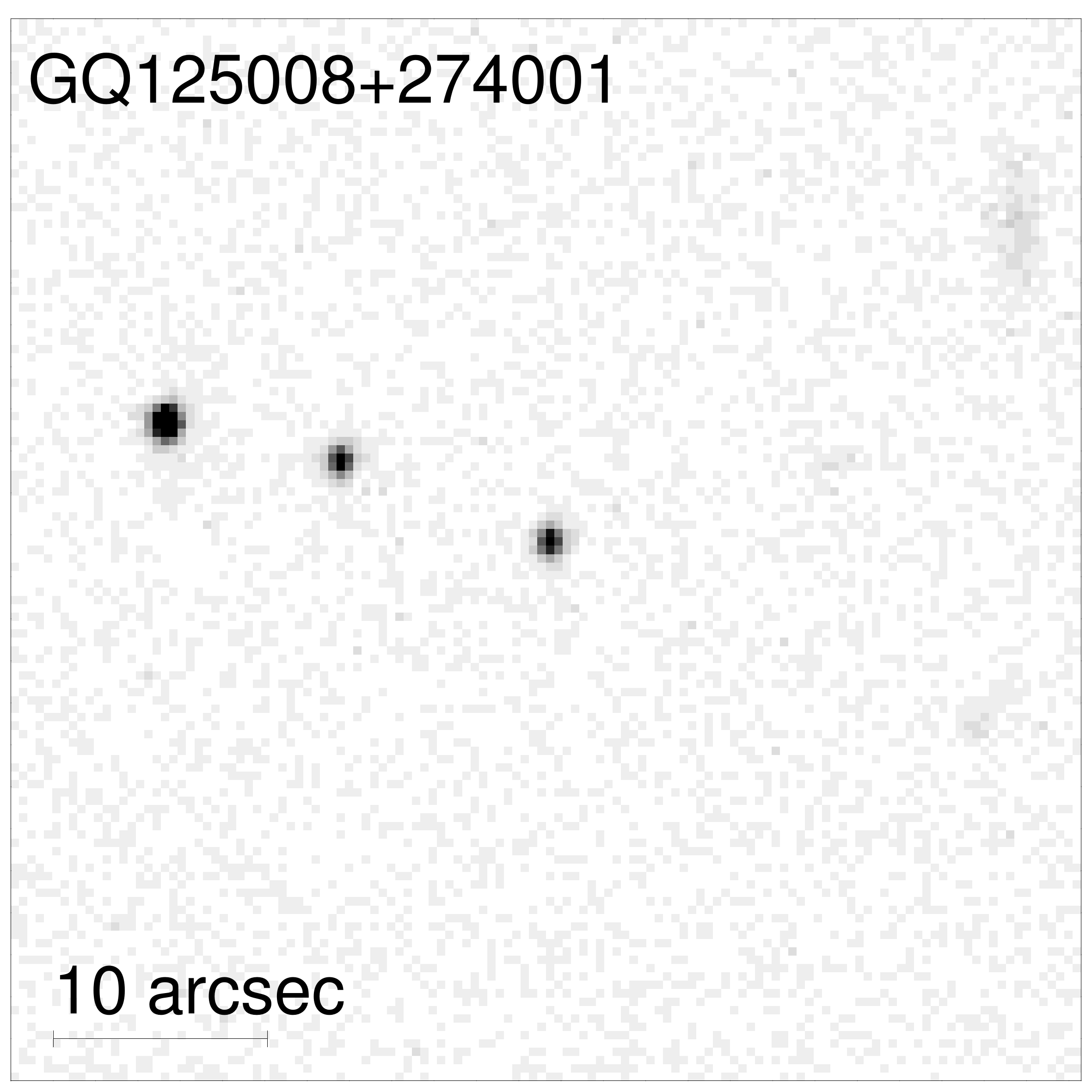,width=3.5cm} 
\epsfig{file=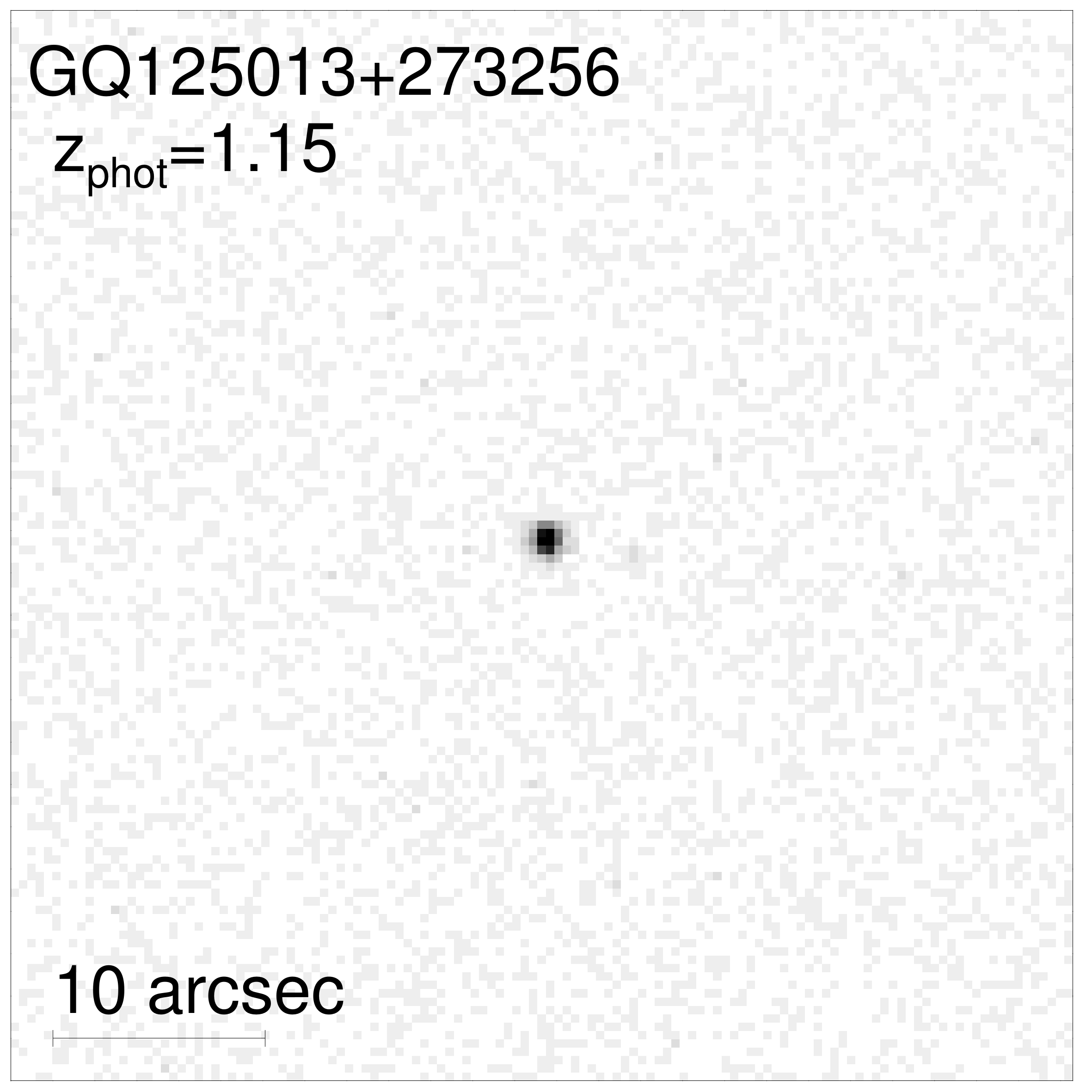,width=3.5cm} 
\epsfig{file=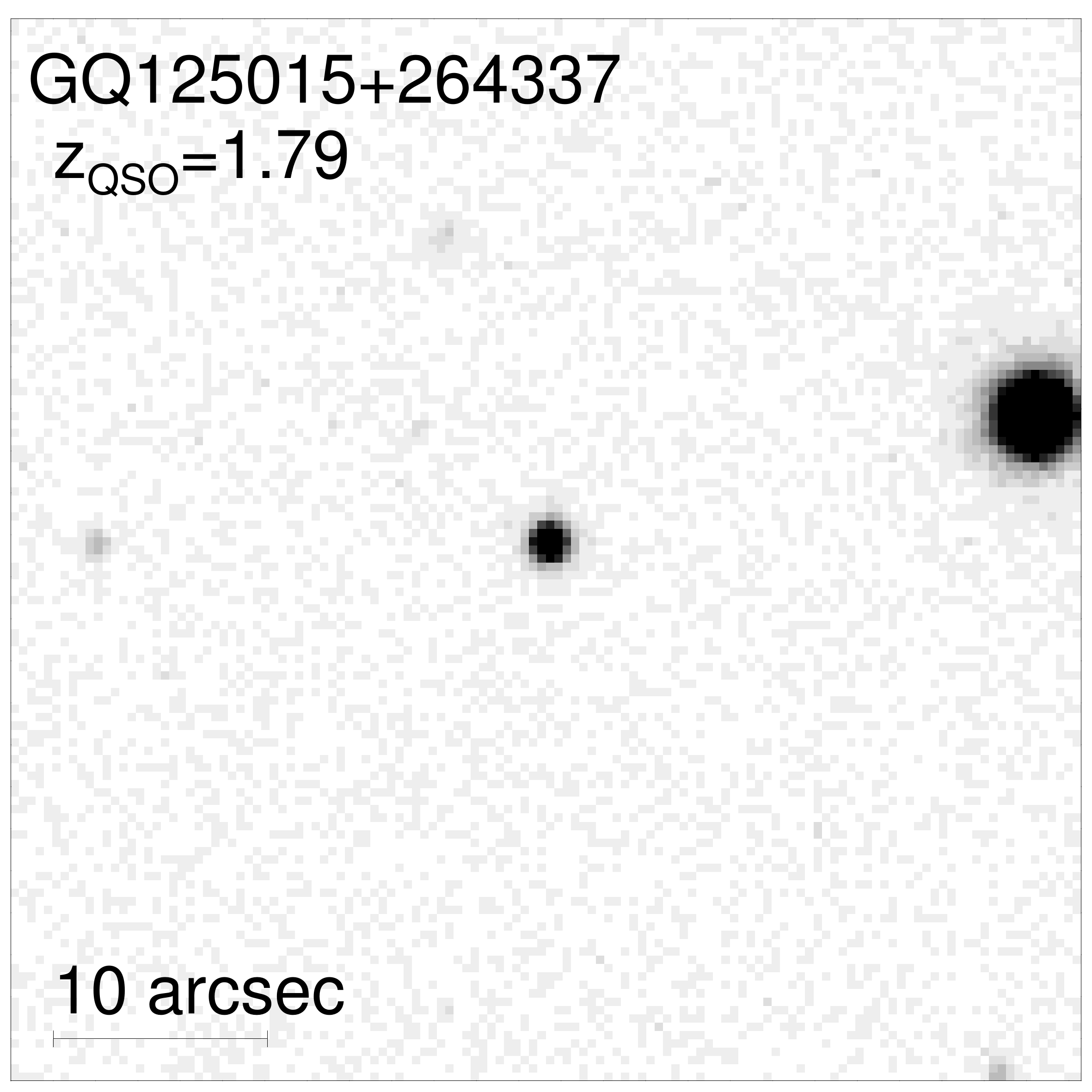,width=3.5cm} 
\epsfig{file=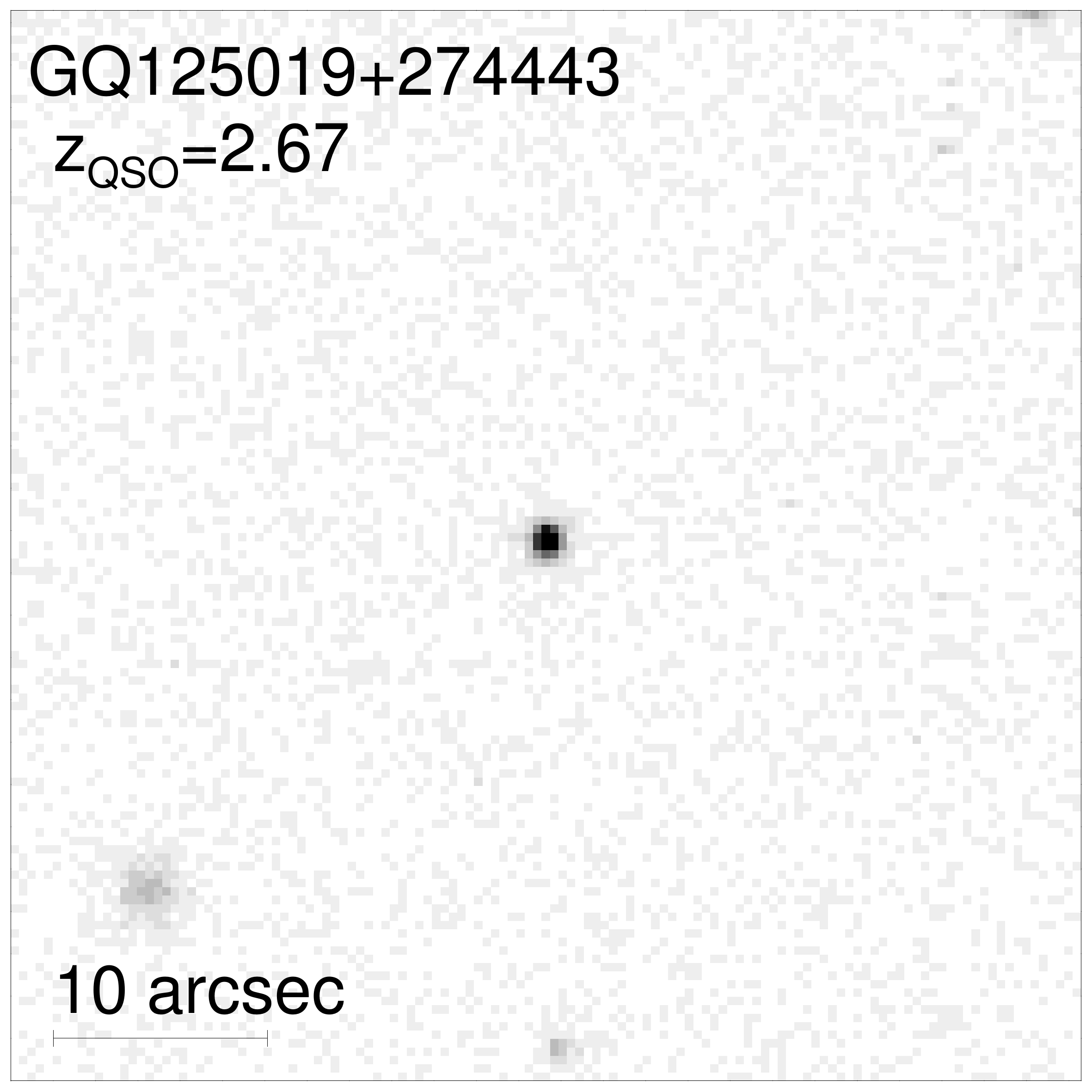,width=3.5cm} 
\epsfig{file=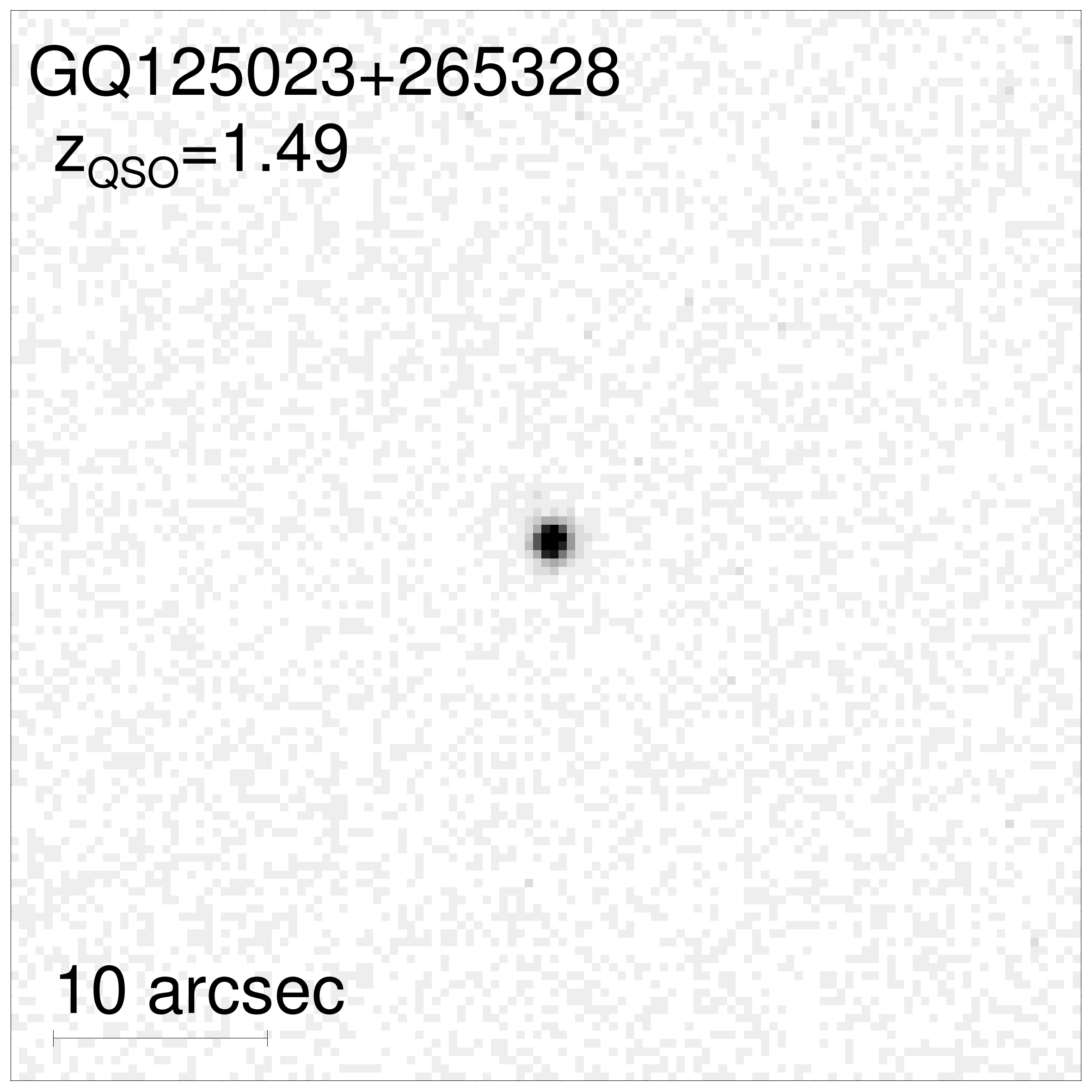,width=3.5cm} 
\epsfig{file=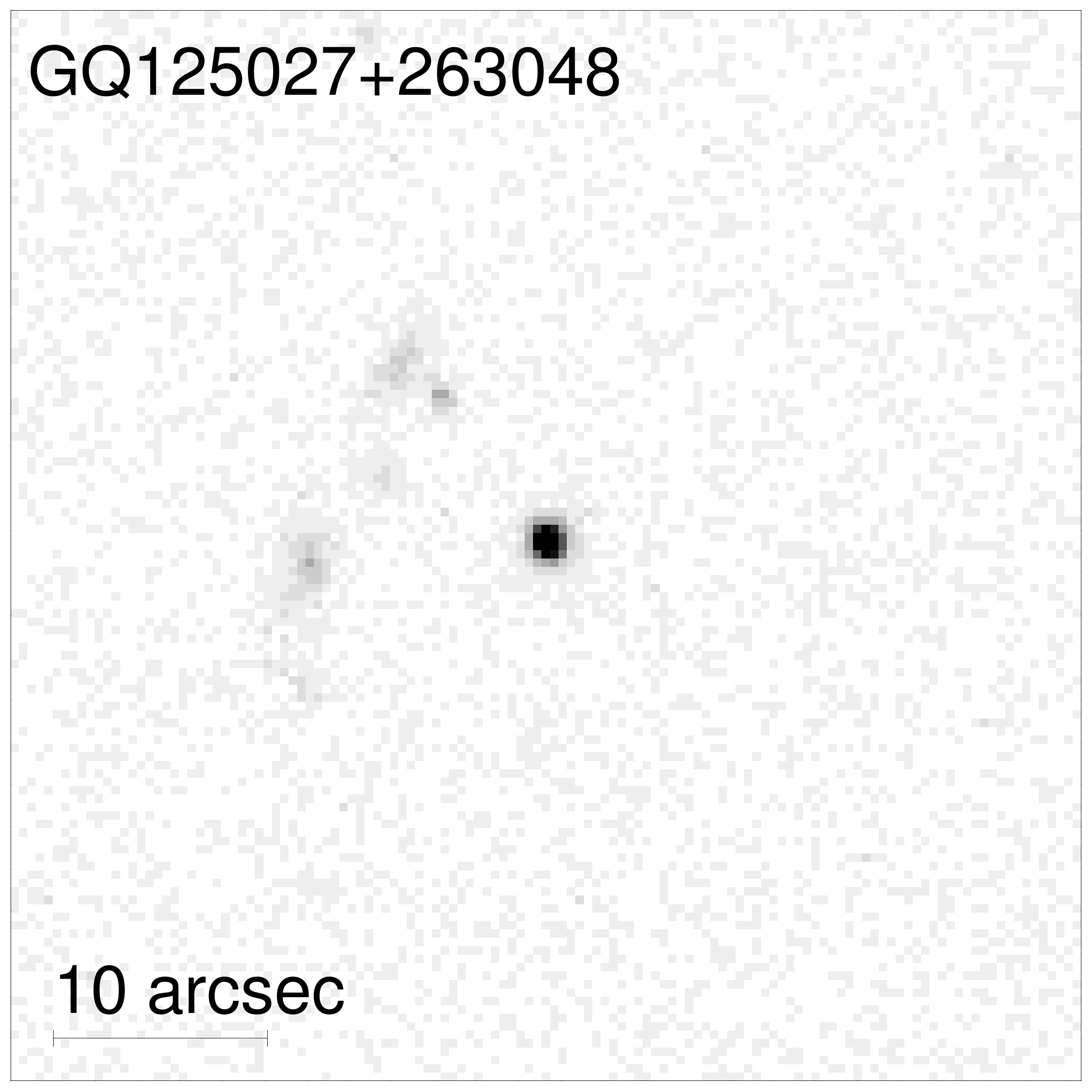,width=3.5cm} 
\epsfig{file=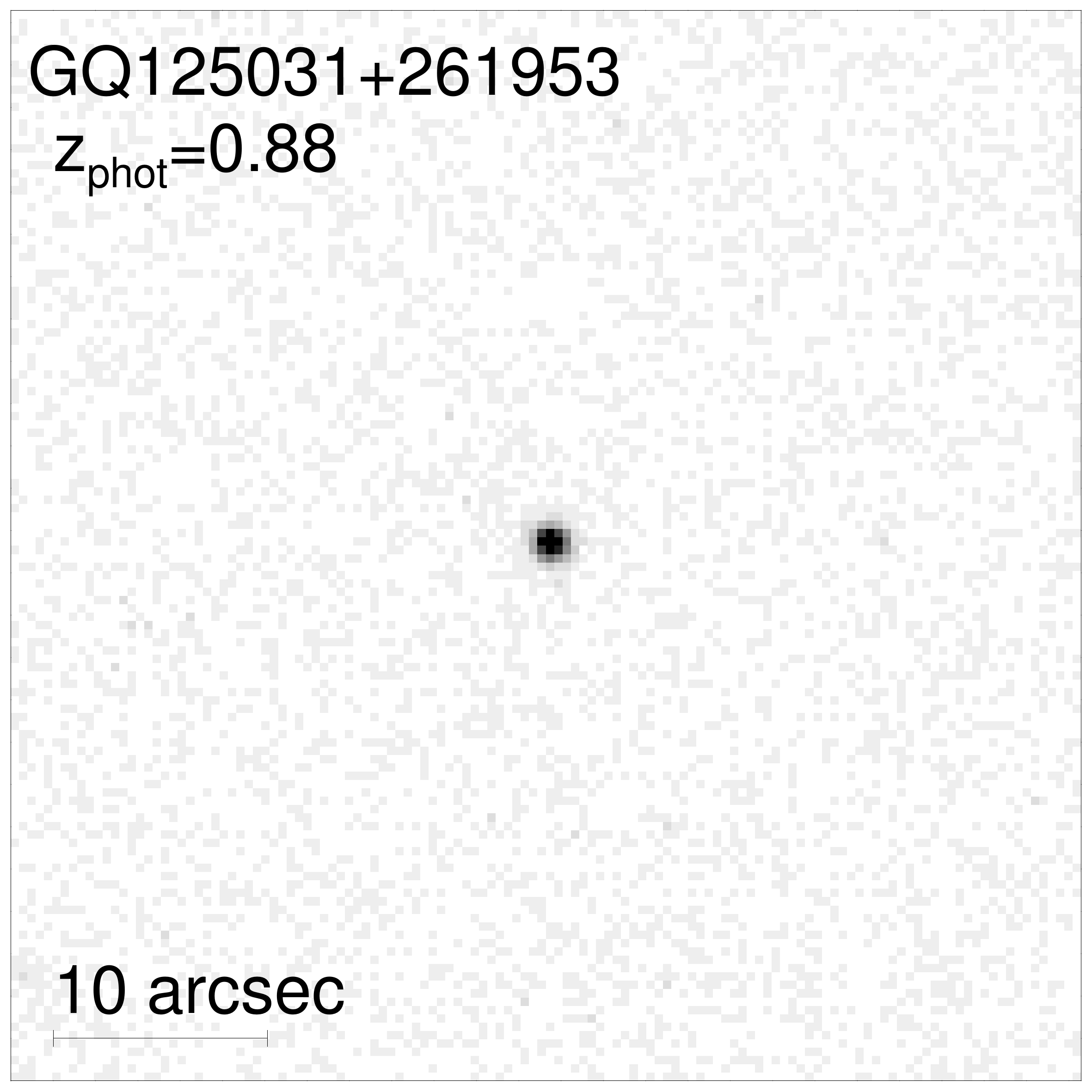,width=3.5cm} 
\epsfig{file=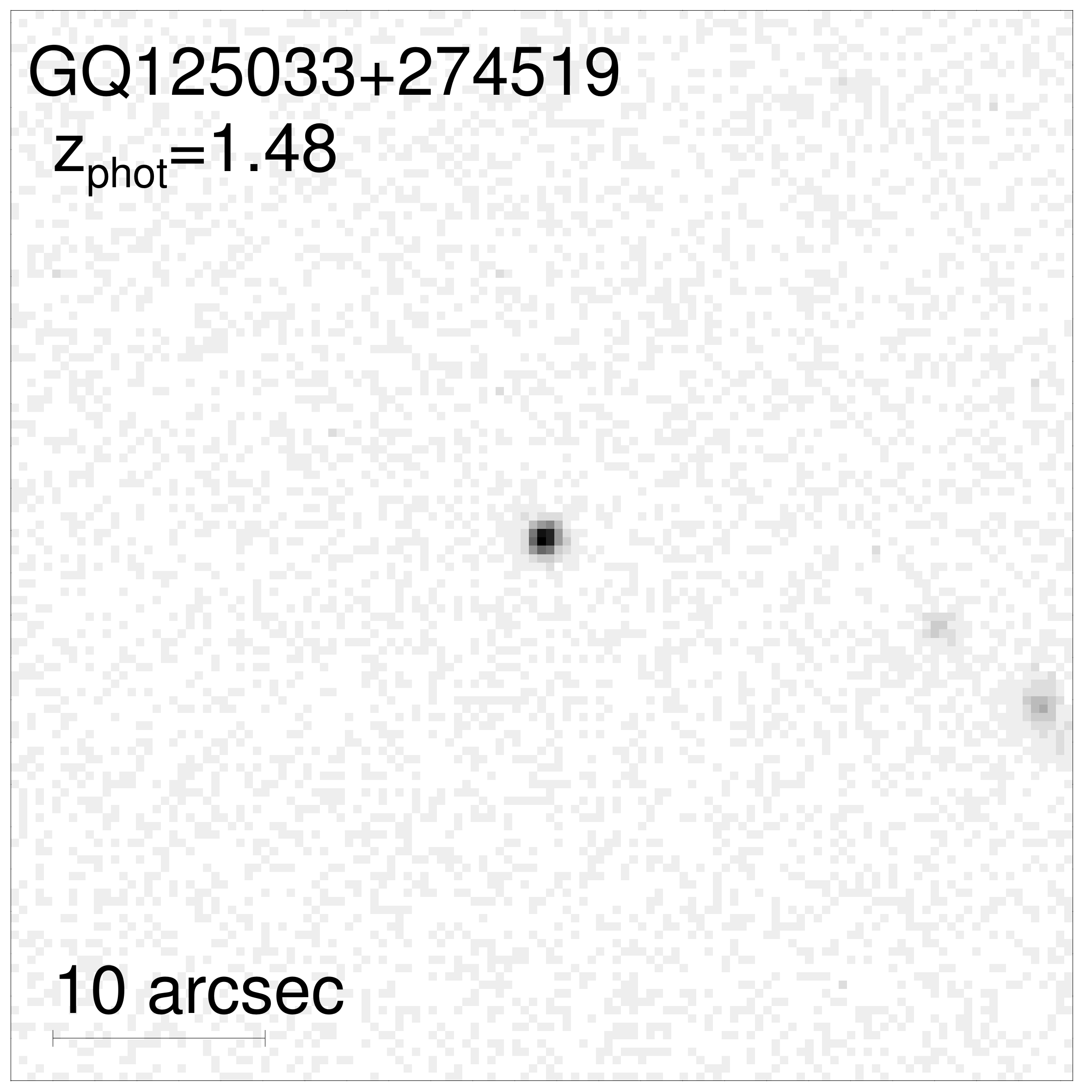,width=3.5cm} 
\epsfig{file=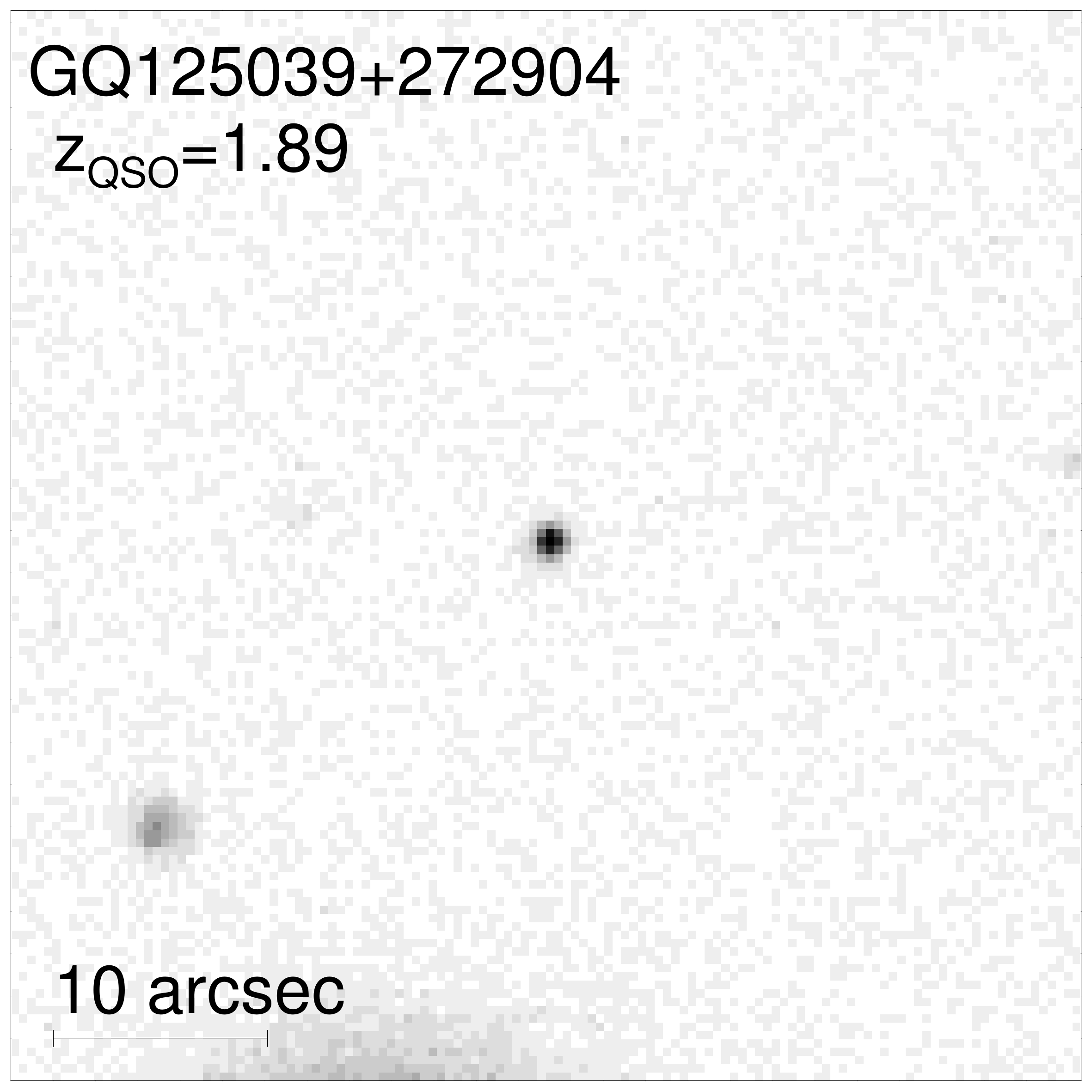,width=3.5cm} 
\epsfig{file=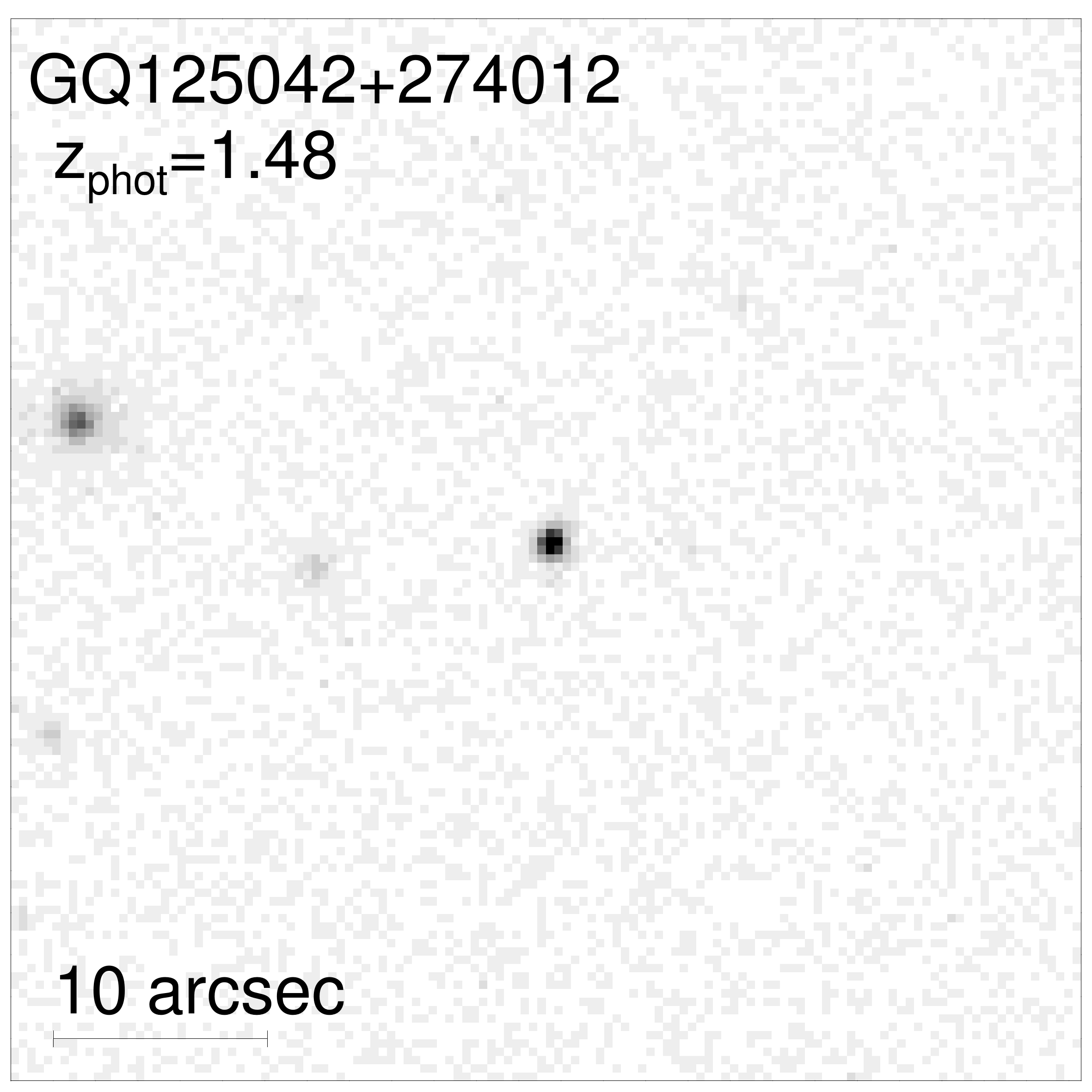,width=3.5cm} 
\epsfig{file=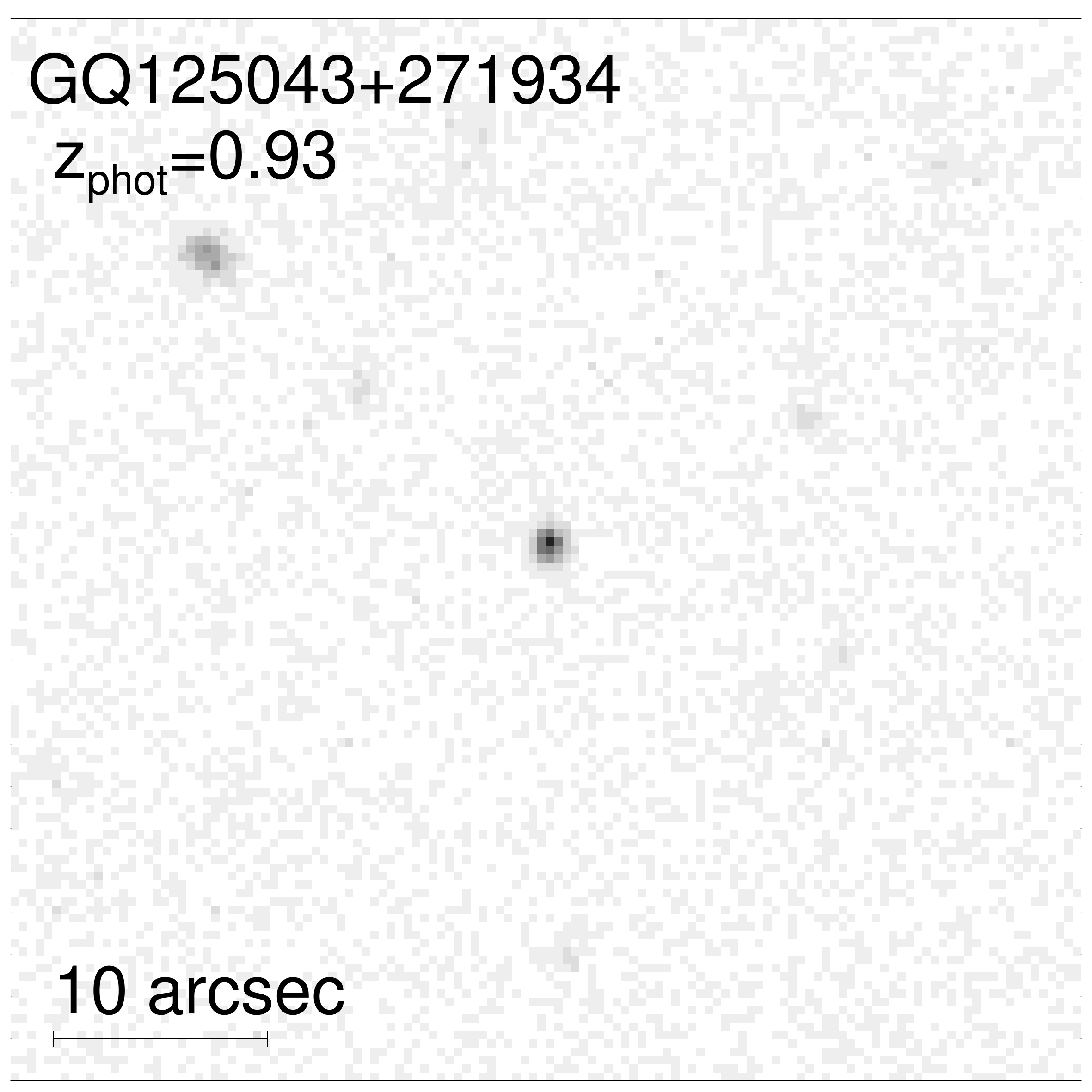,width=3.5cm} 
\epsfig{file=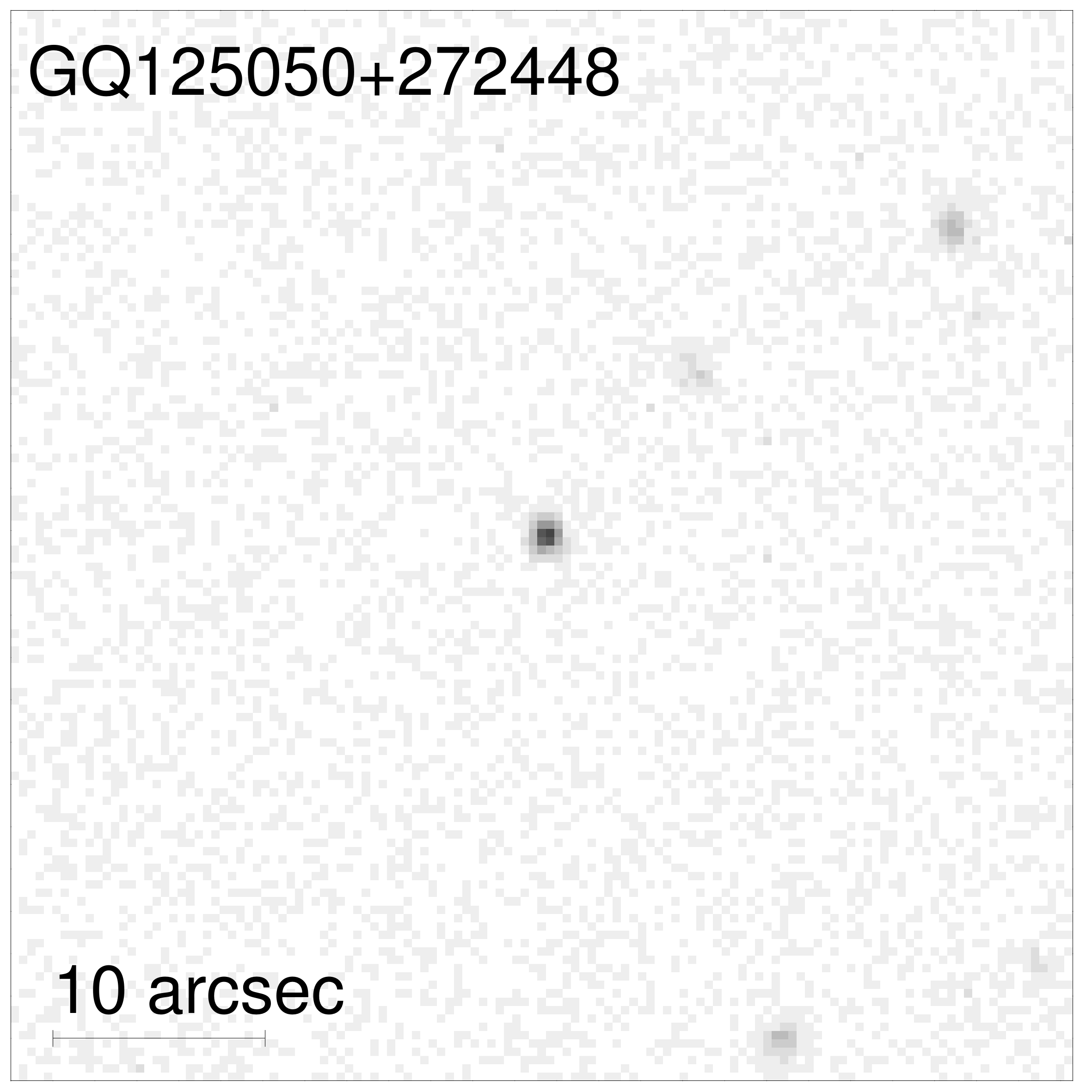,width=3.5cm} 
\epsfig{file=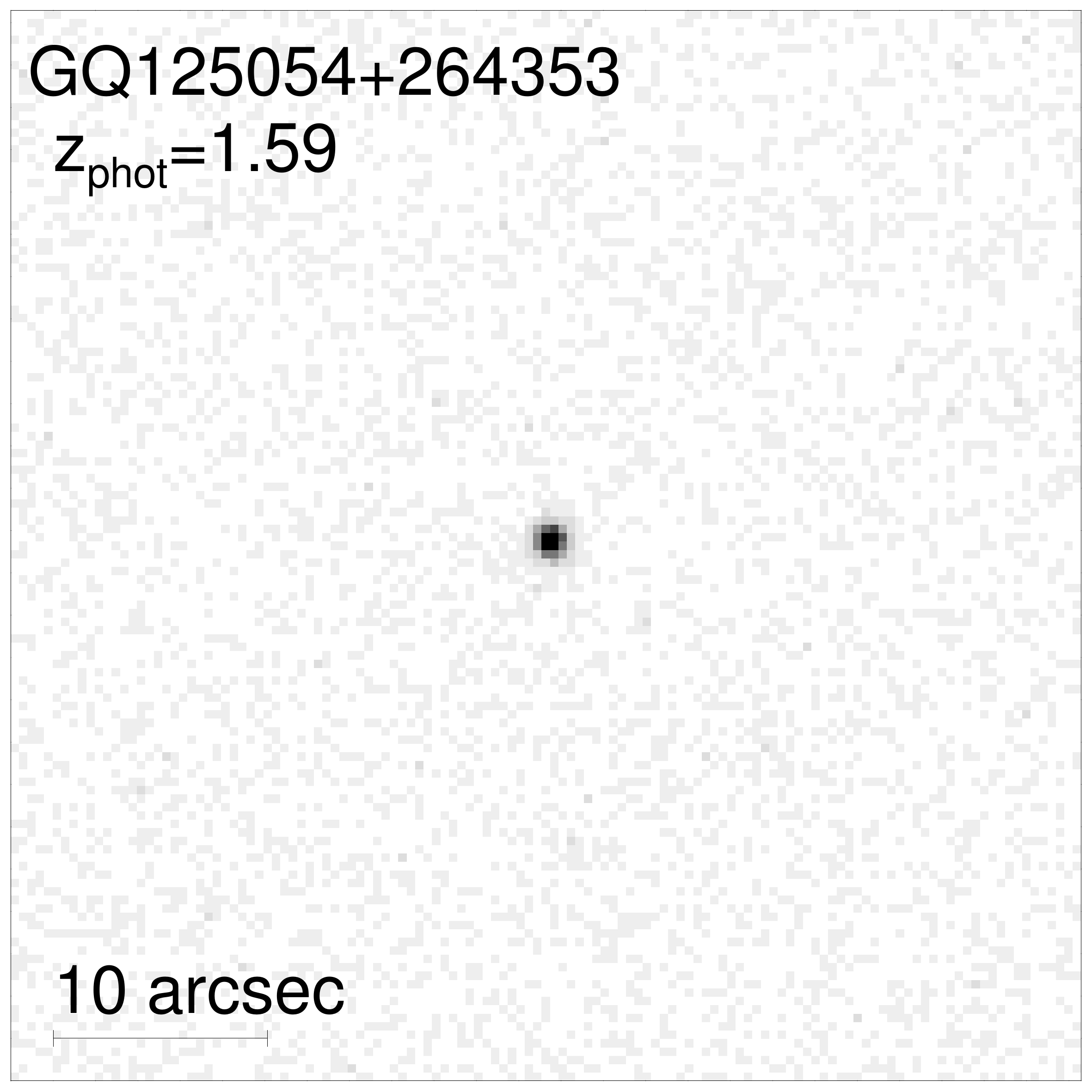,width=3.5cm} 
\epsfig{file=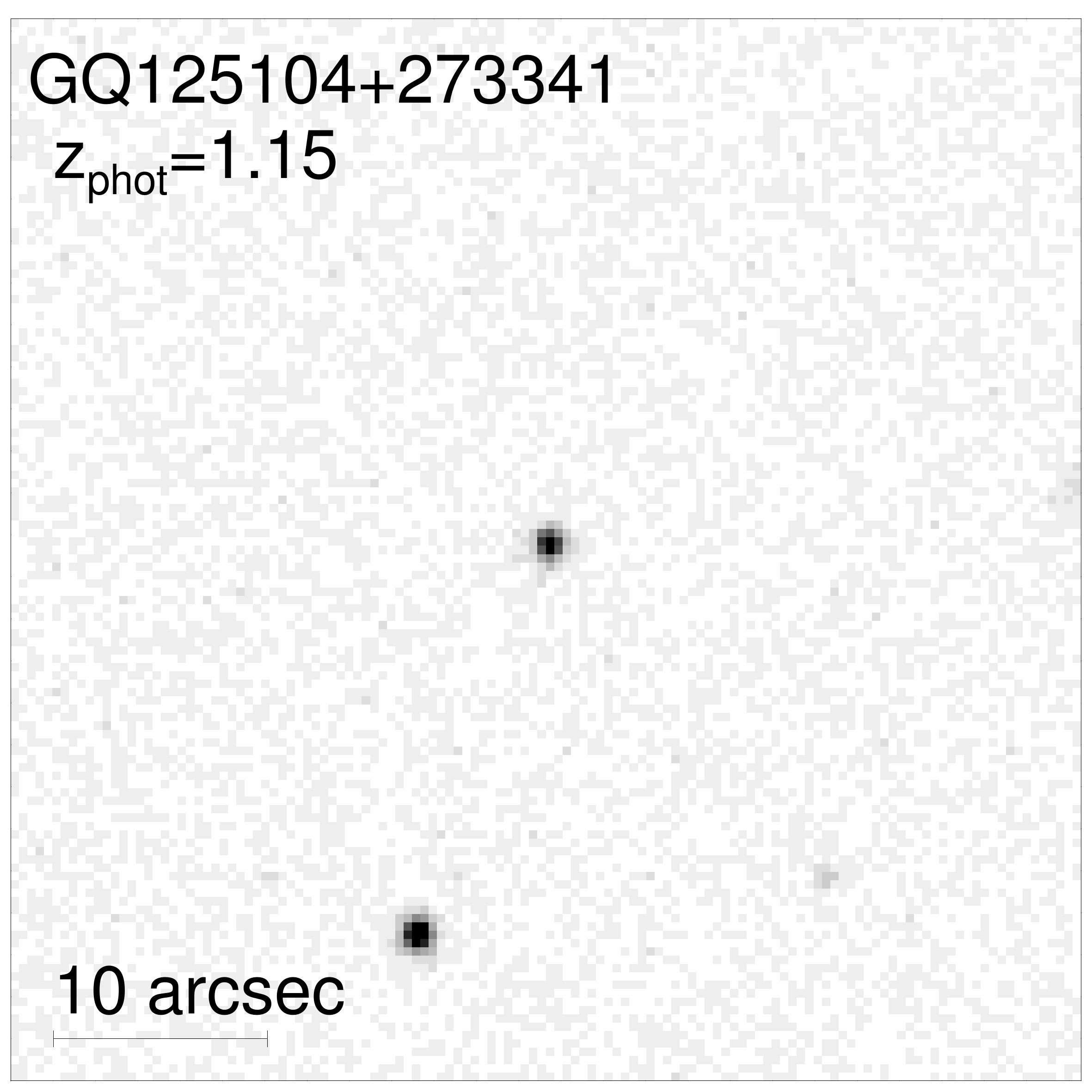,width=3.5cm} 
\epsfig{file=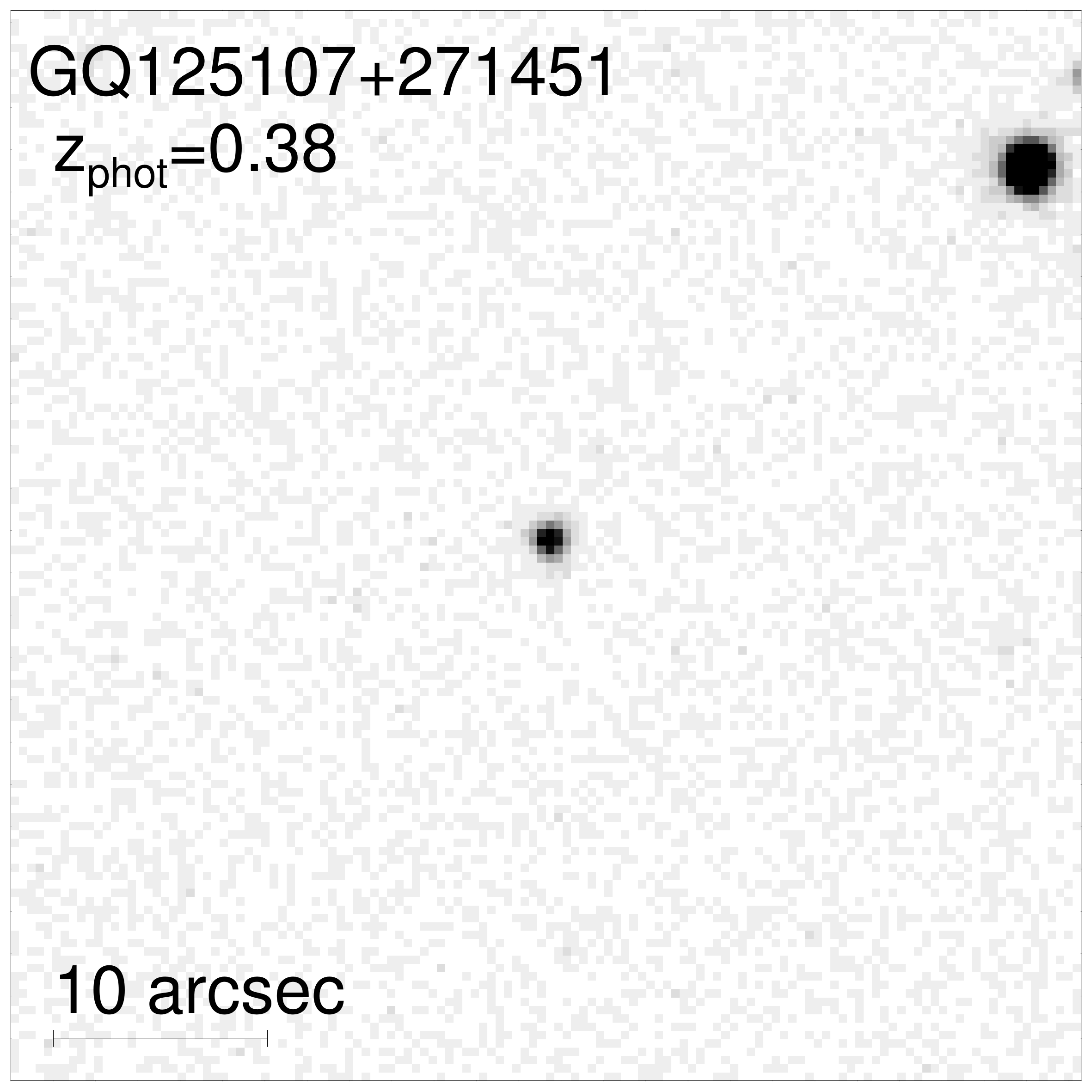,width=3.5cm} 
\epsfig{file=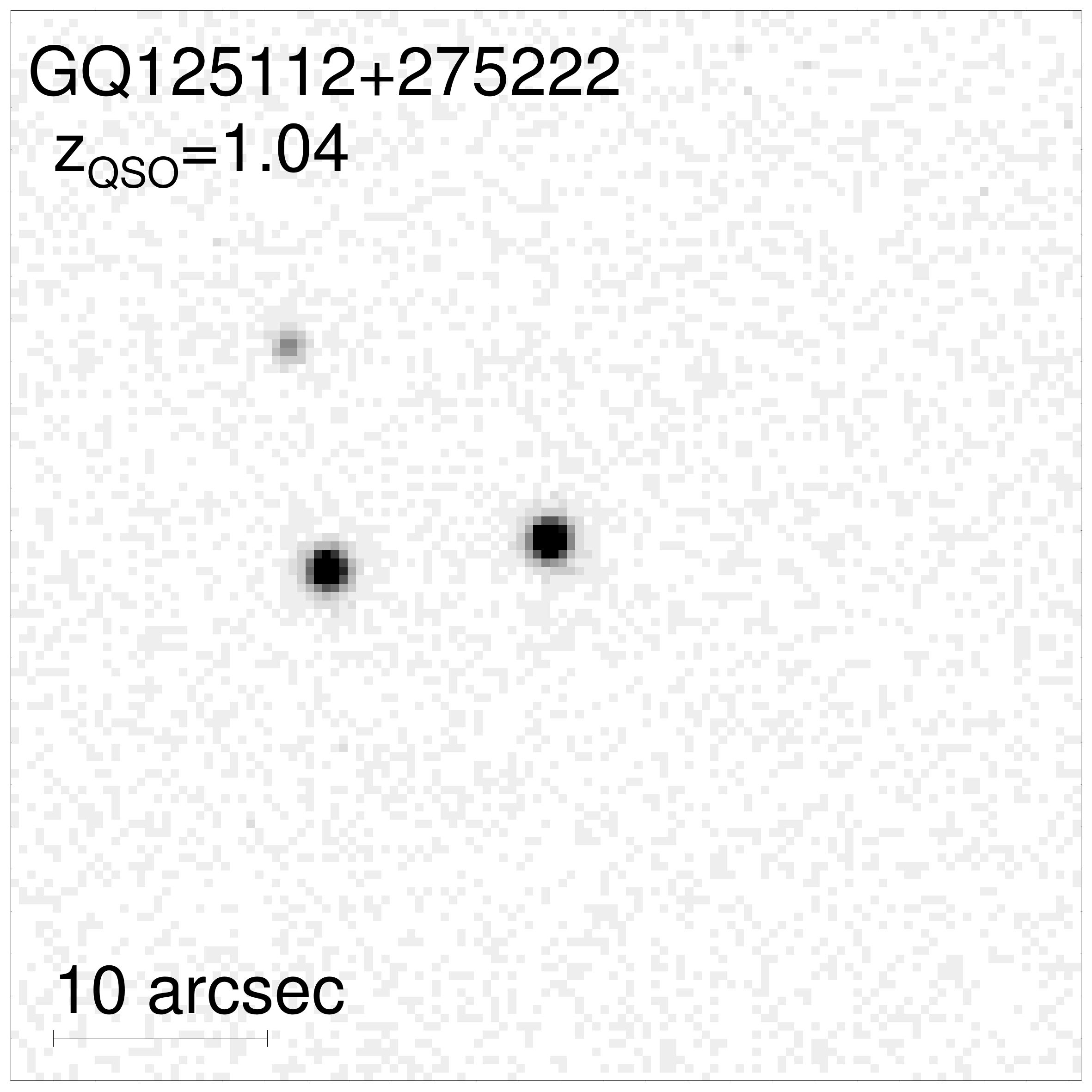,width=3.5cm} 
\epsfig{file=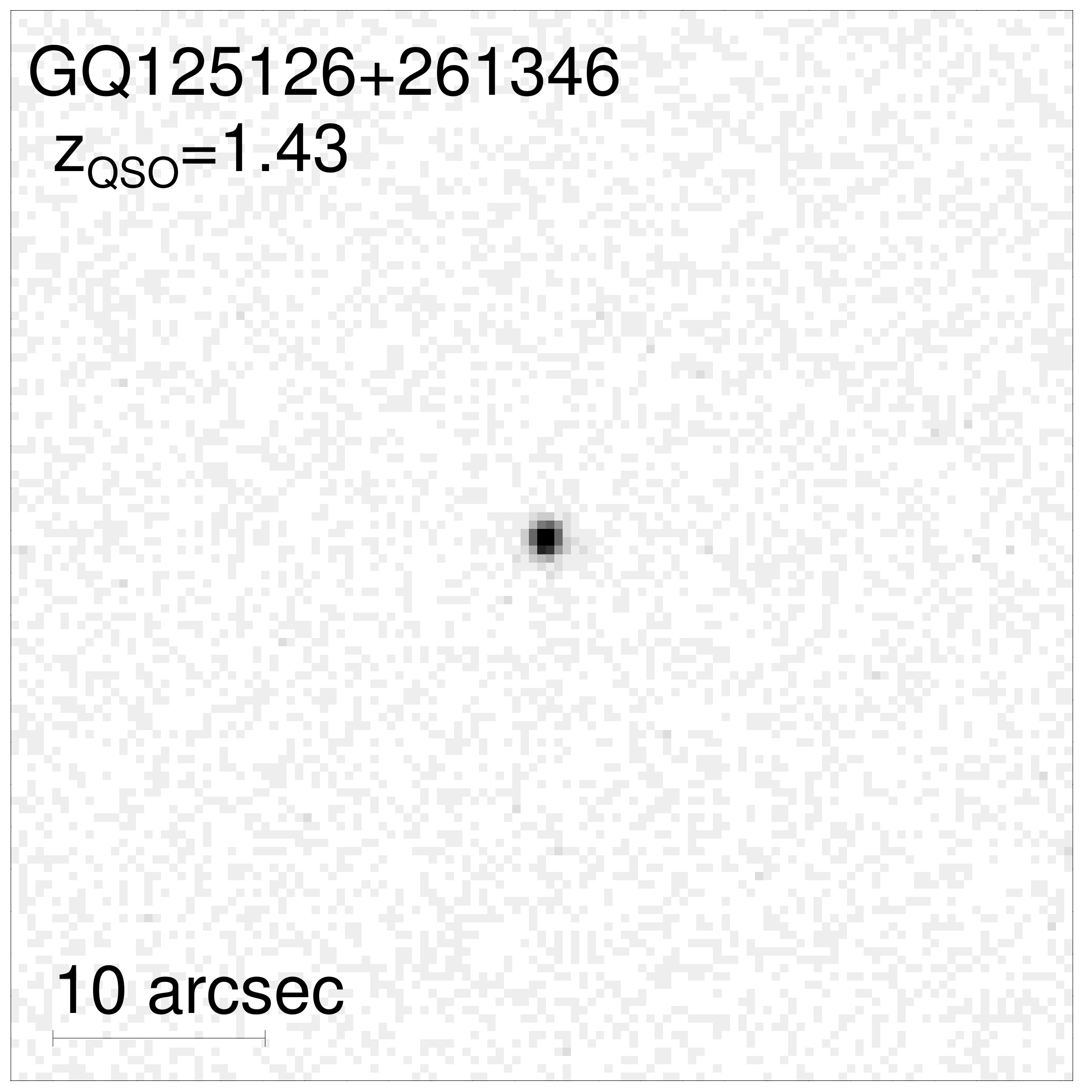,width=3.5cm} 
\epsfig{file=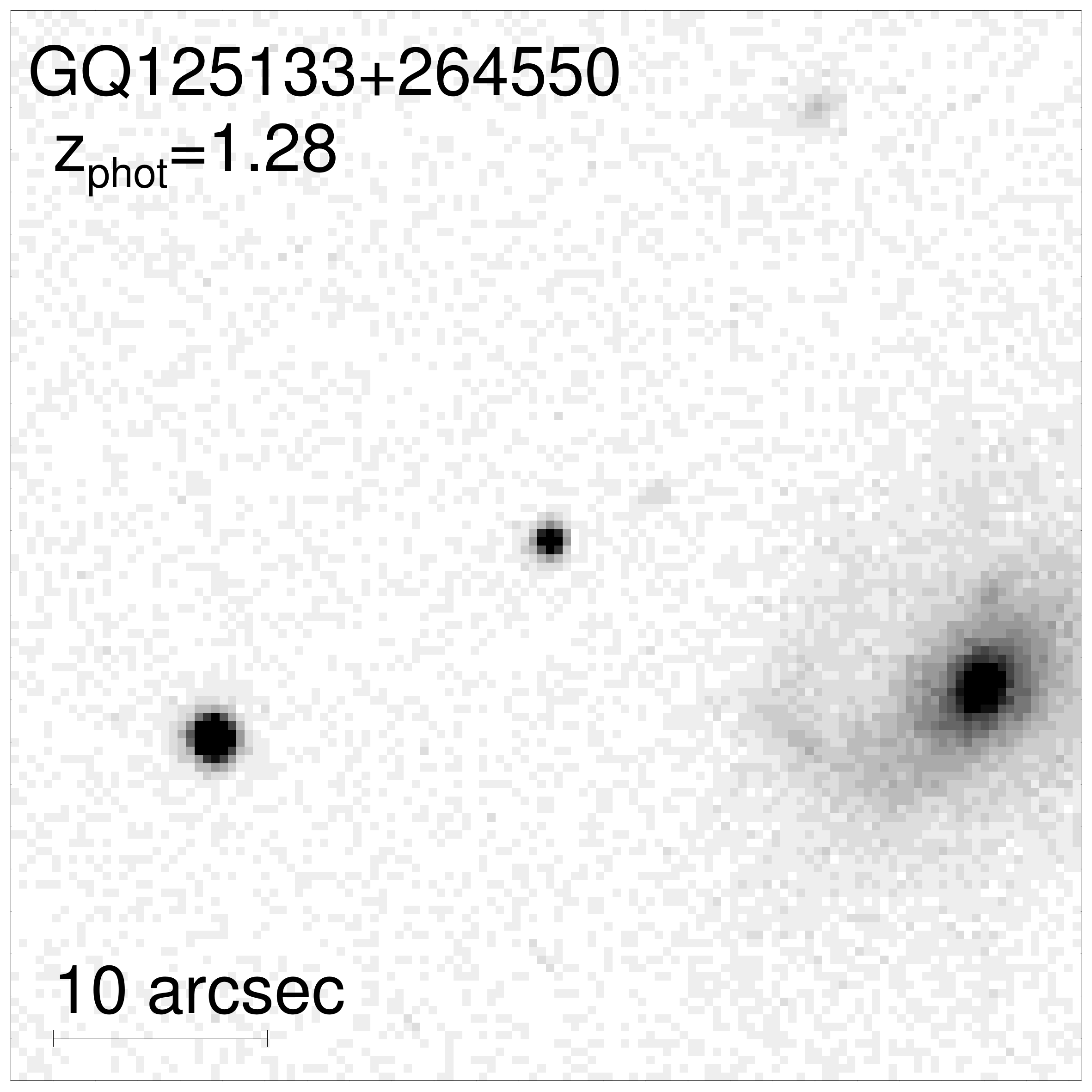,width=3.5cm} 
\epsfig{file=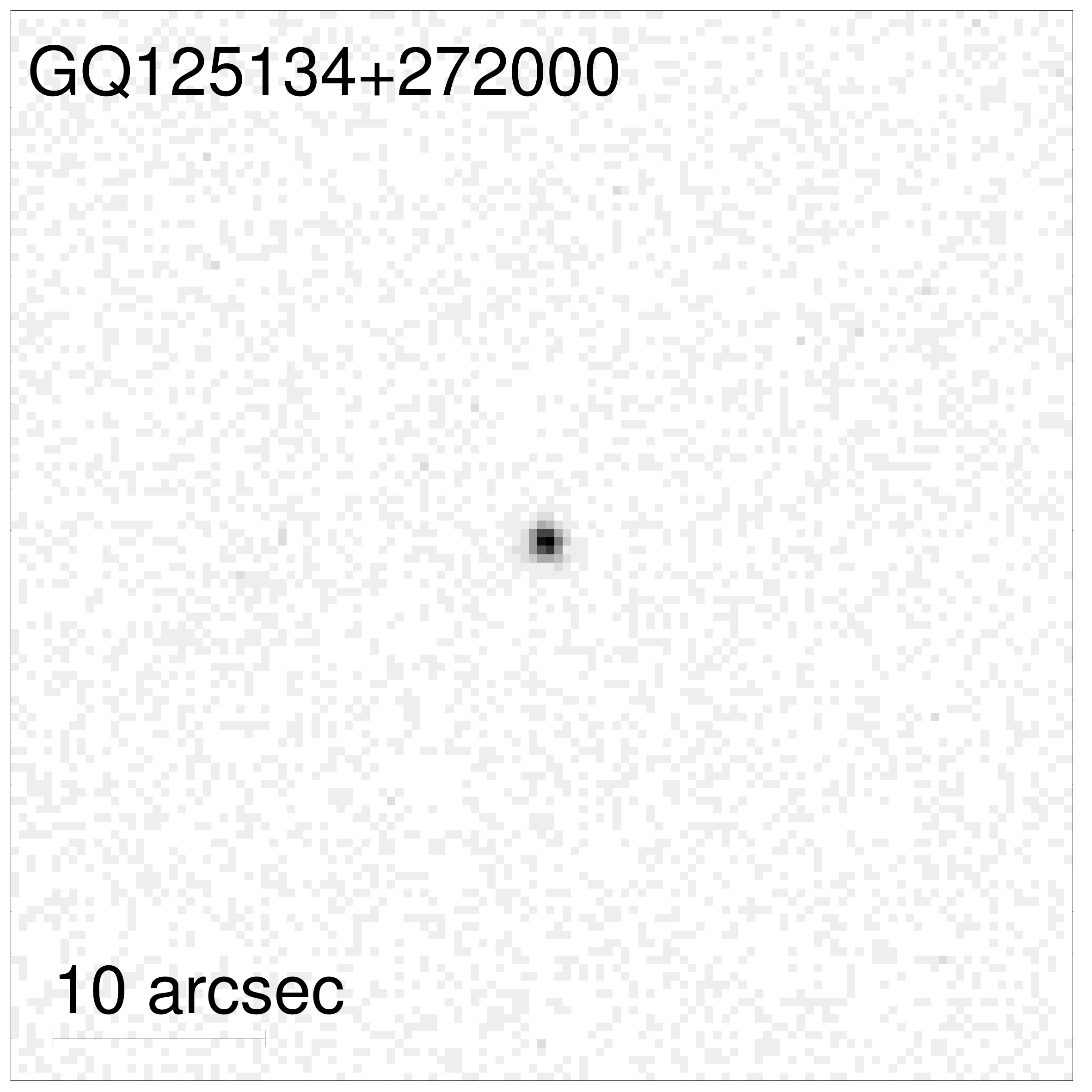,width=3.5cm} 
\epsfig{file=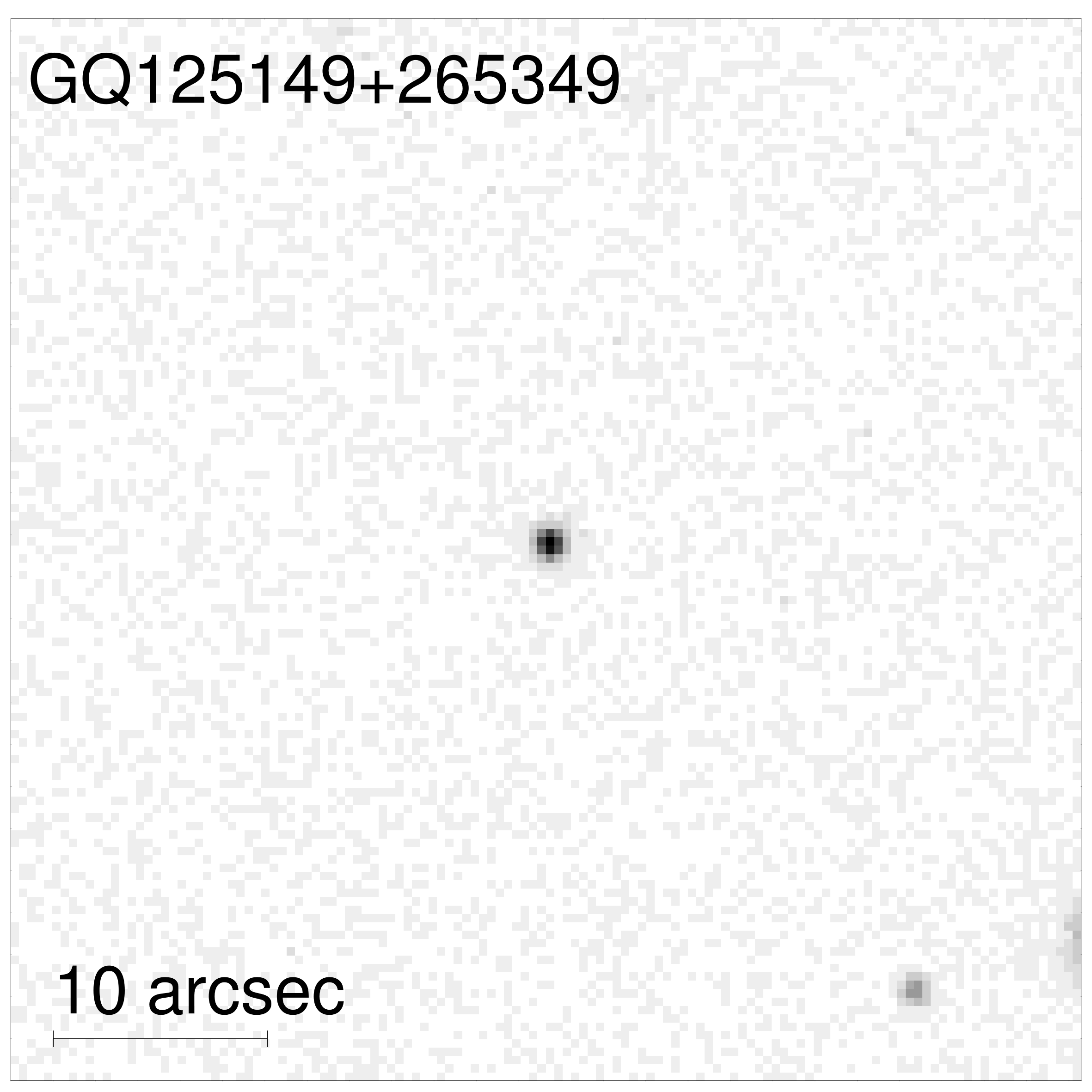,width=3.5cm} 
\epsfig{file=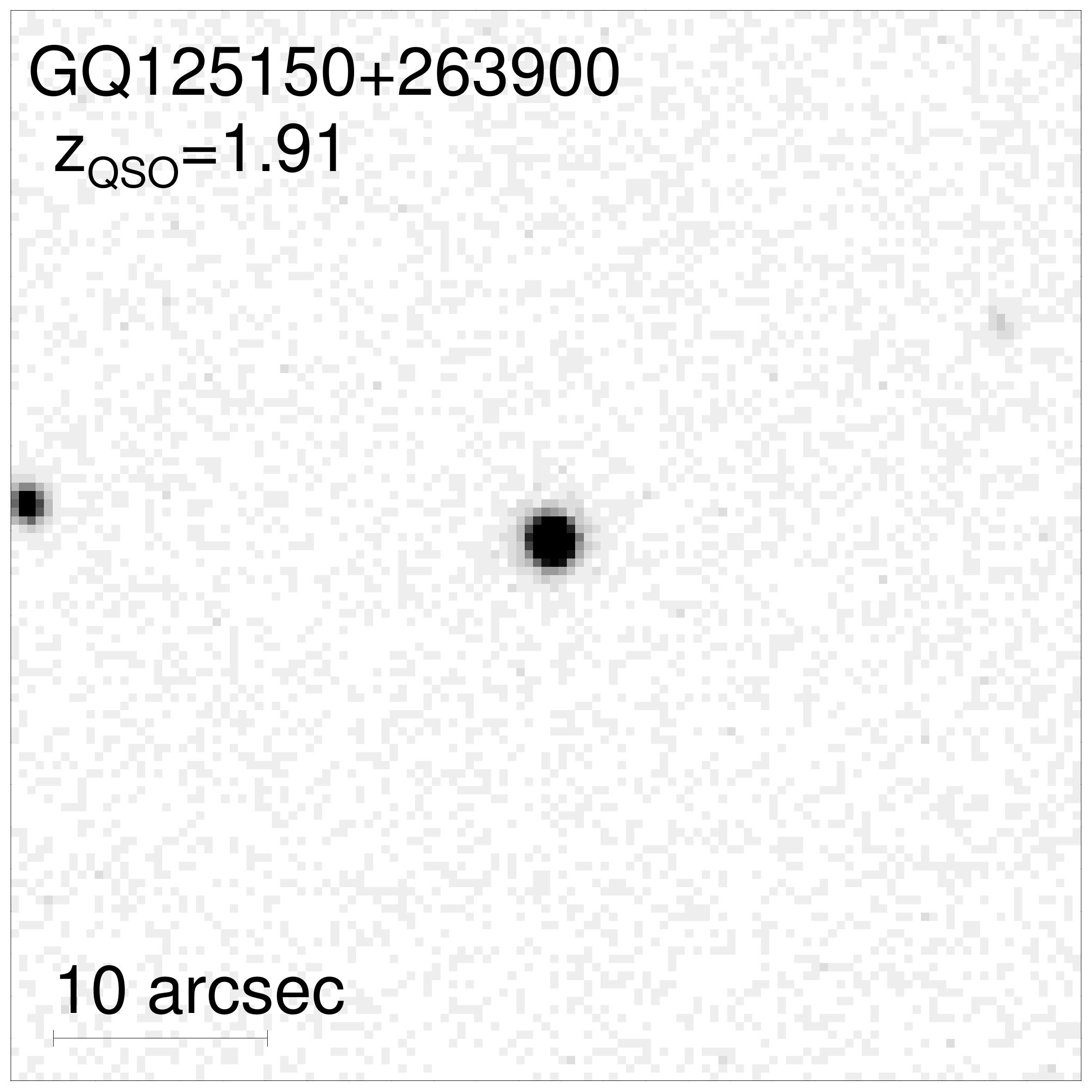,width=3.5cm} 
\epsfig{file=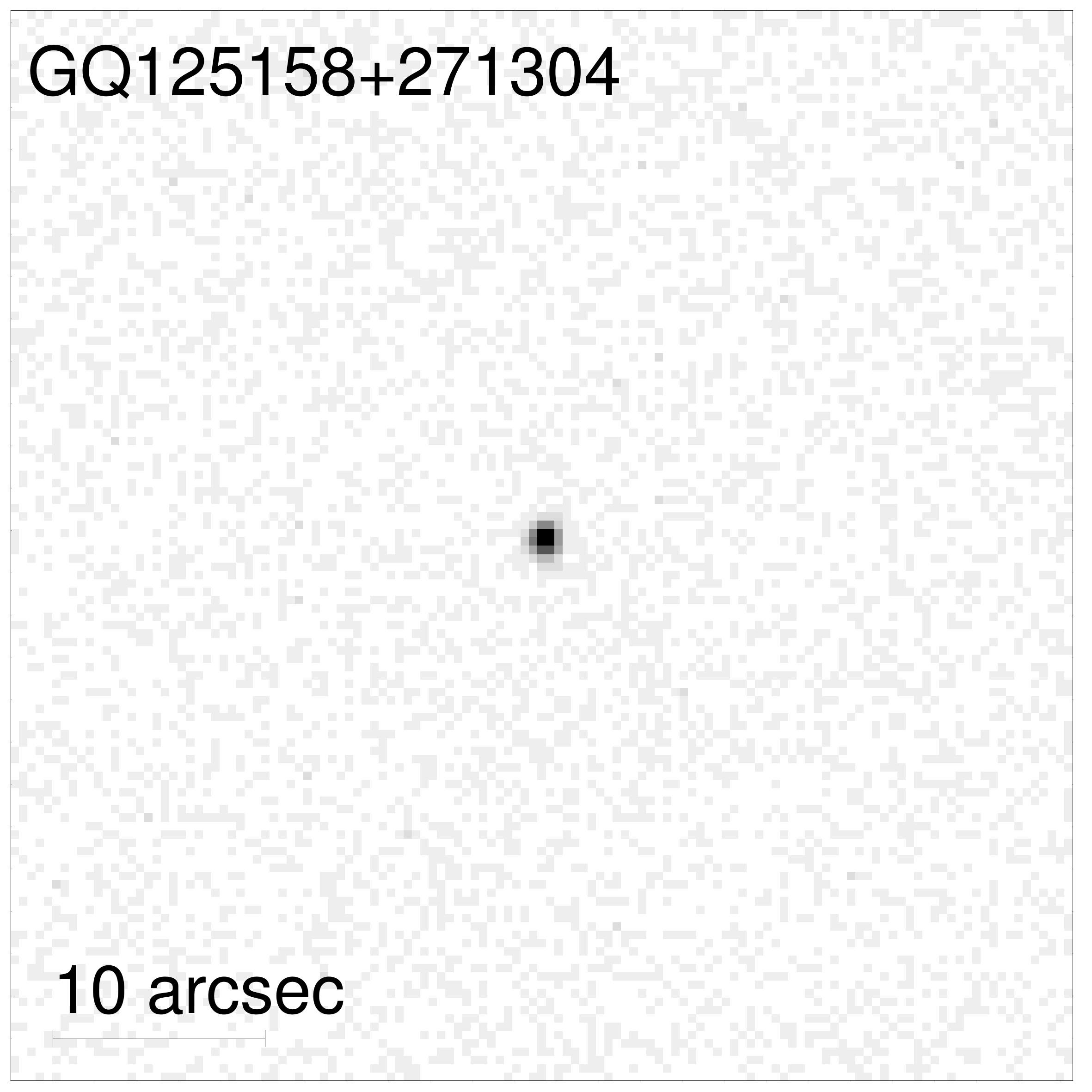,width=3.5cm} 
\epsfig{file=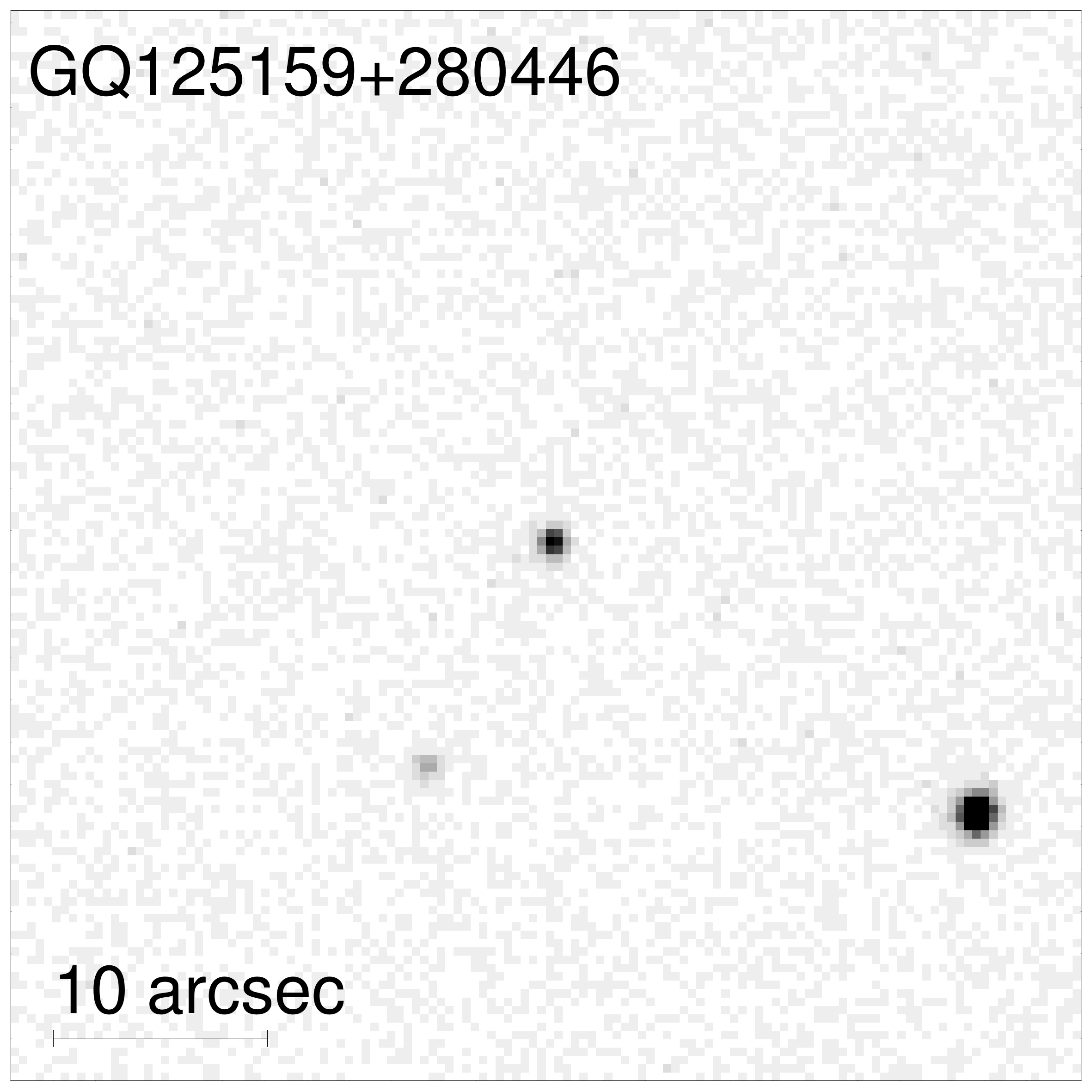,width=3.5cm} 
\epsfig{file=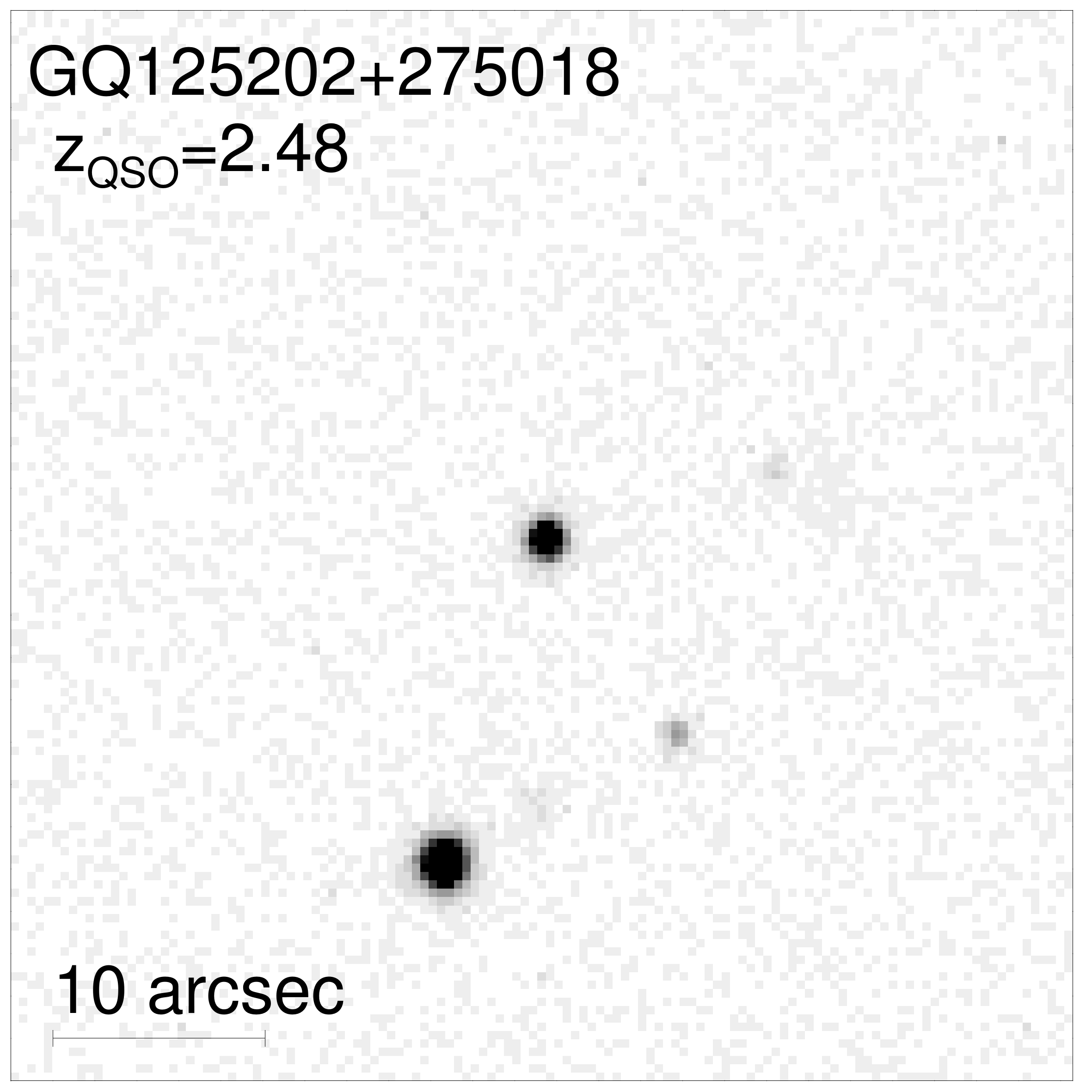,width=3.5cm} 
\epsfig{file=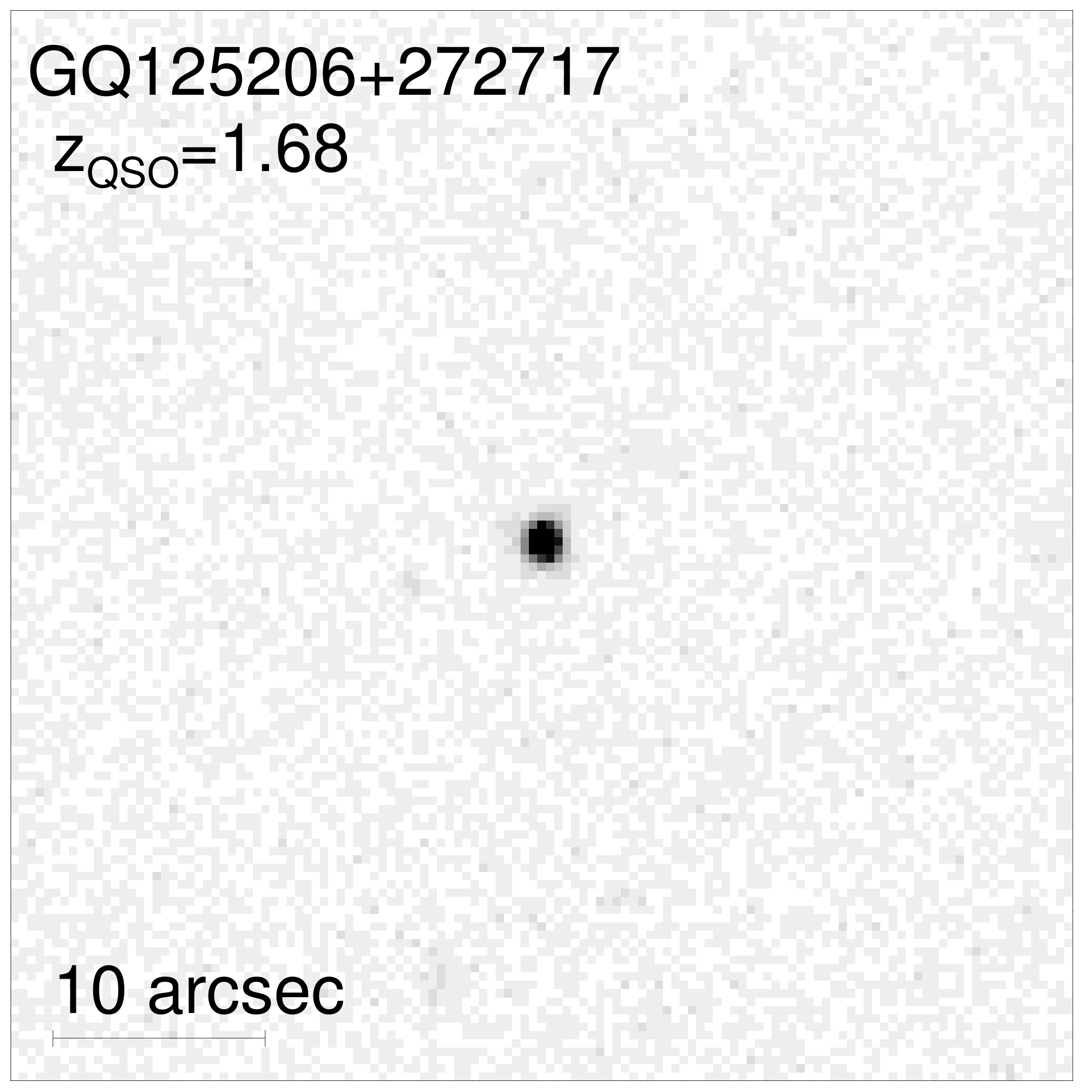,width=3.5cm} 
\epsfig{file=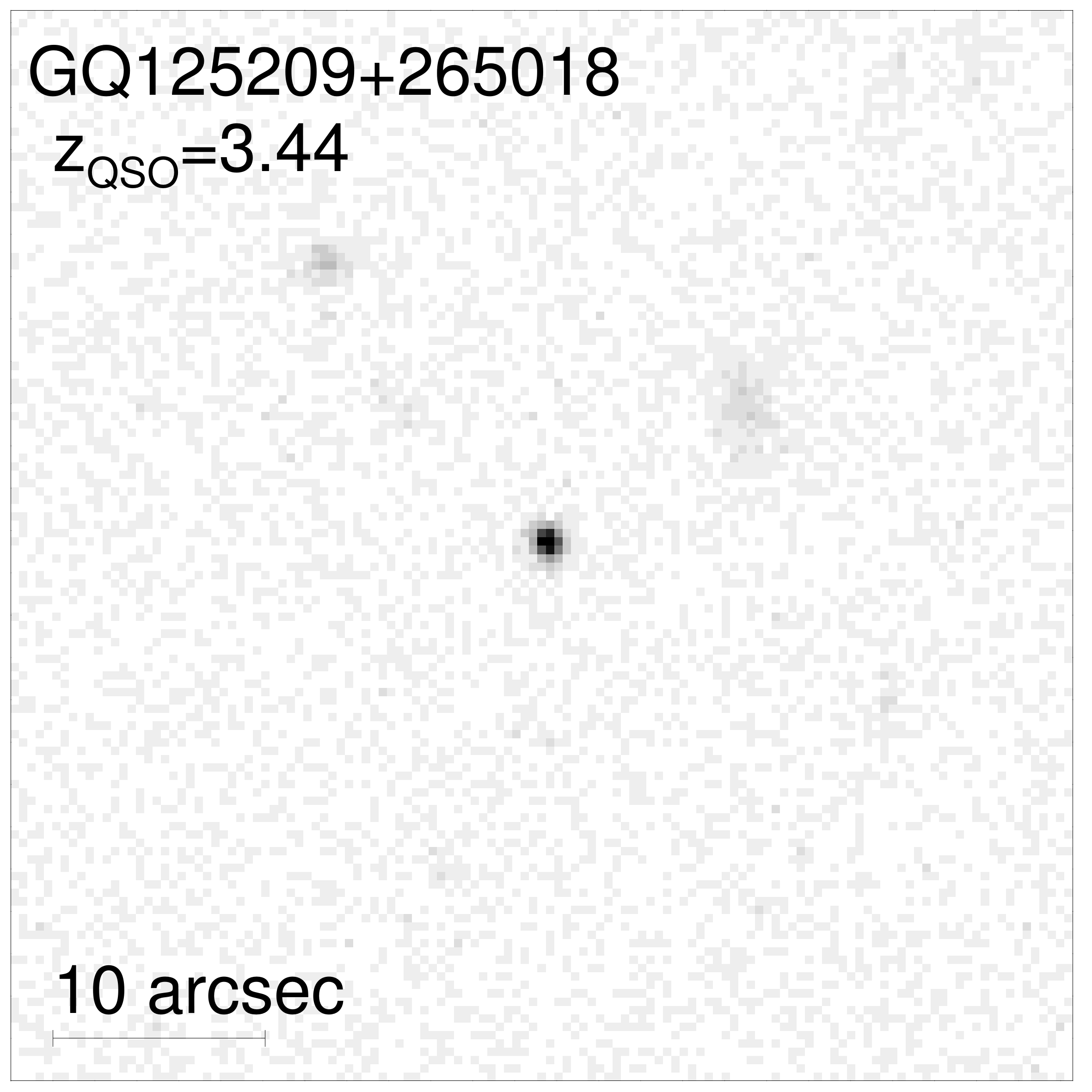,width=3.5cm} 
\epsfig{file=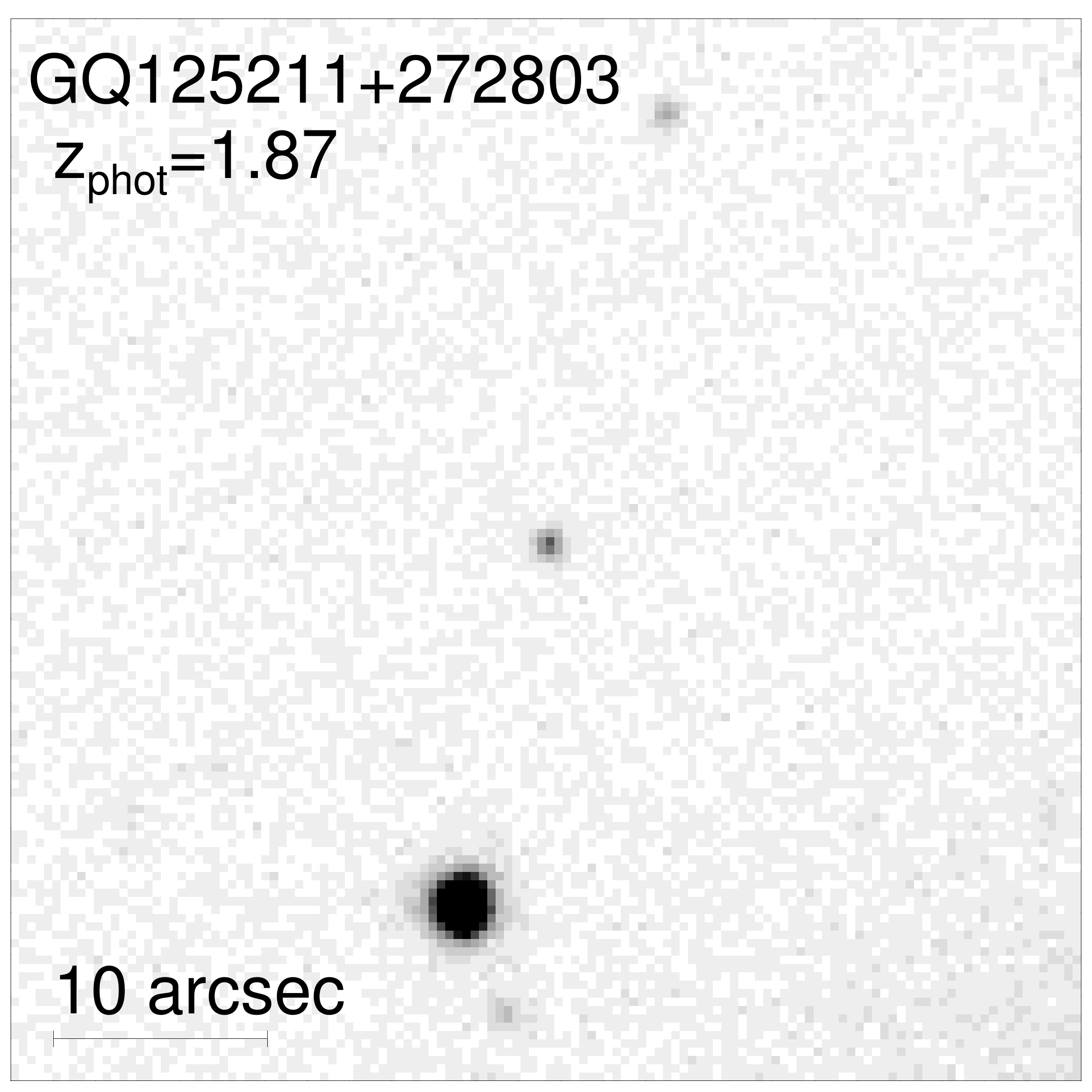,width=3.5cm} 
\epsfig{file=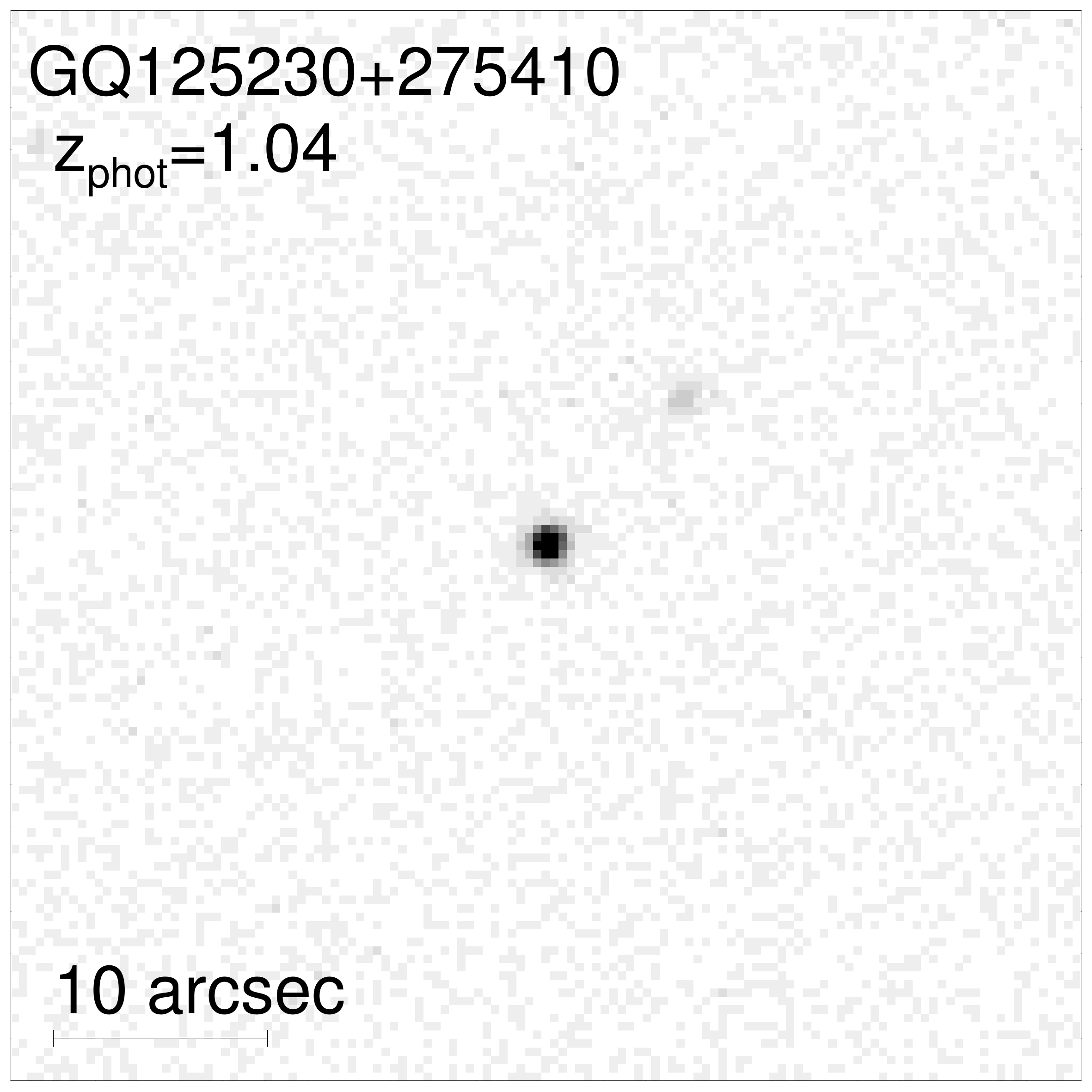,width=3.5cm} 
    \caption{50$\times$50 arcsec$^2$ thumbnails around each stationary source.}
	\label{fig:mosaic2}
\end{figure*}

\begin{figure*} [!t]
\centering
\epsfig{file=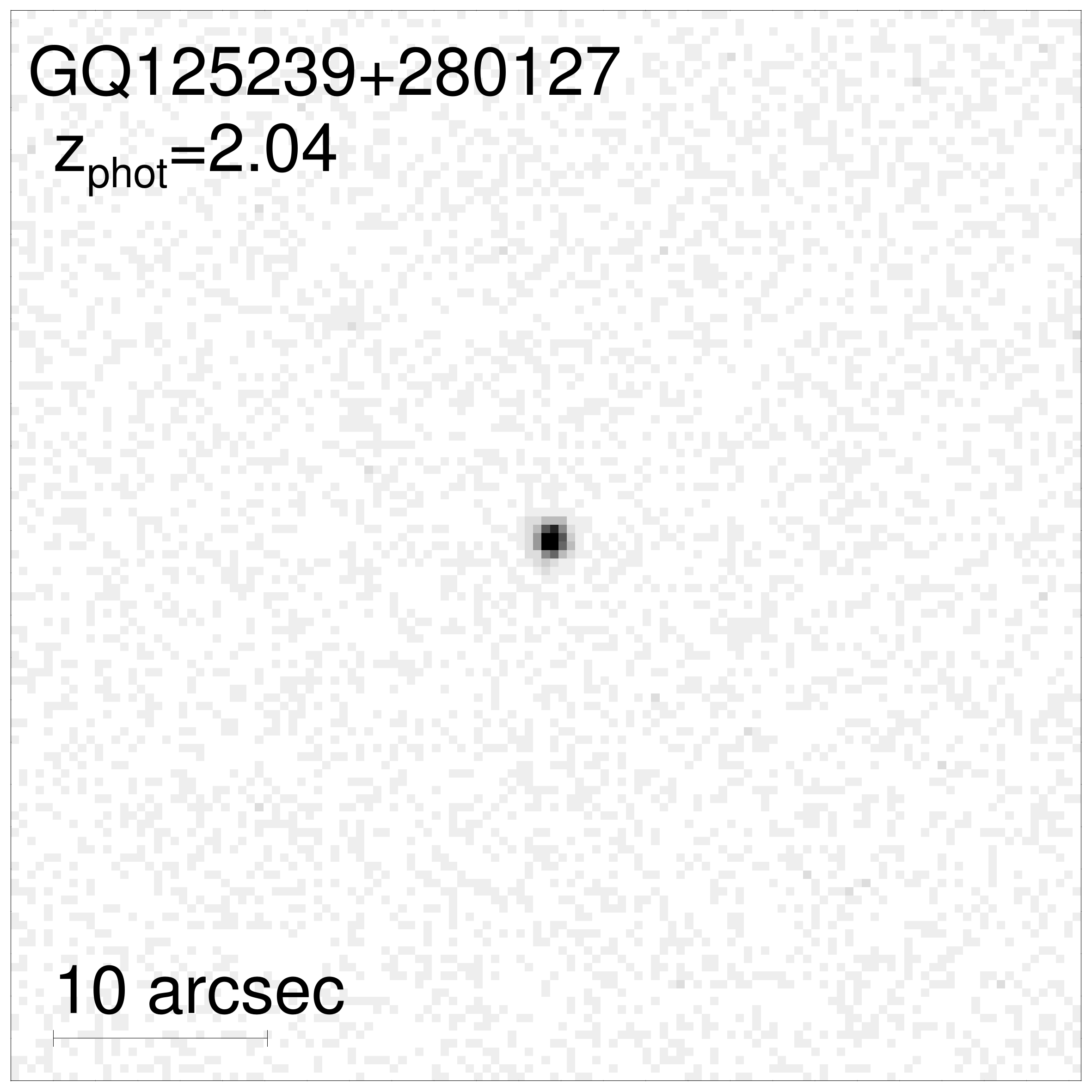,width=3.5cm} 
\epsfig{file=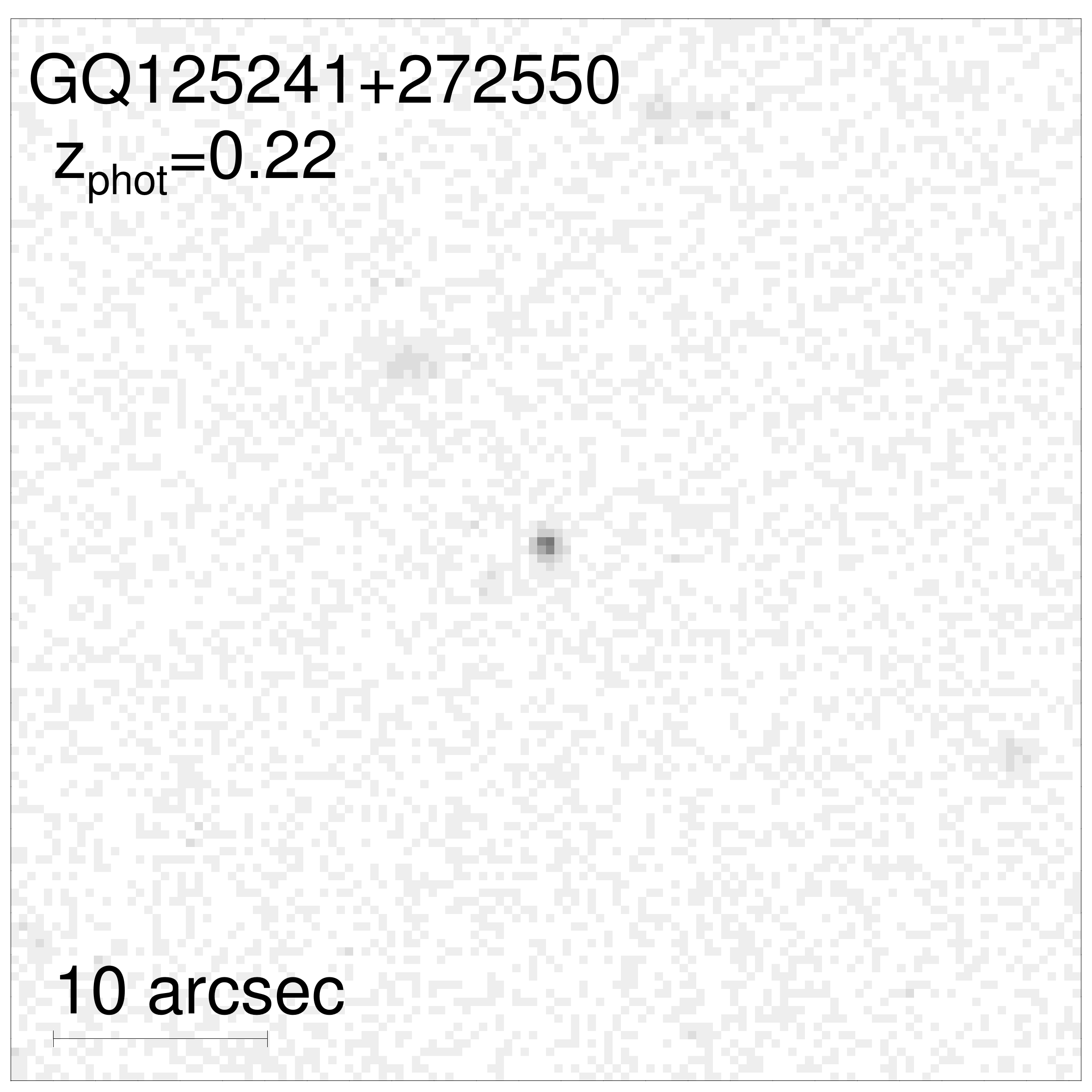,width=3.5cm} 
\epsfig{file=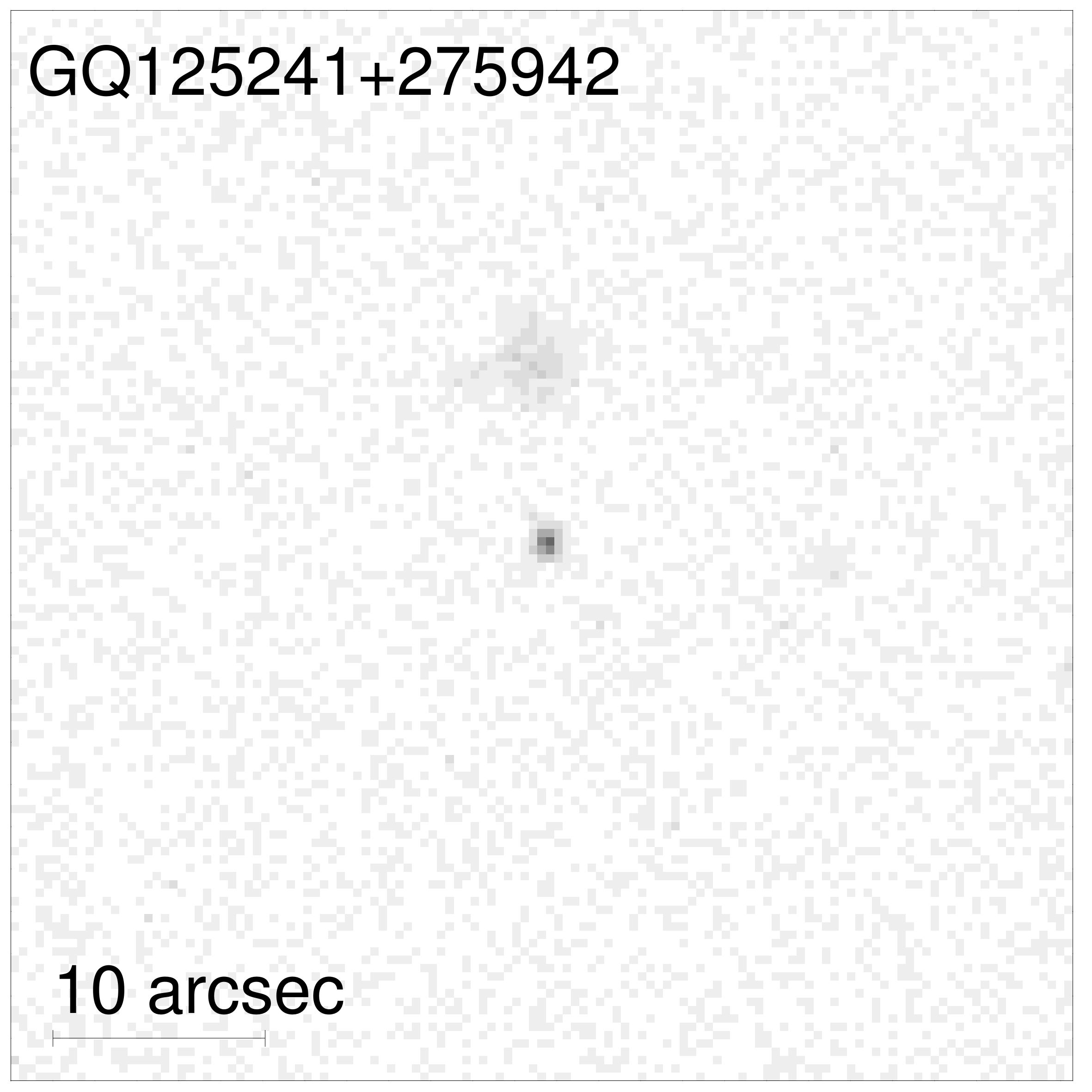,width=3.5cm} 
\epsfig{file=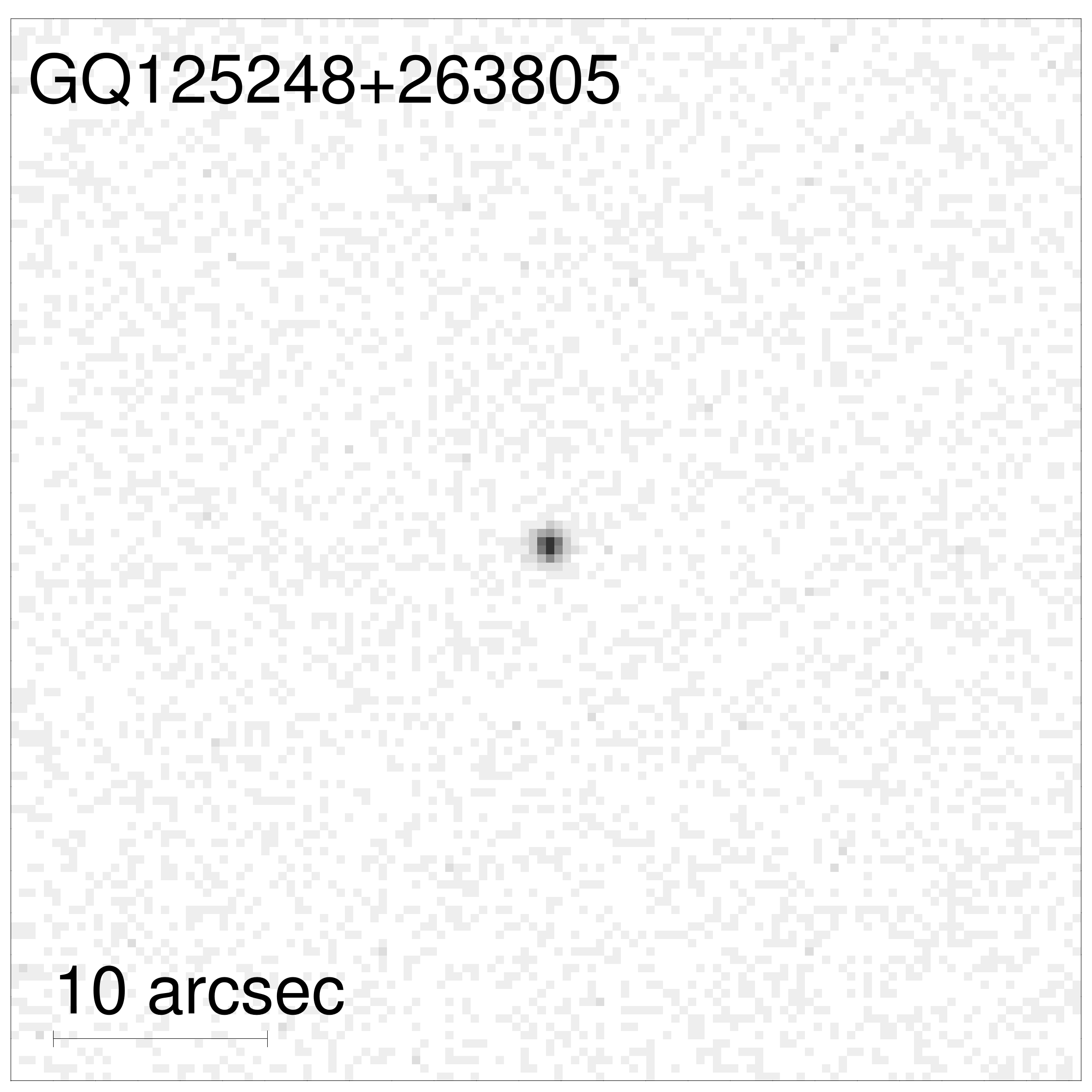,width=3.5cm} 
\epsfig{file=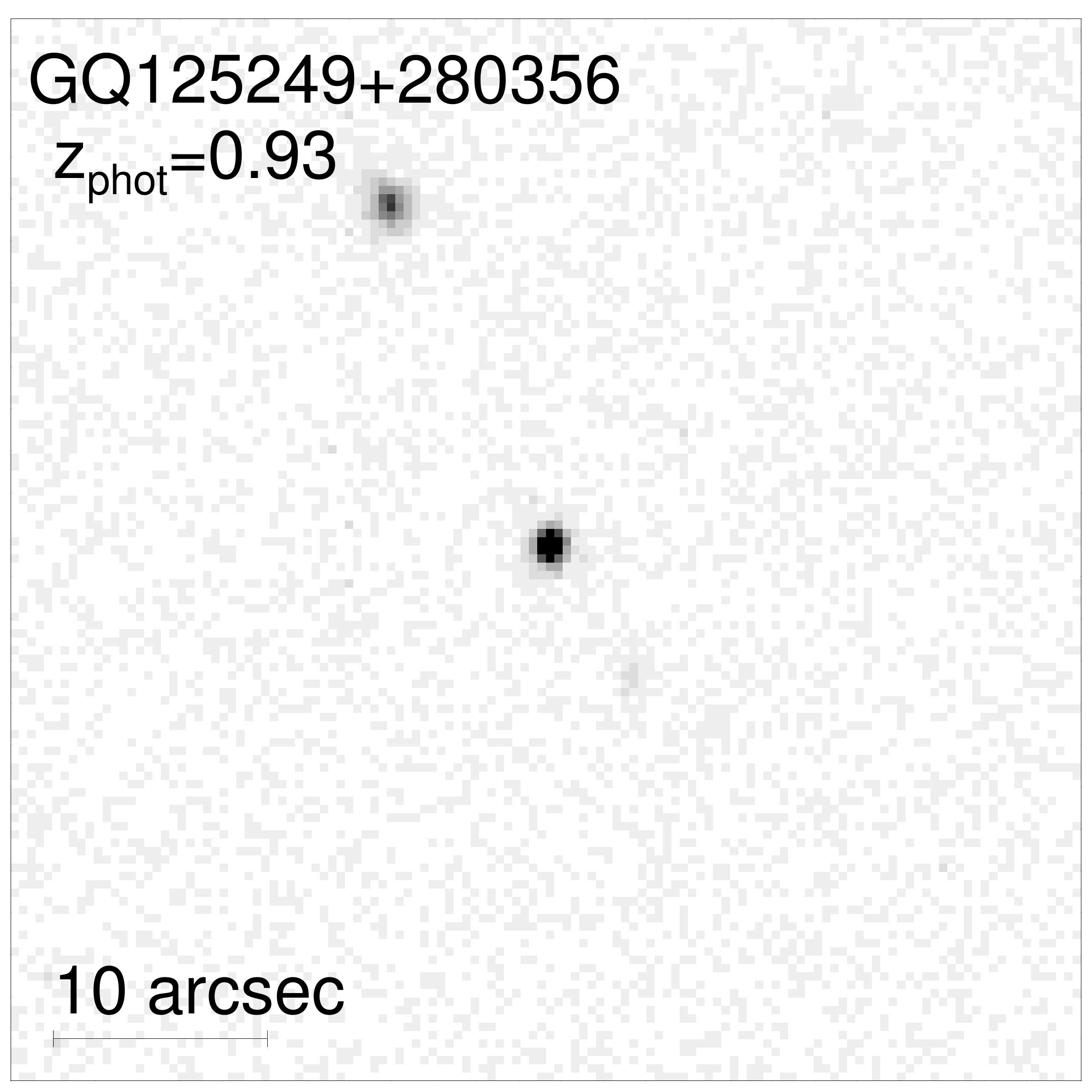,width=3.5cm} 
\epsfig{file=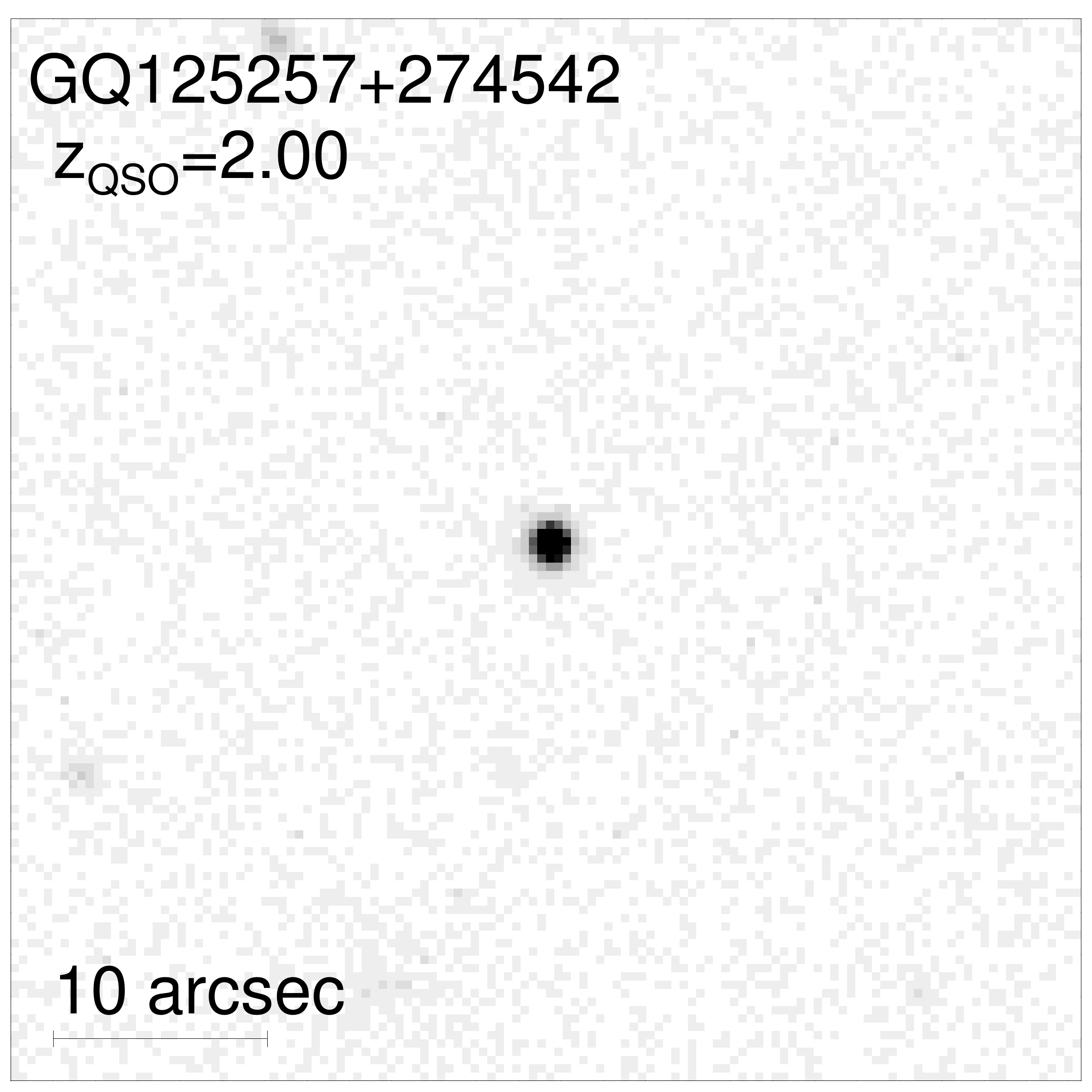,width=3.5cm} 
\epsfig{file=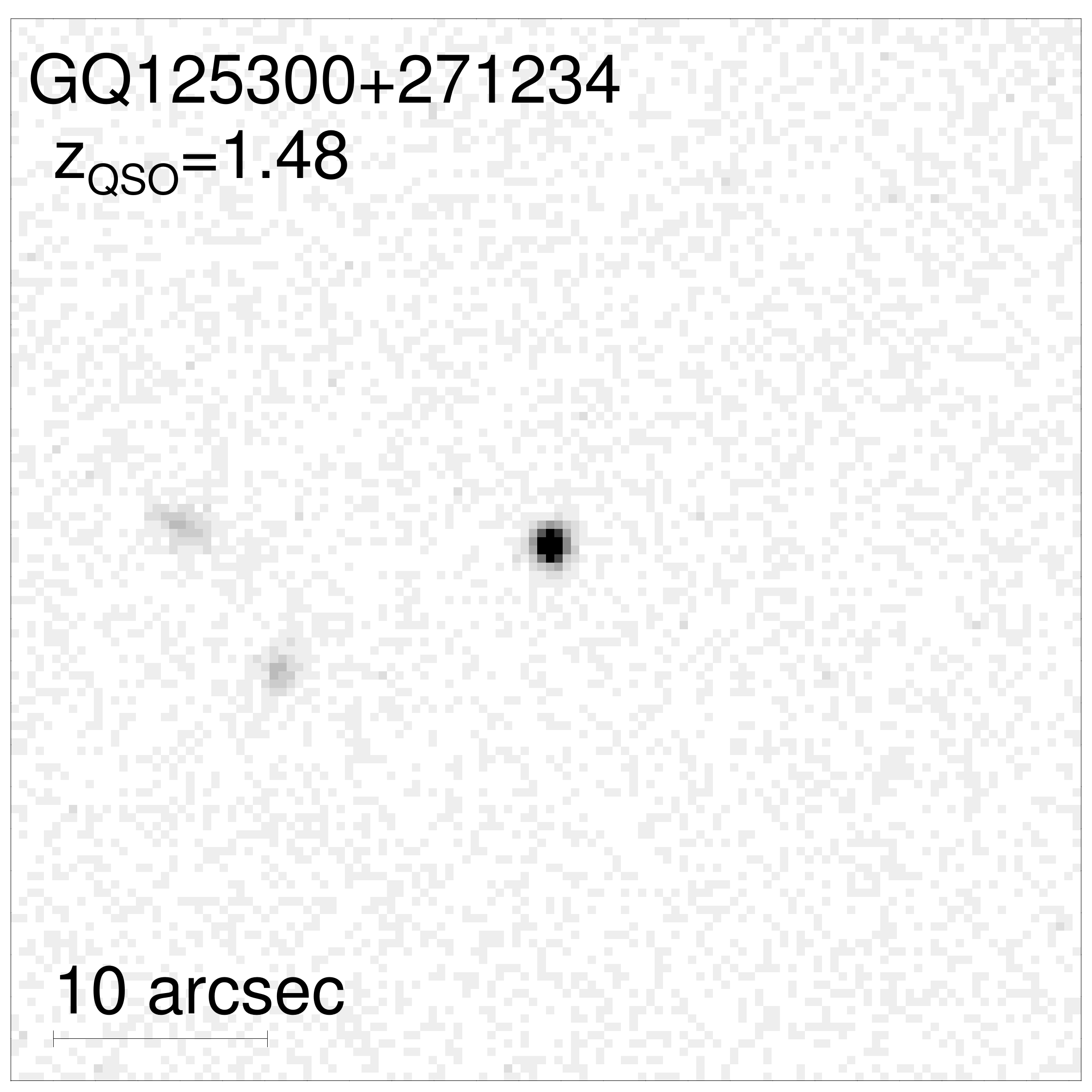,width=3.5cm} 
\epsfig{file=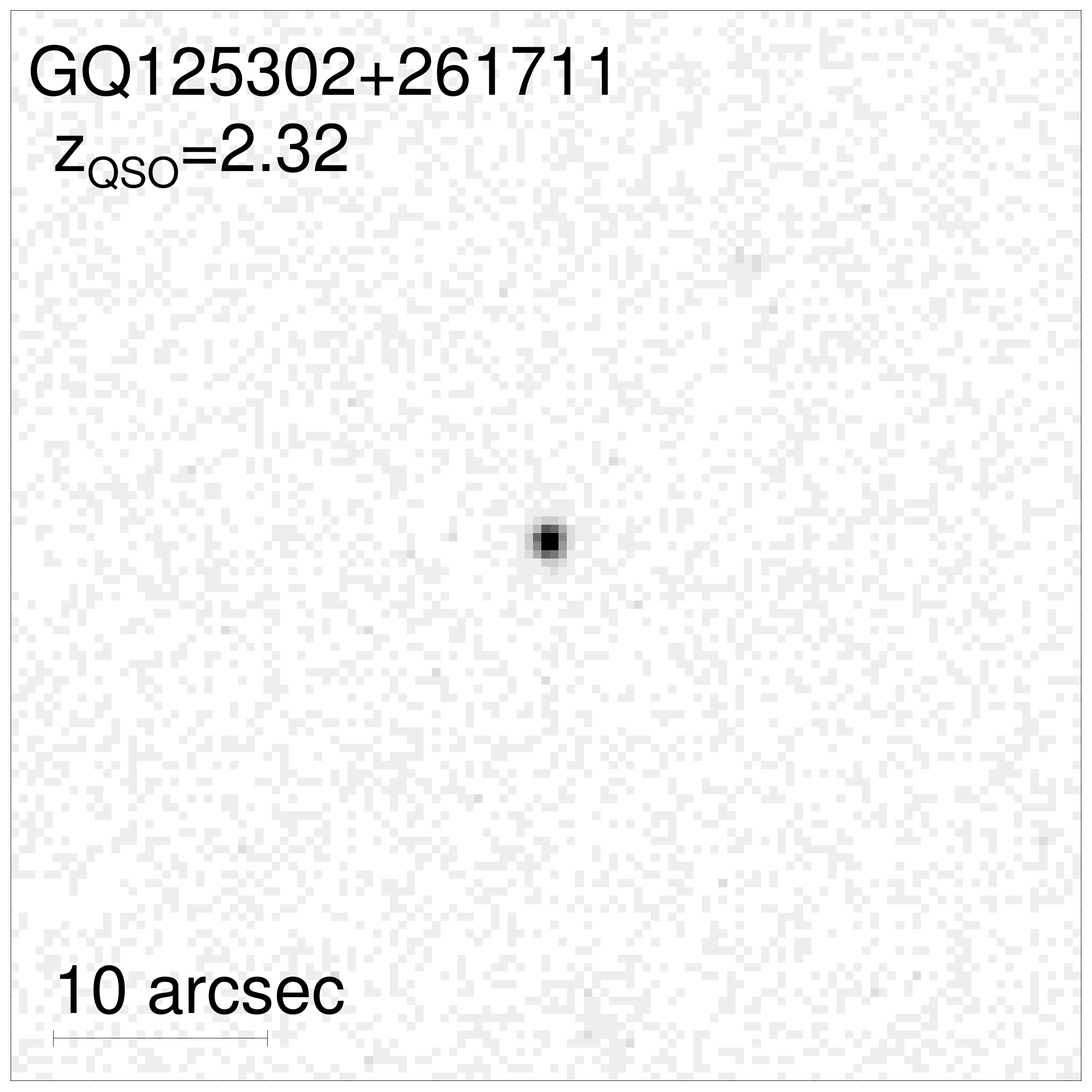,width=3.5cm} 
\epsfig{file=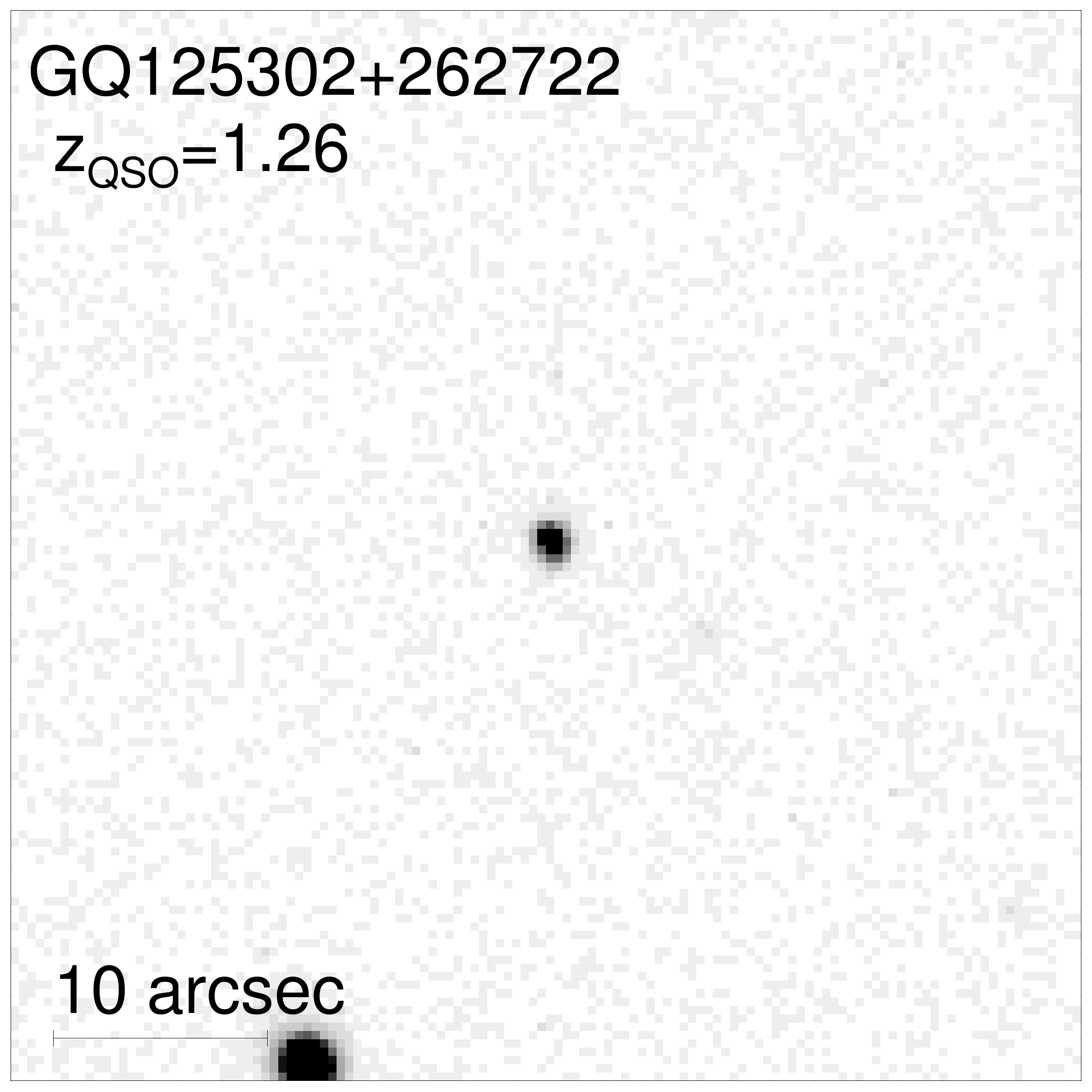,width=3.5cm} 
\epsfig{file=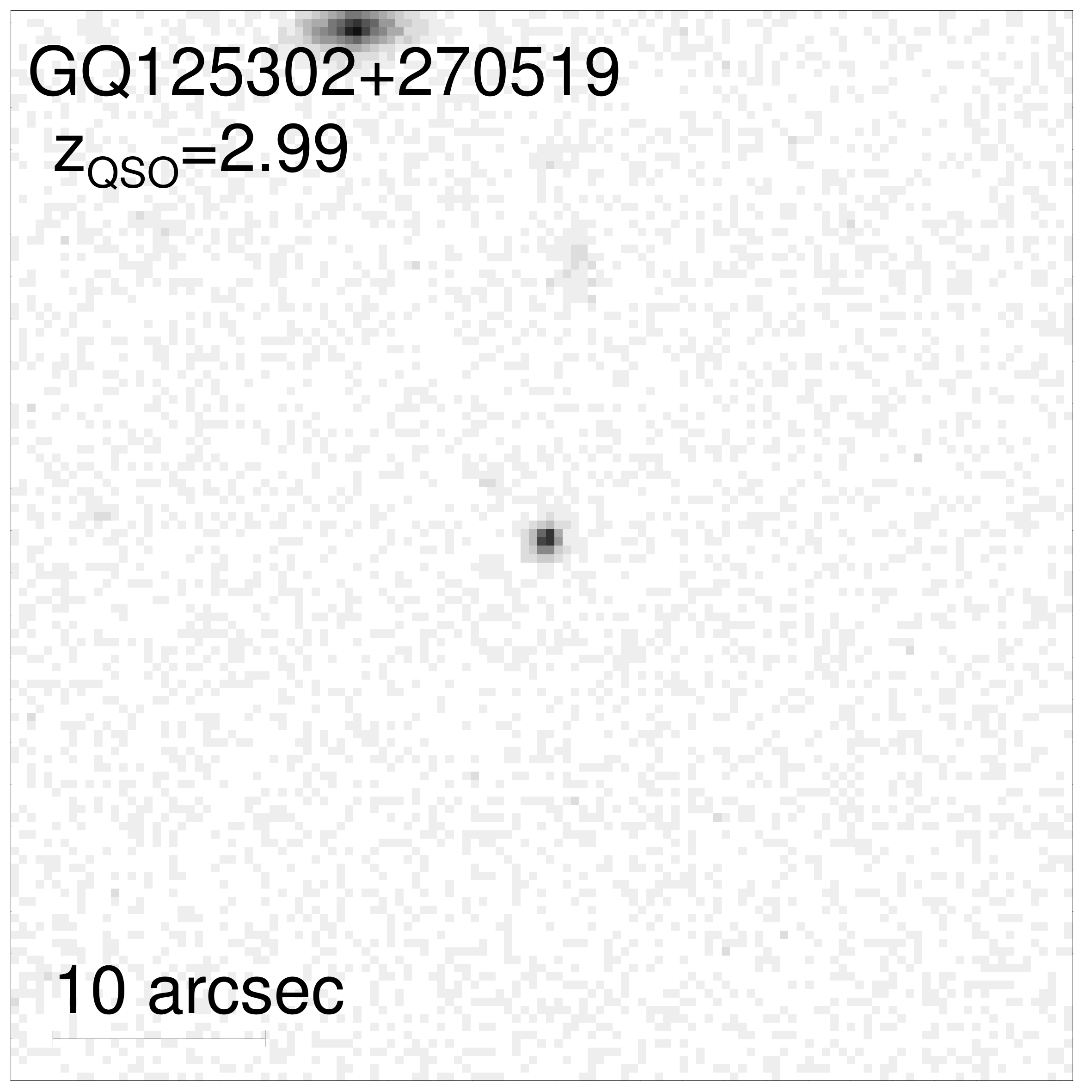,width=3.5cm} 
\epsfig{file=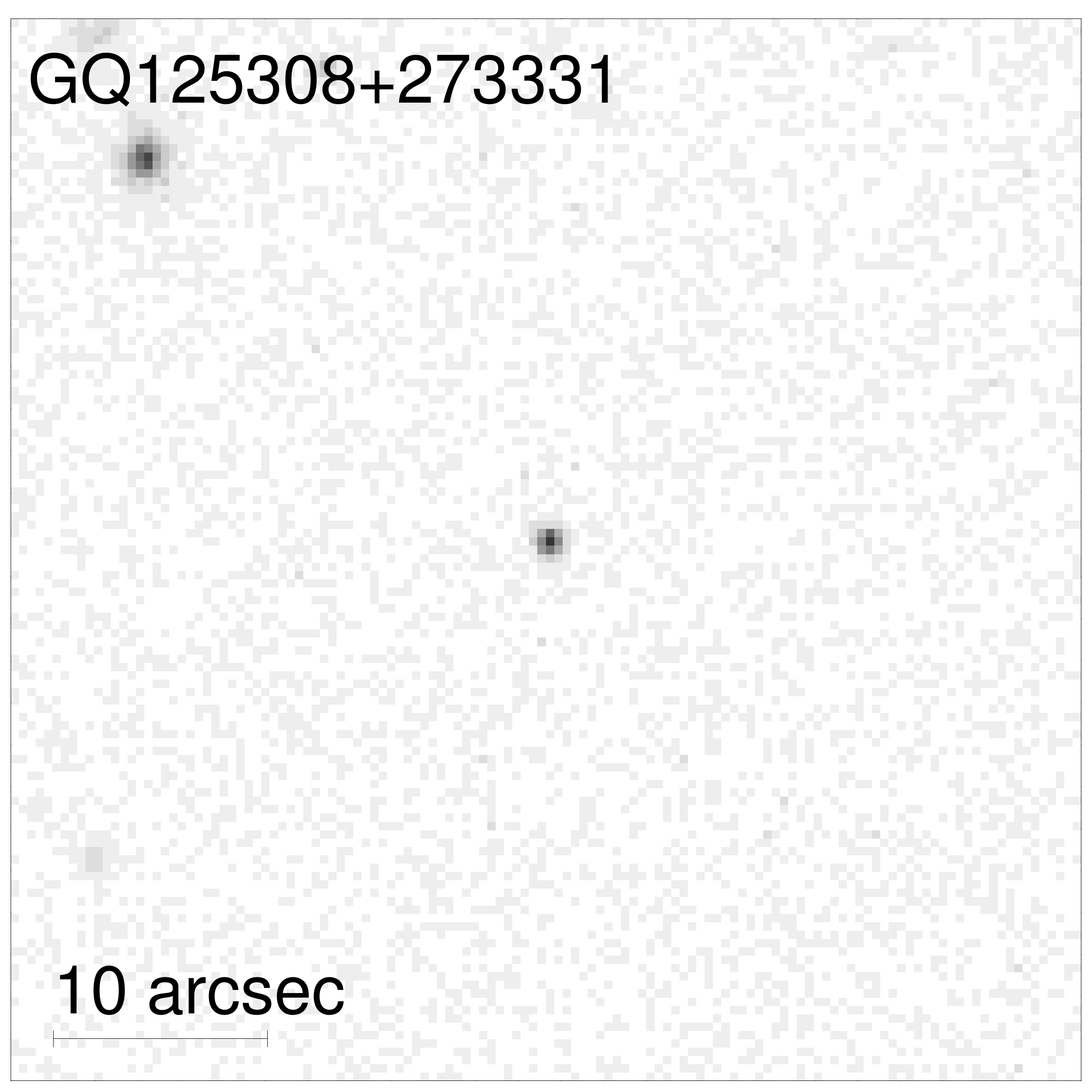,width=3.5cm} 
\epsfig{file=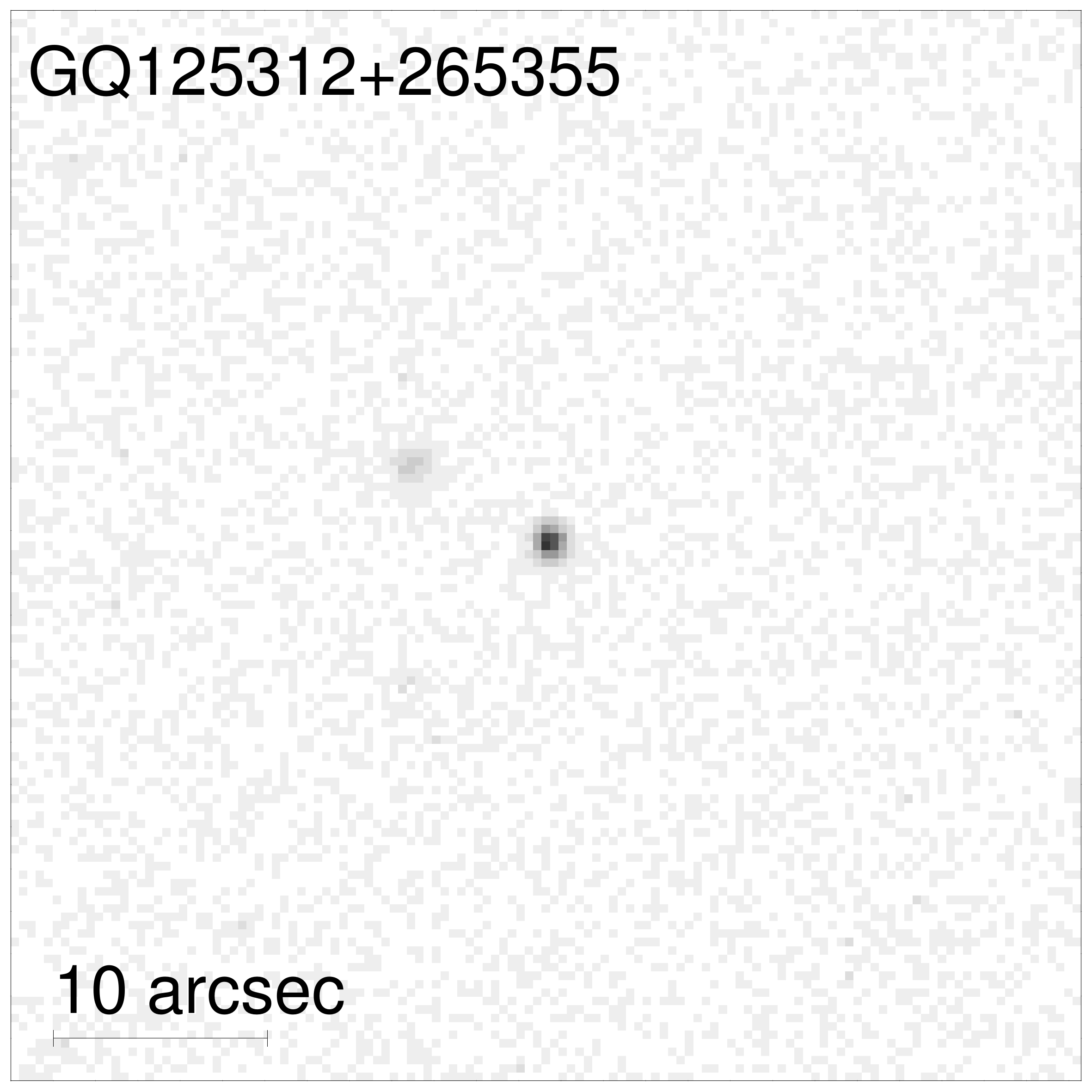,width=3.5cm} 
\epsfig{file=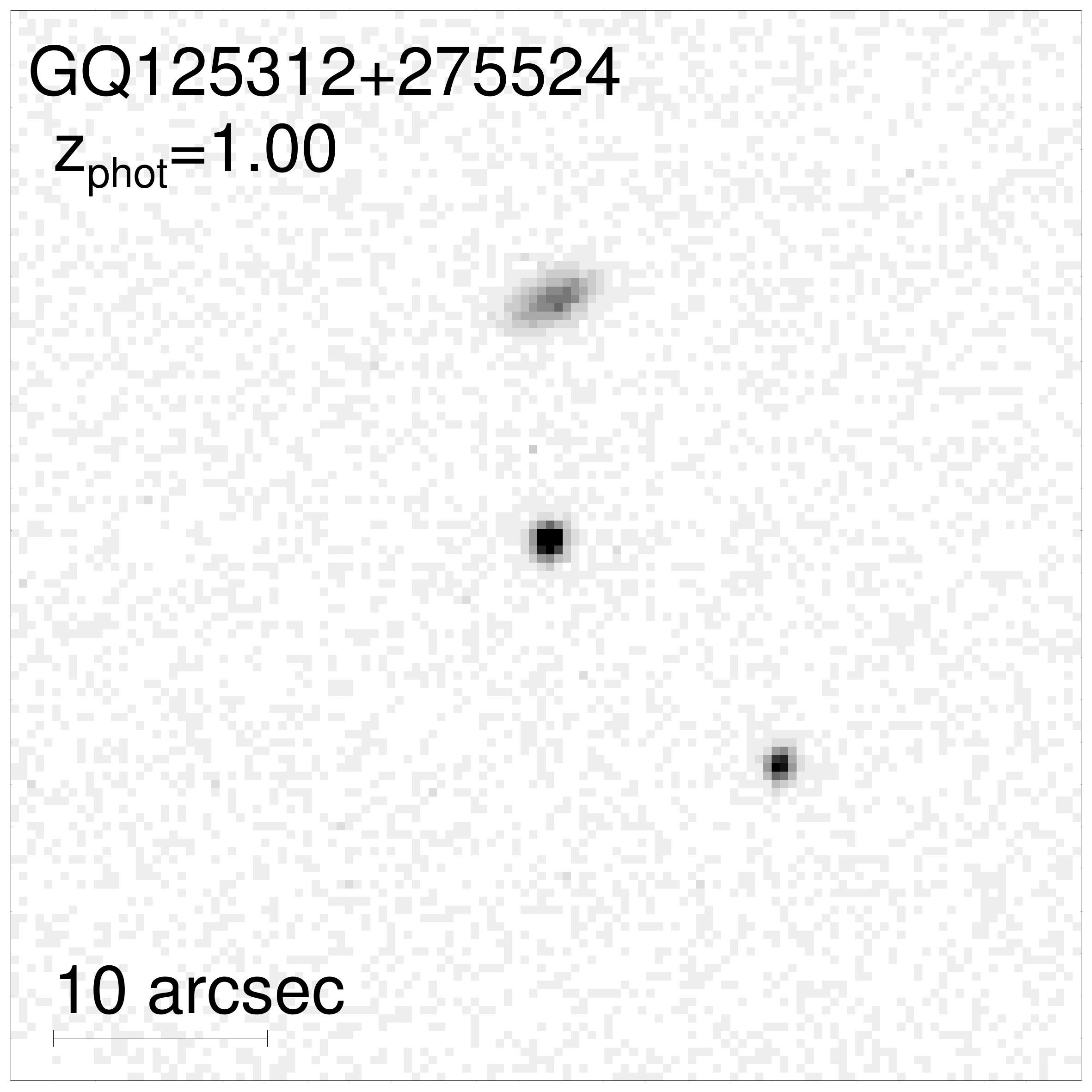,width=3.5cm} 
\epsfig{file=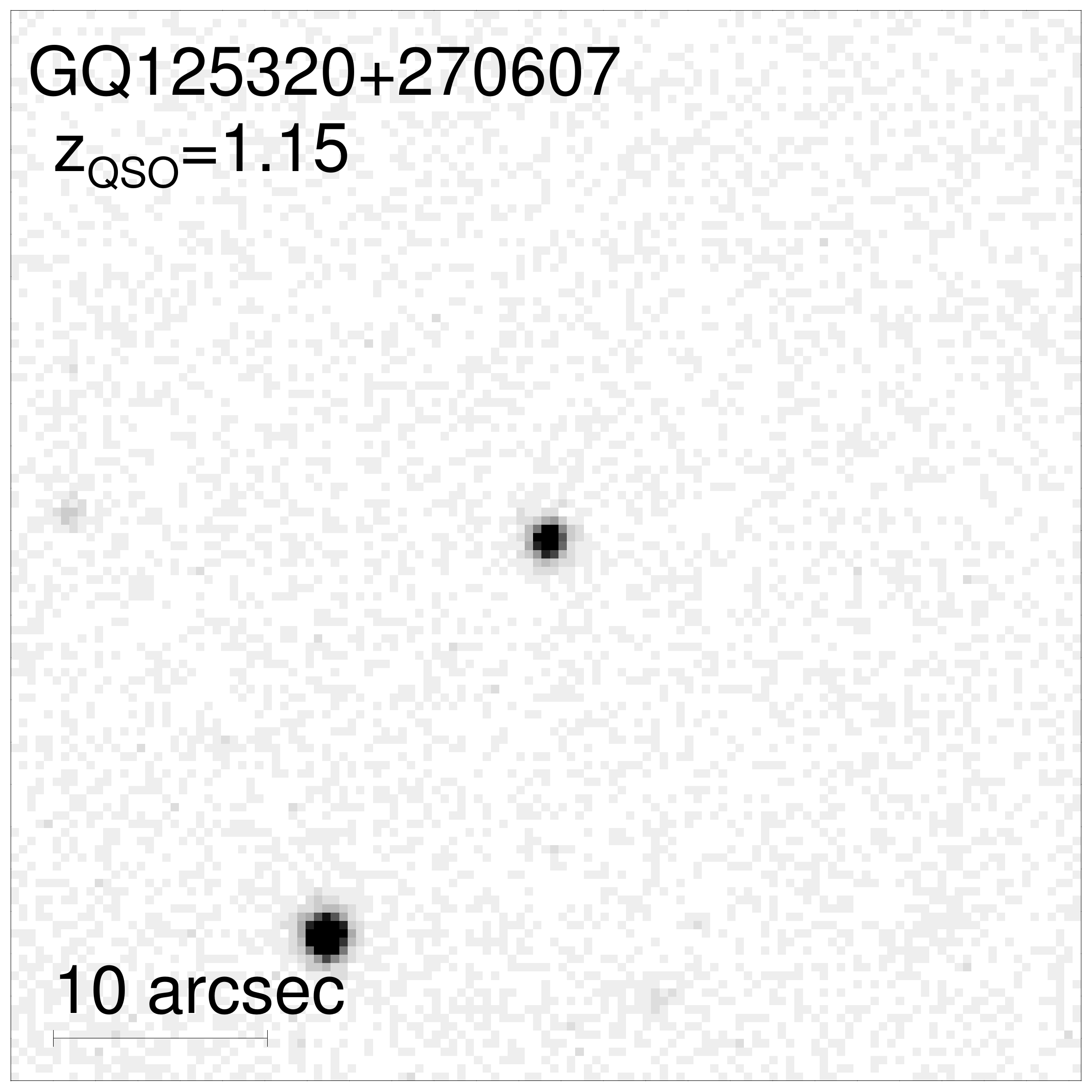,width=3.5cm} 
\epsfig{file=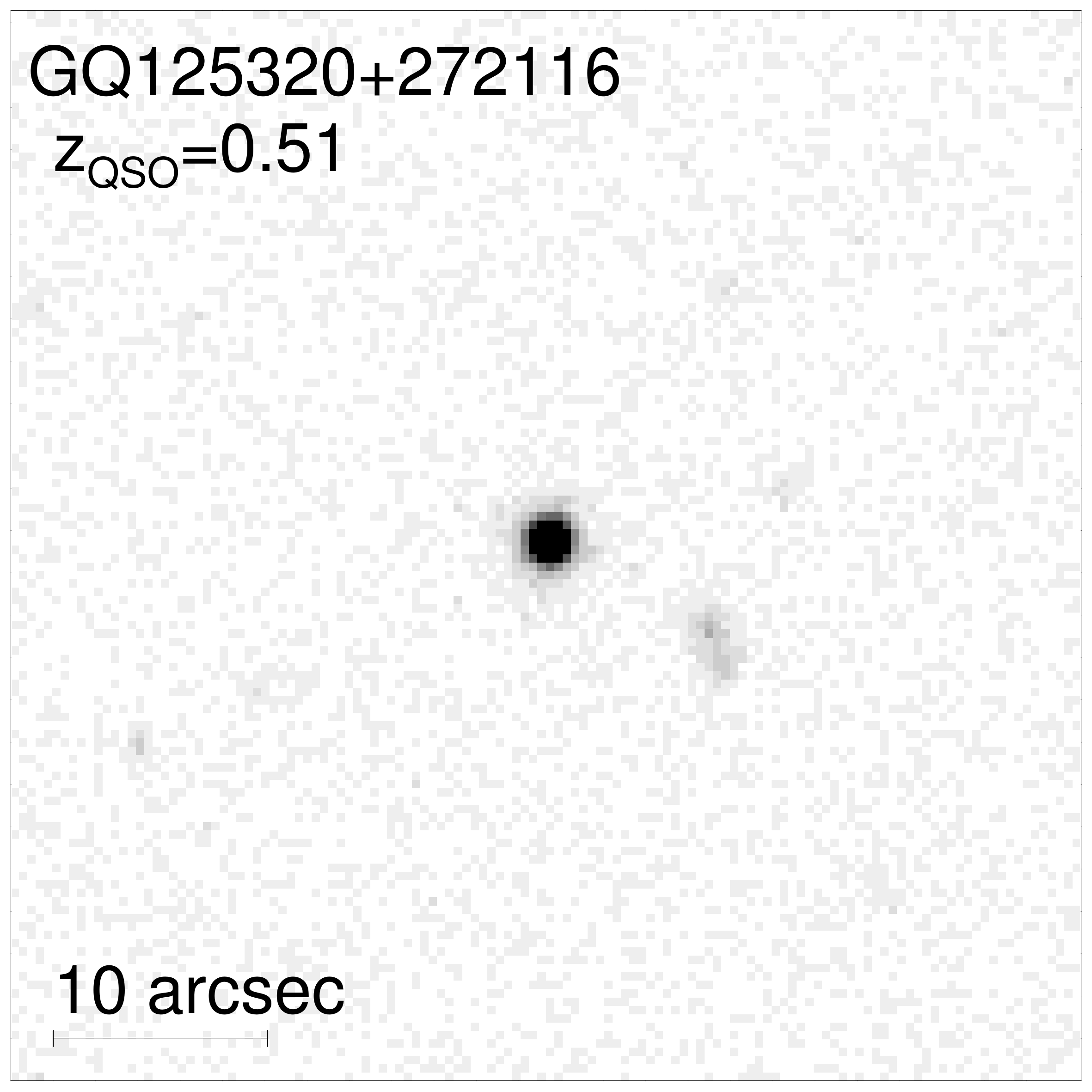,width=3.5cm} 
\epsfig{file=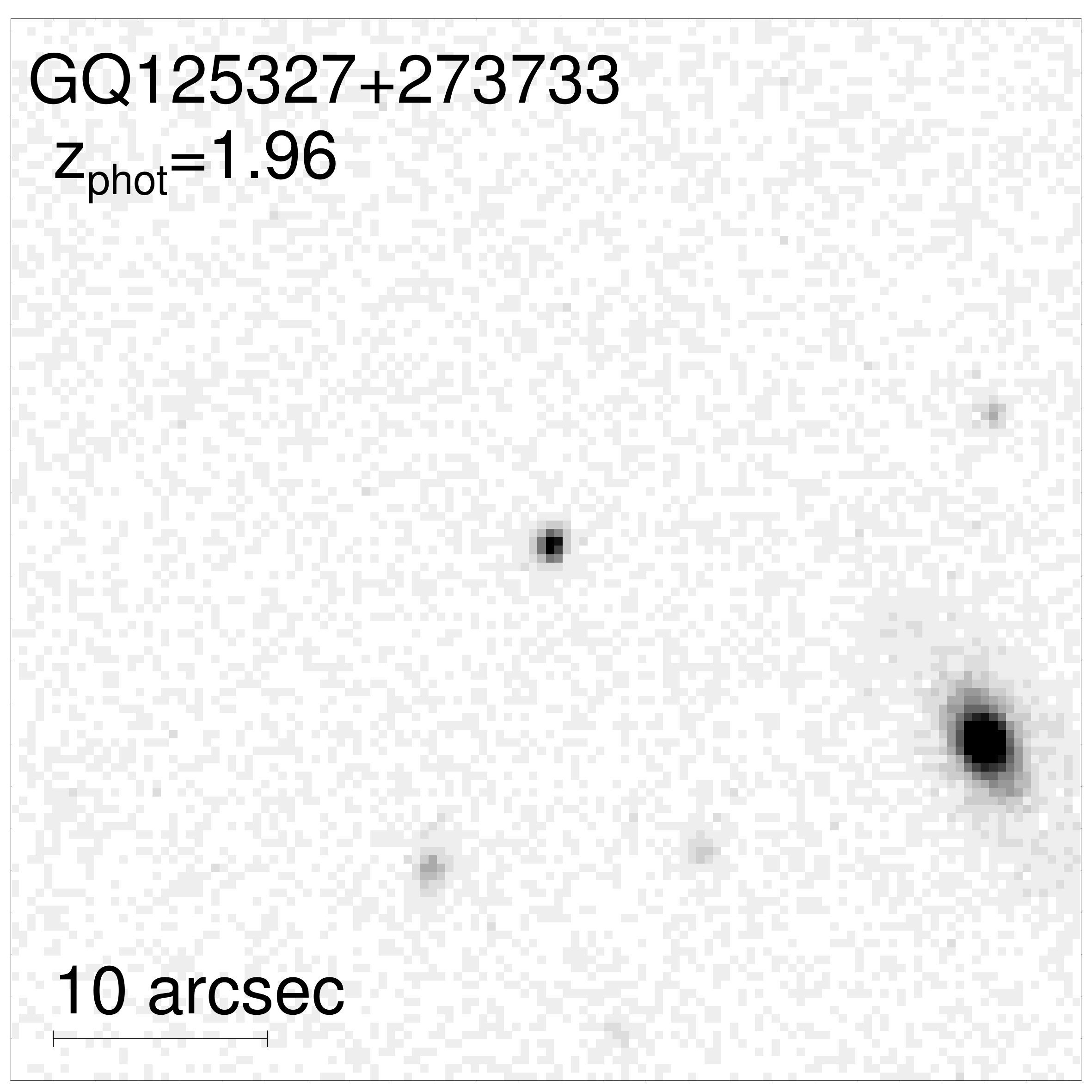,width=3.5cm} 
\epsfig{file=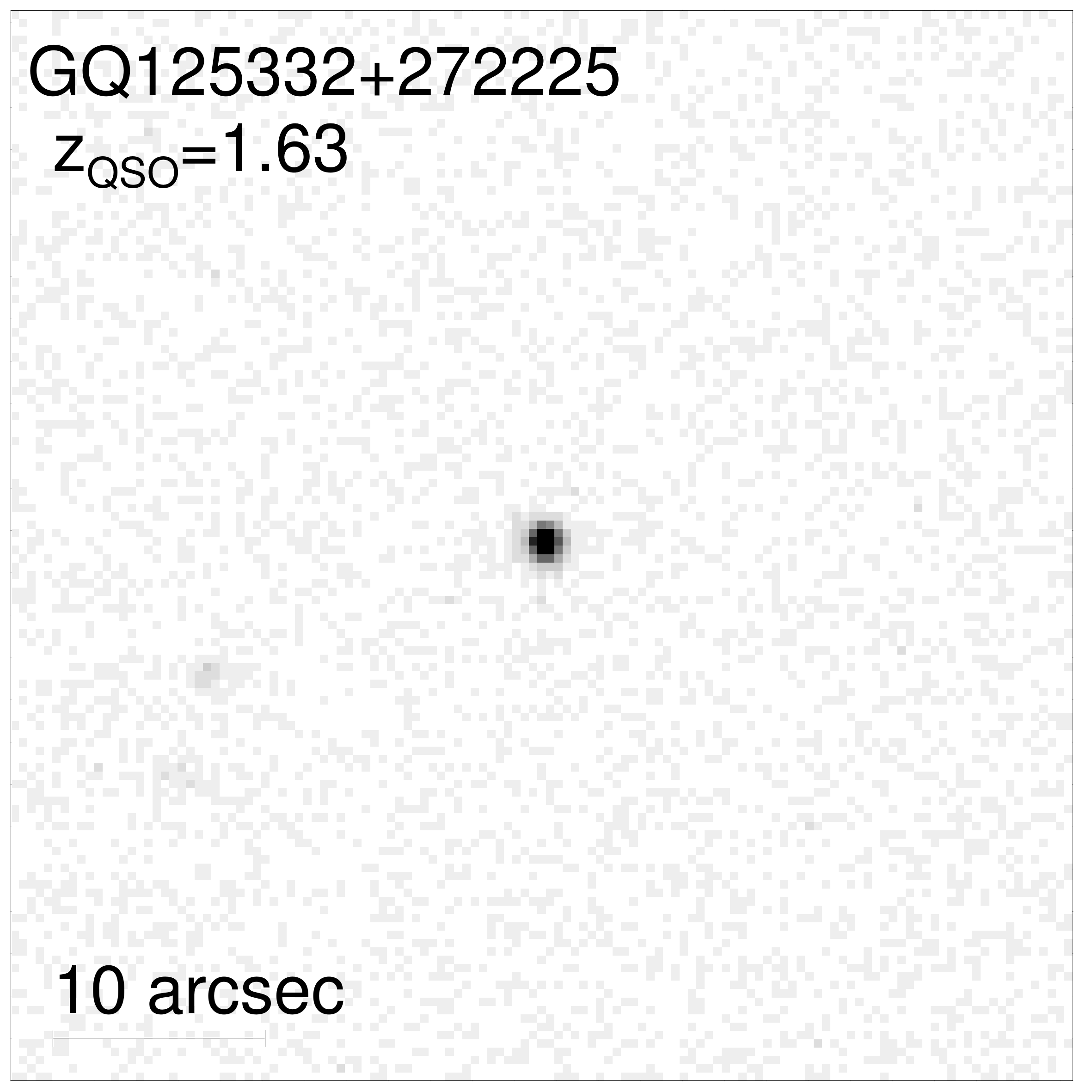,width=3.5cm} 
\epsfig{file=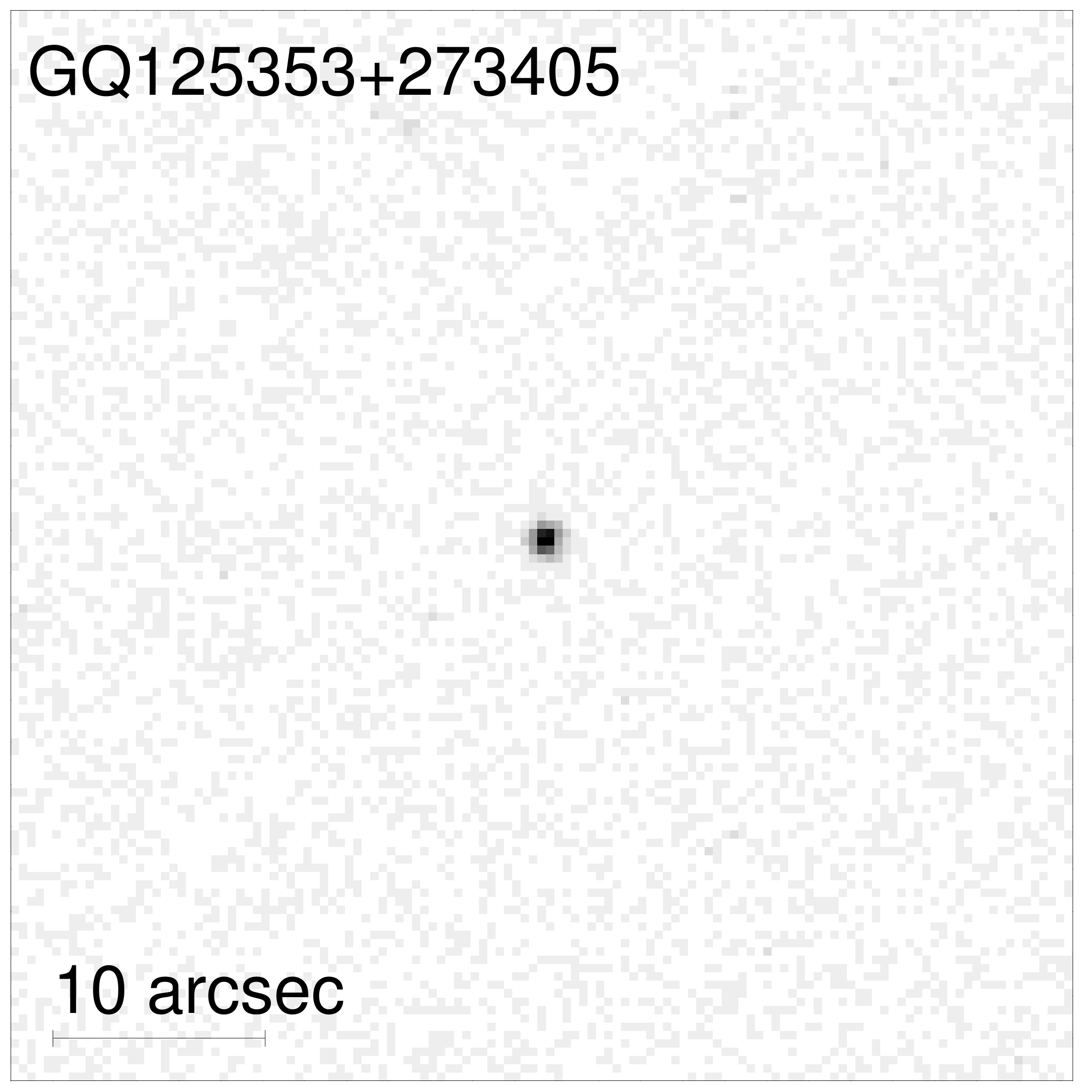,width=3.5cm} 
\epsfig{file=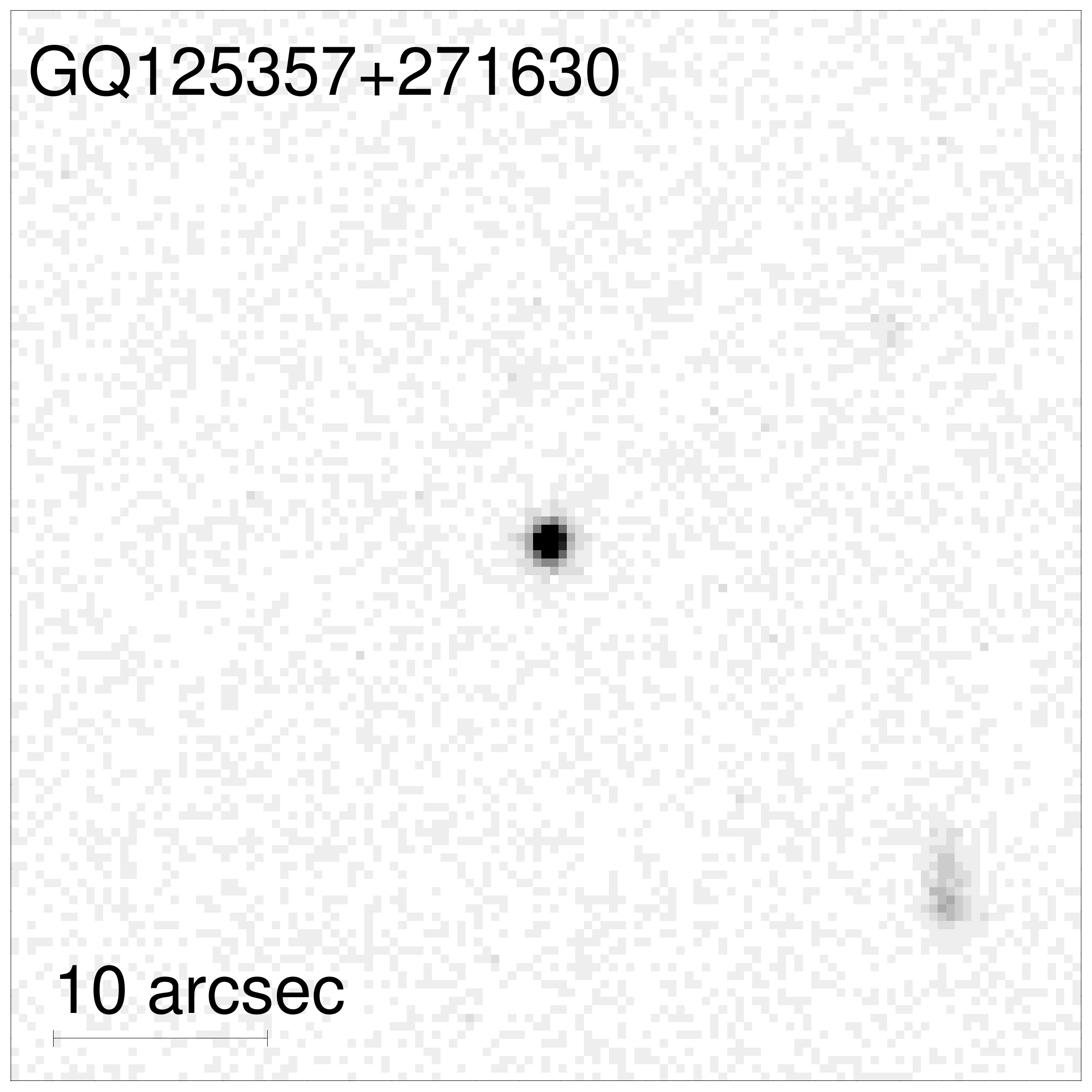,width=3.5cm} 
\epsfig{file=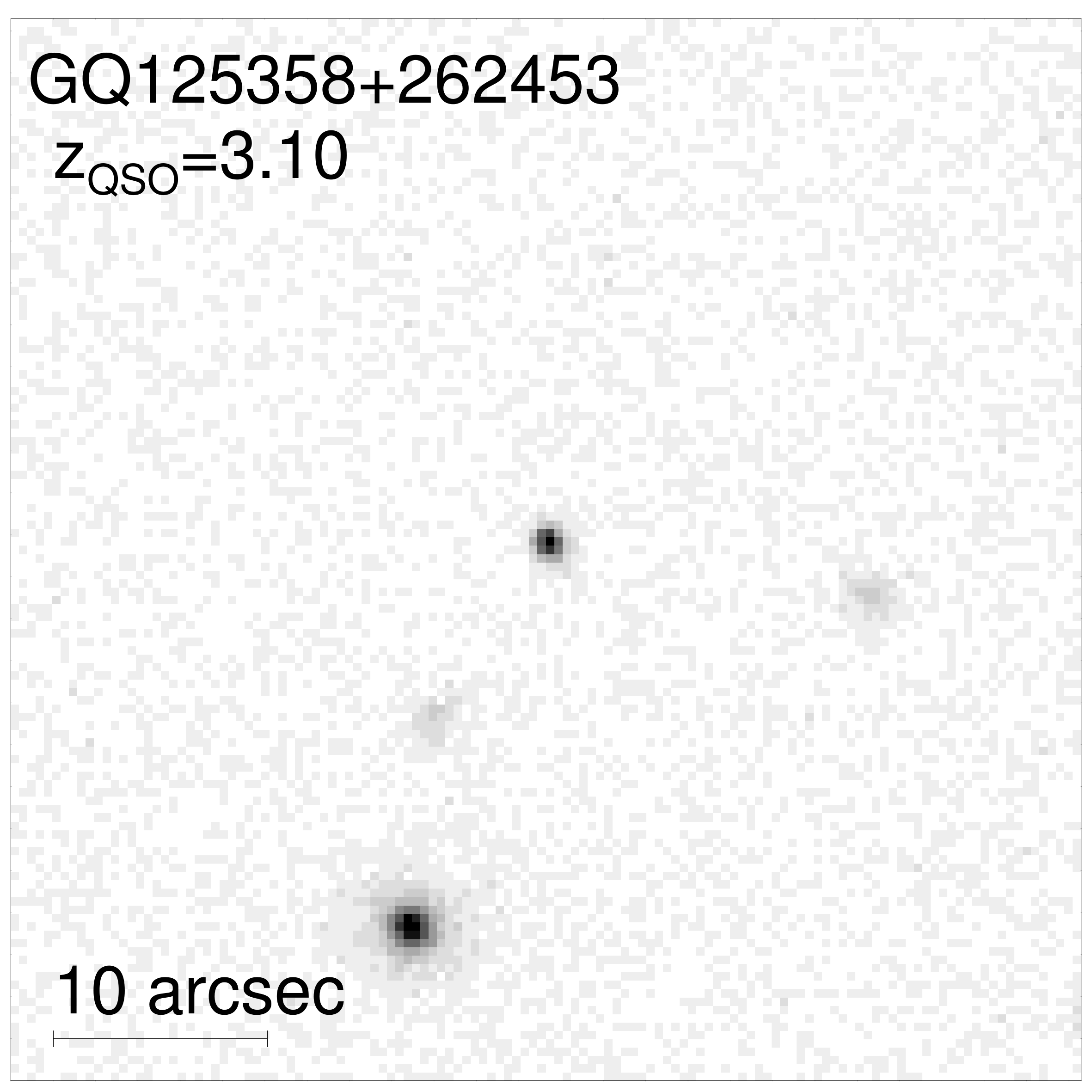,width=3.5cm} 
\epsfig{file=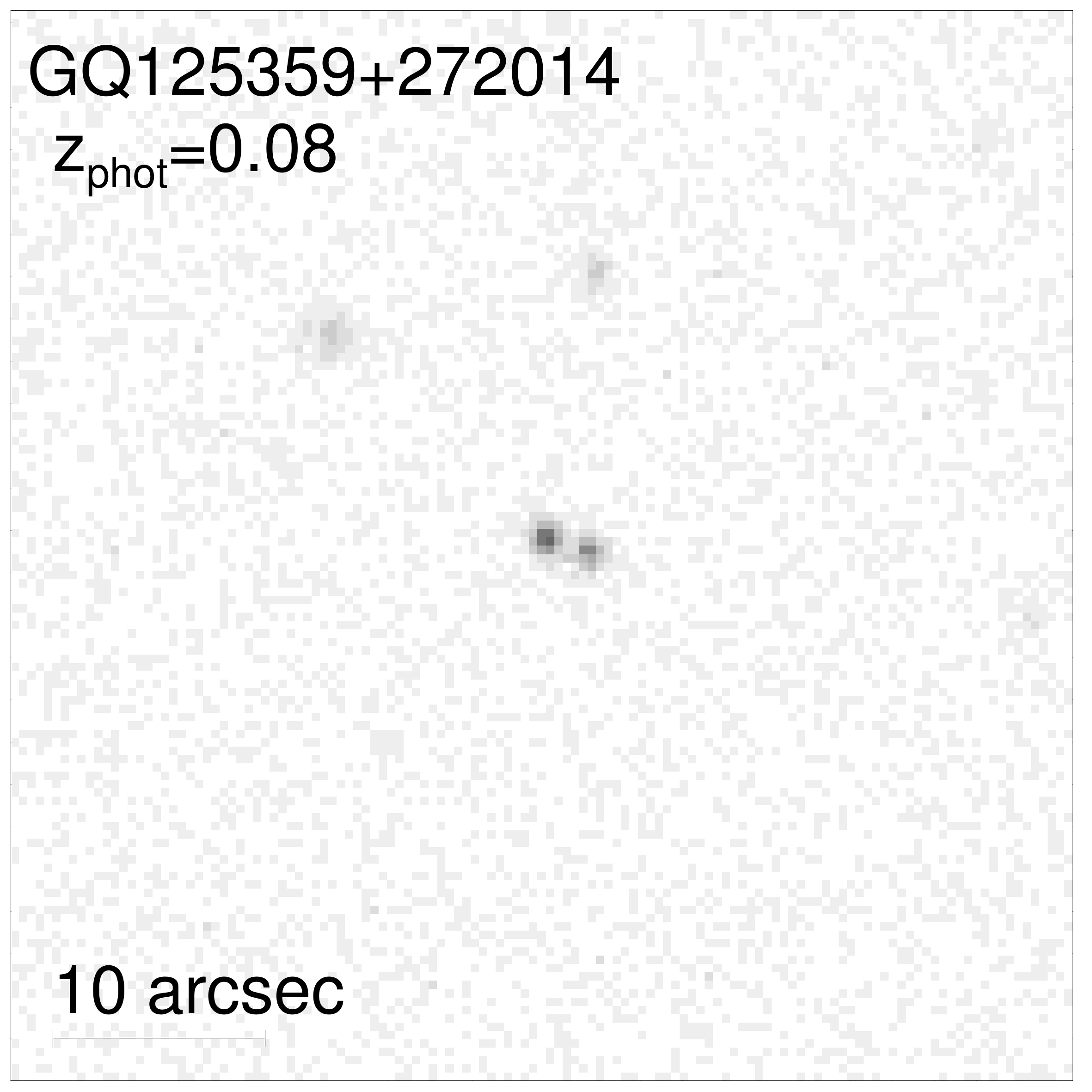,width=3.5cm} 
\epsfig{file=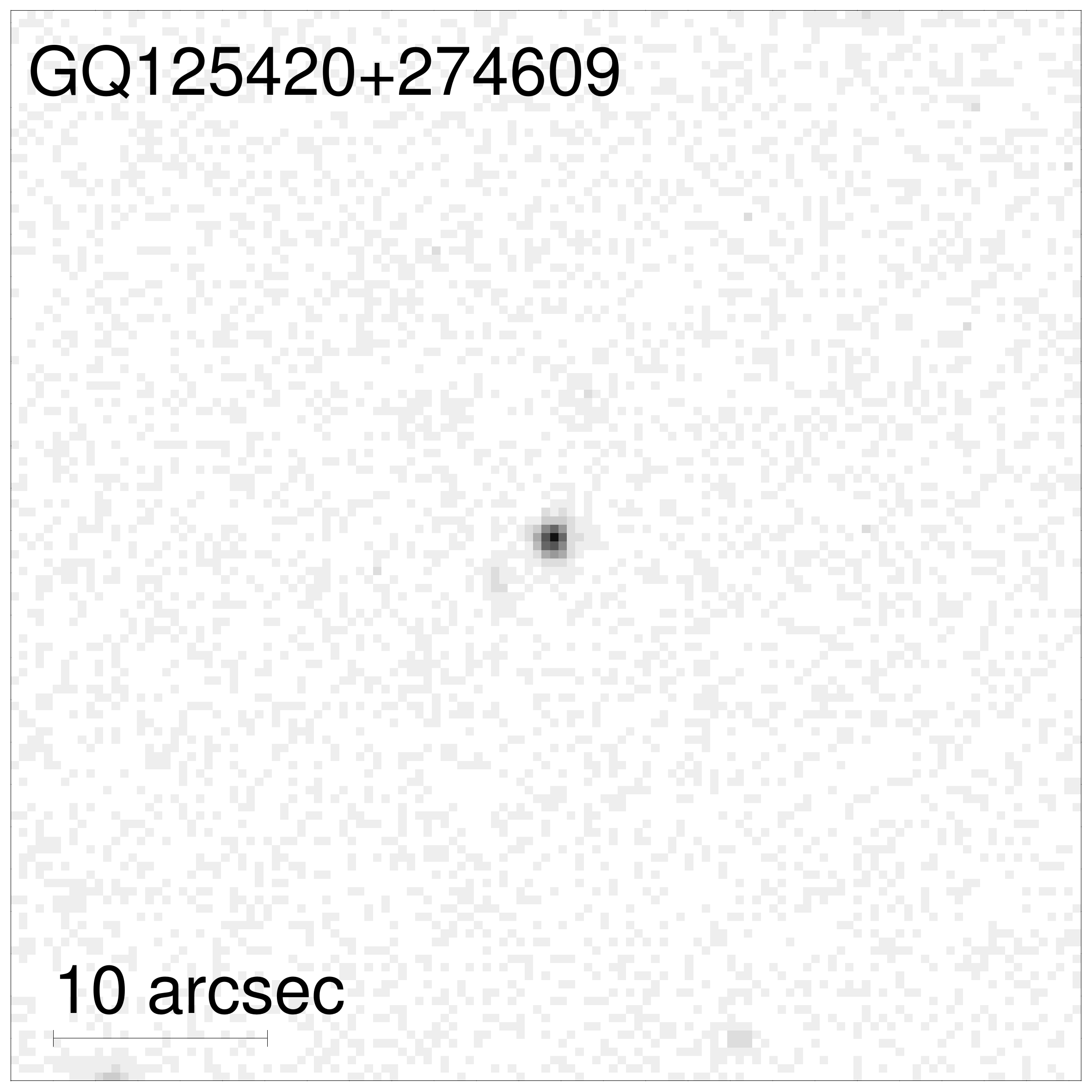,width=3.5cm} 
\epsfig{file=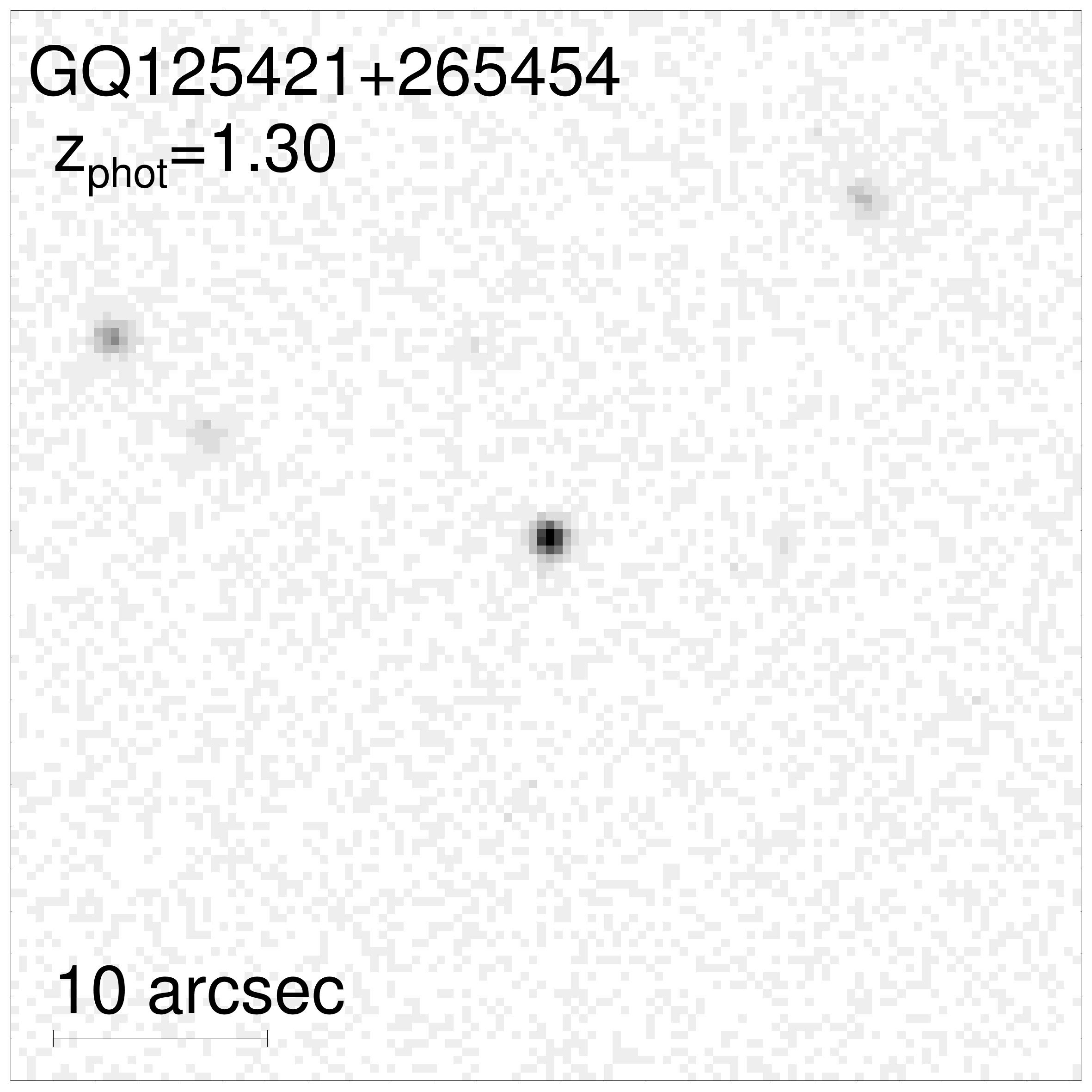,width=3.5cm} 
\epsfig{file=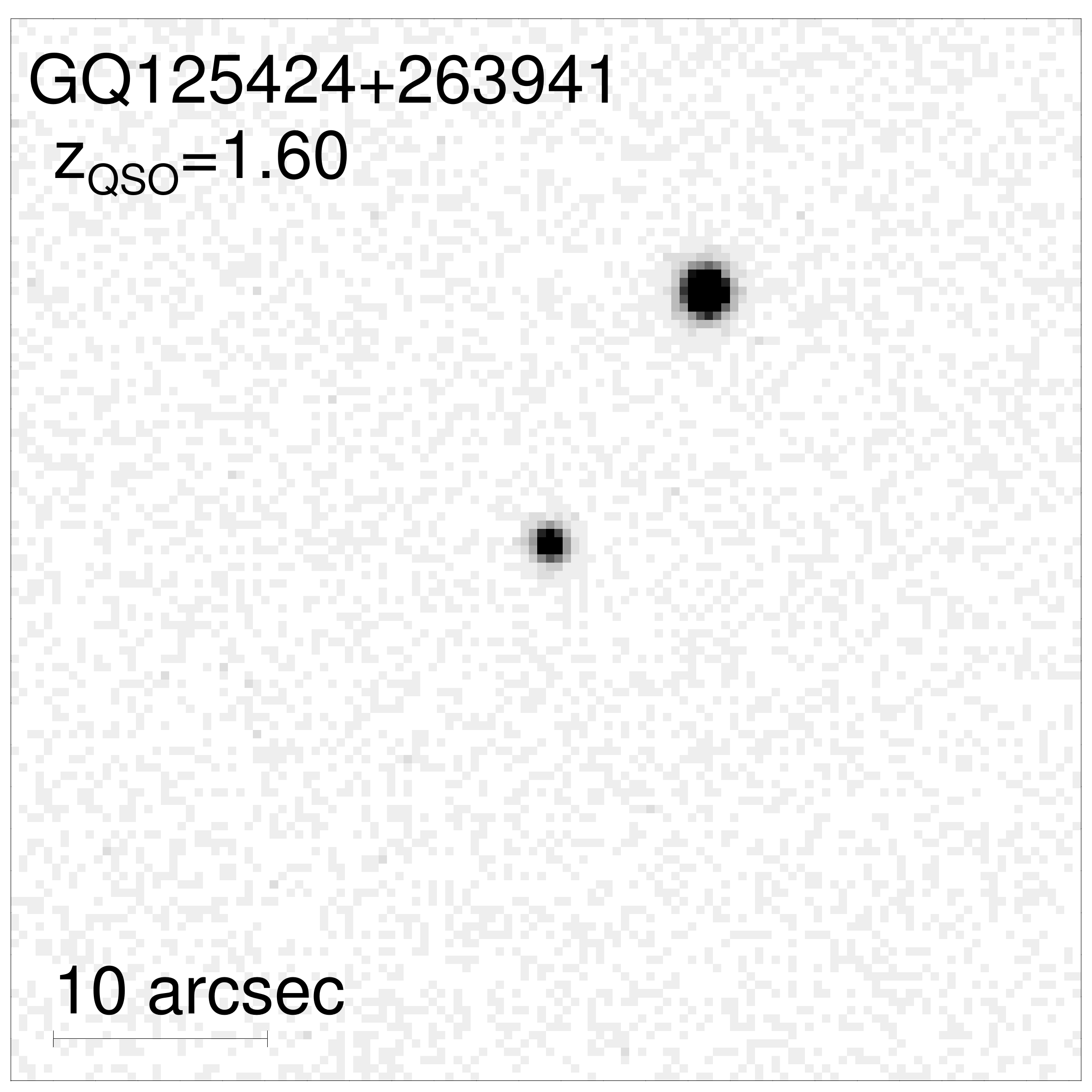,width=3.5cm} 
\epsfig{file=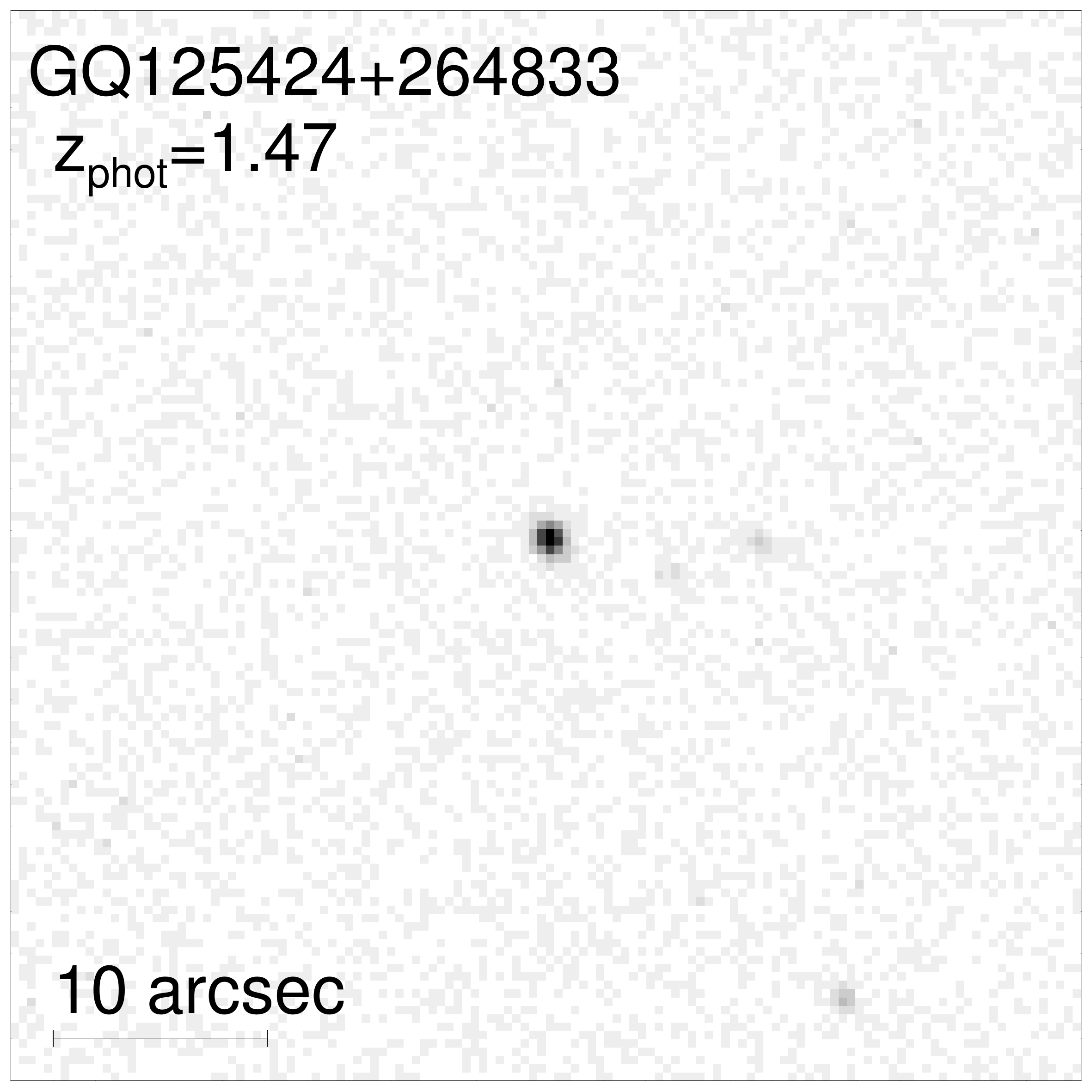,width=3.5cm} 
\epsfig{file=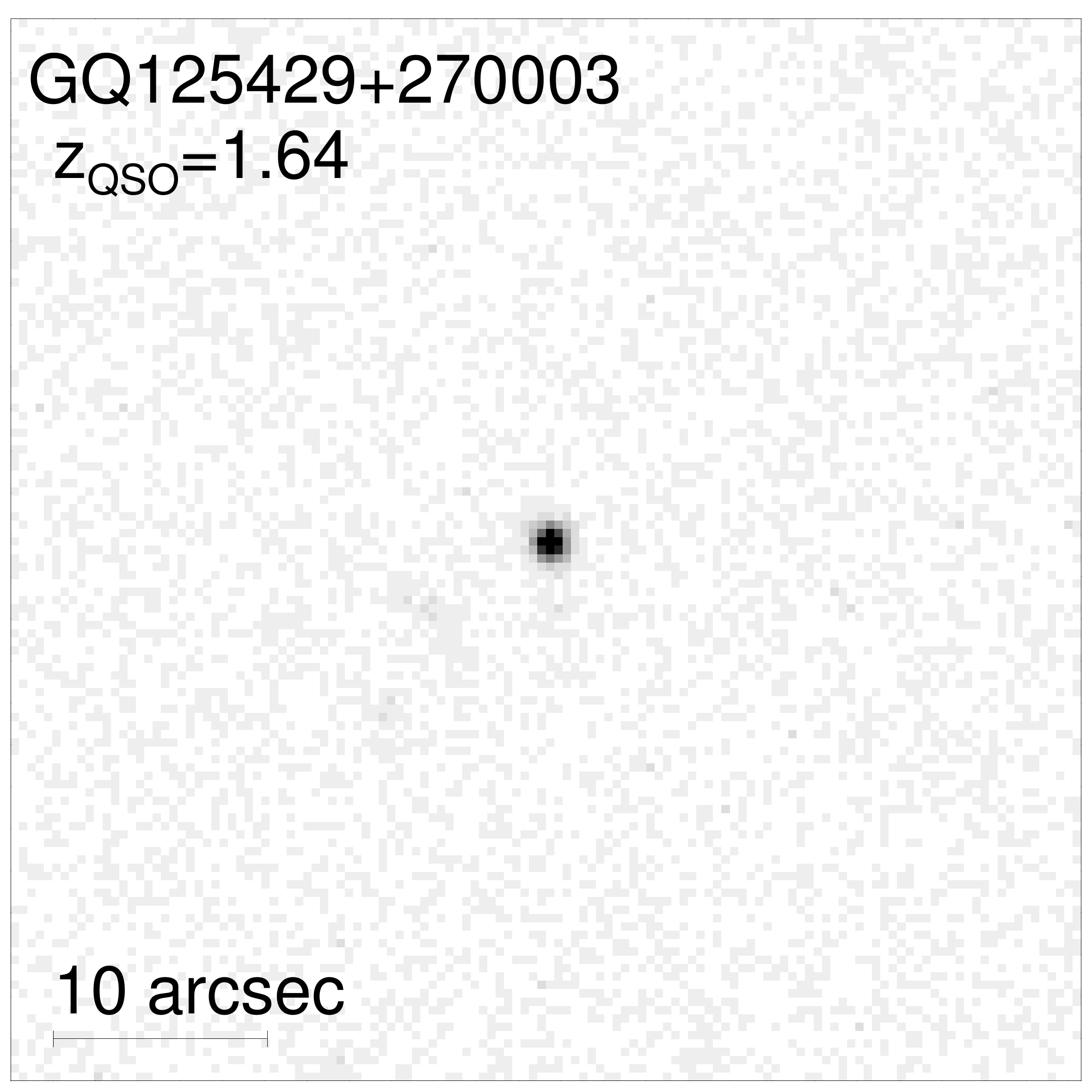,width=3.5cm} 
\epsfig{file=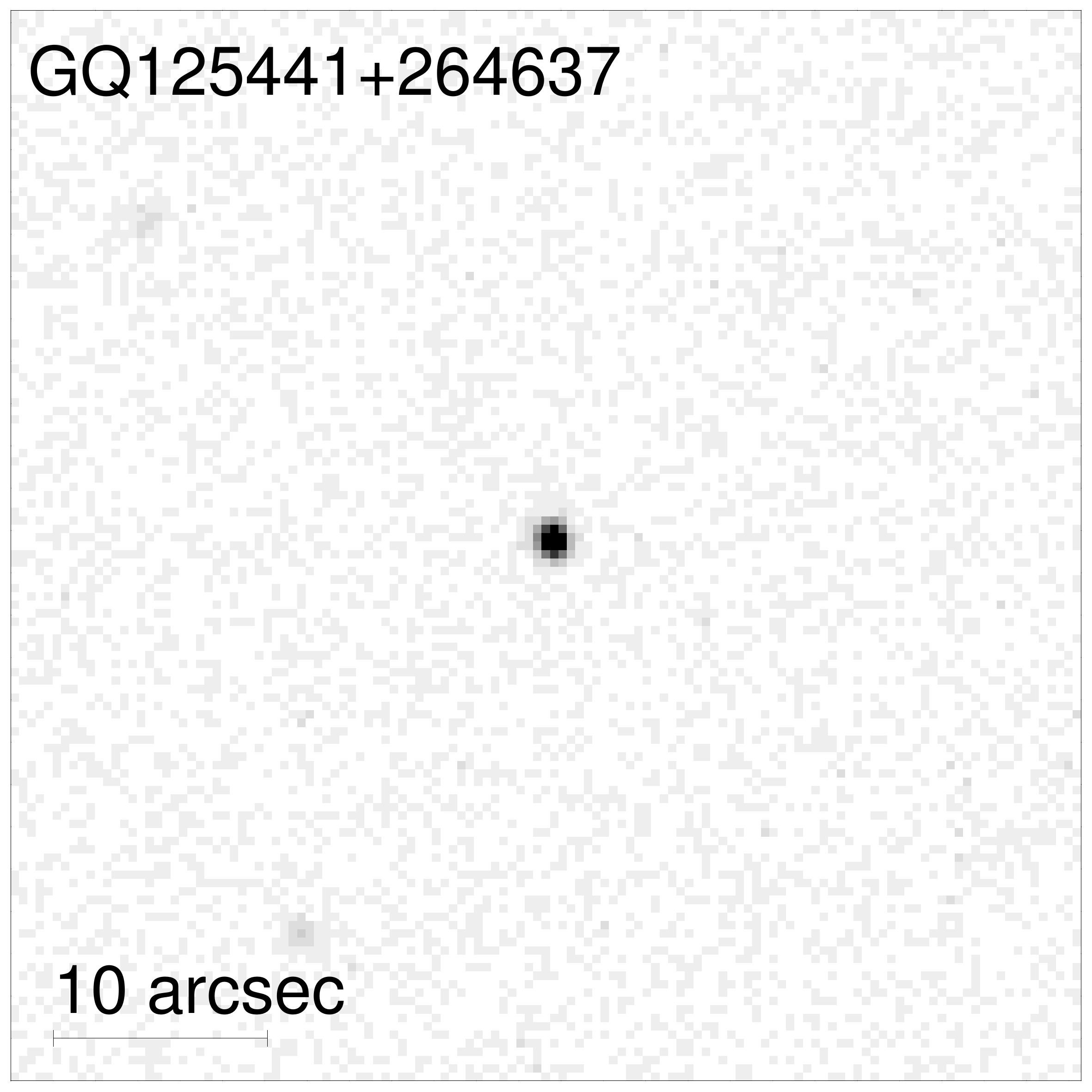,width=3.5cm} 
\epsfig{file=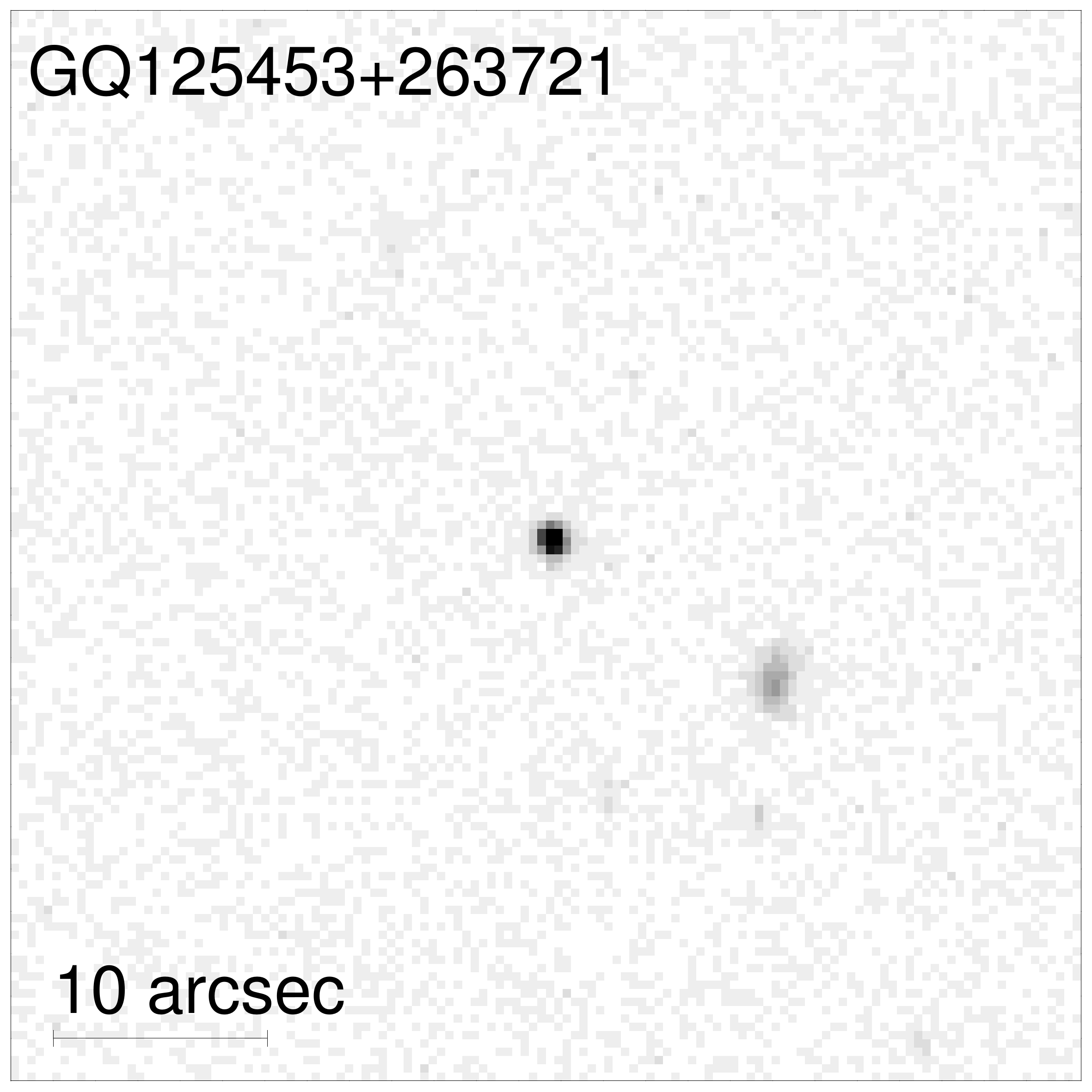,width=3.5cm} 
\epsfig{file=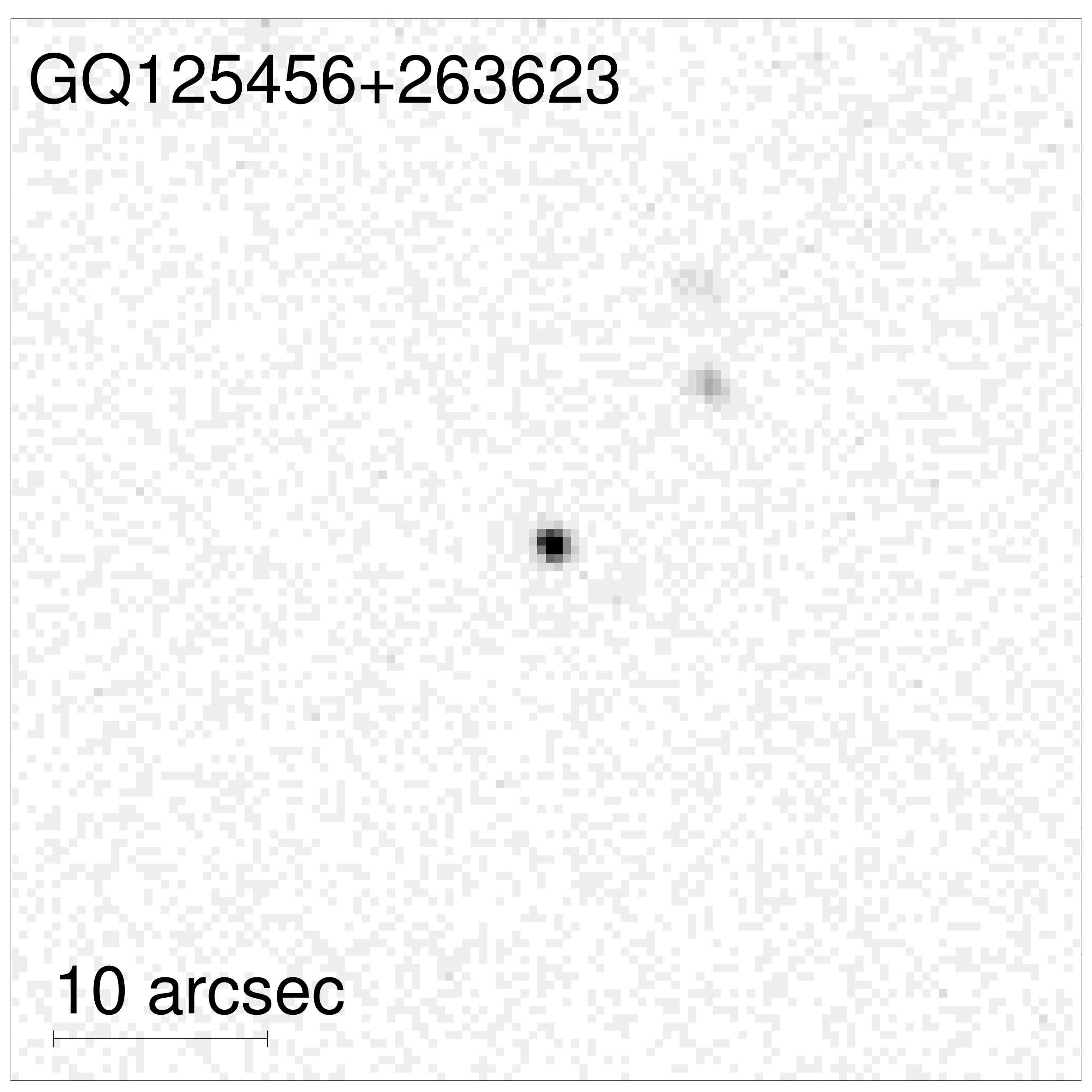,width=3.5cm} 
\epsfig{file=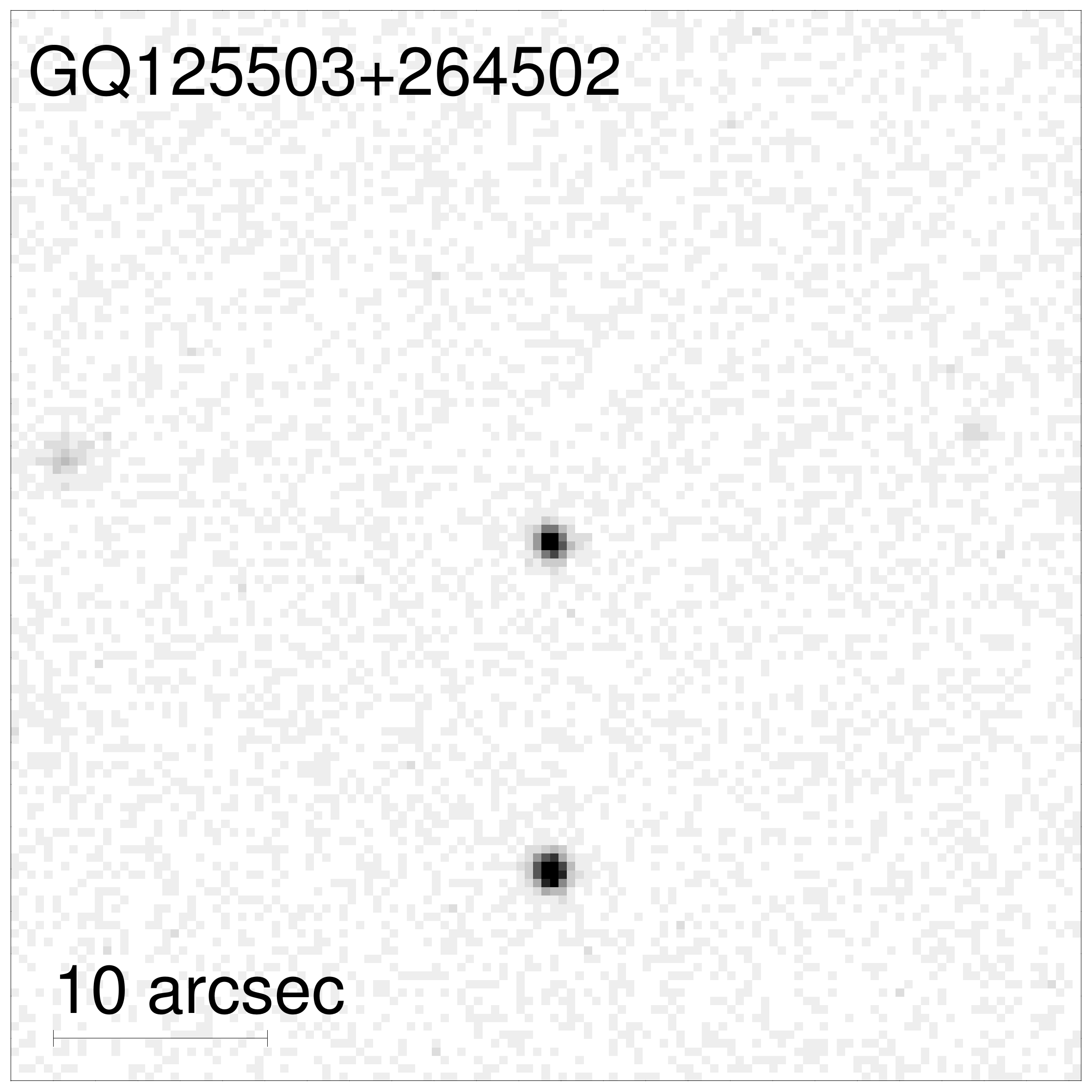,width=3.5cm} 
\epsfig{file=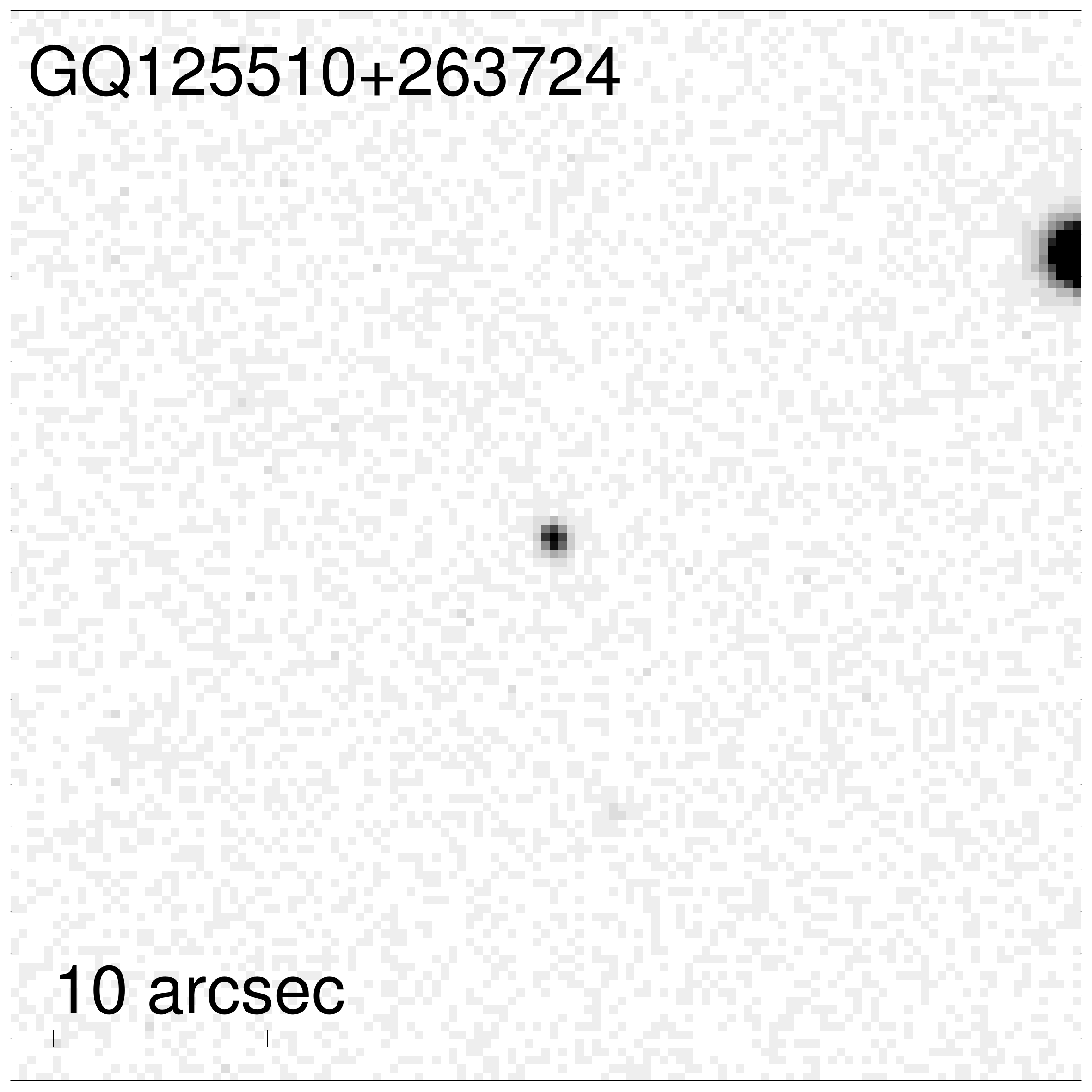,width=3.5cm} 
\epsfig{file=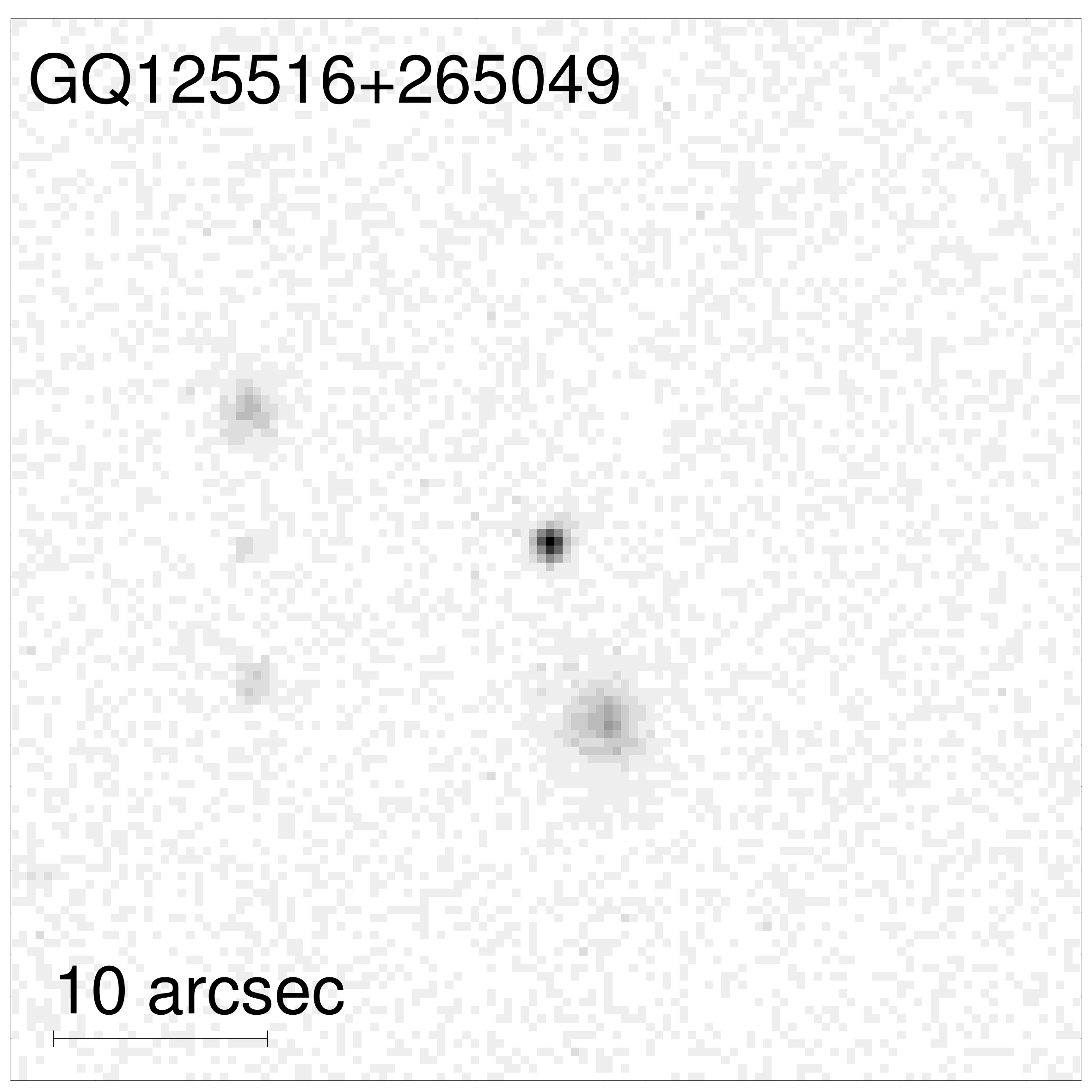,width=3.5cm} 
\epsfig{file=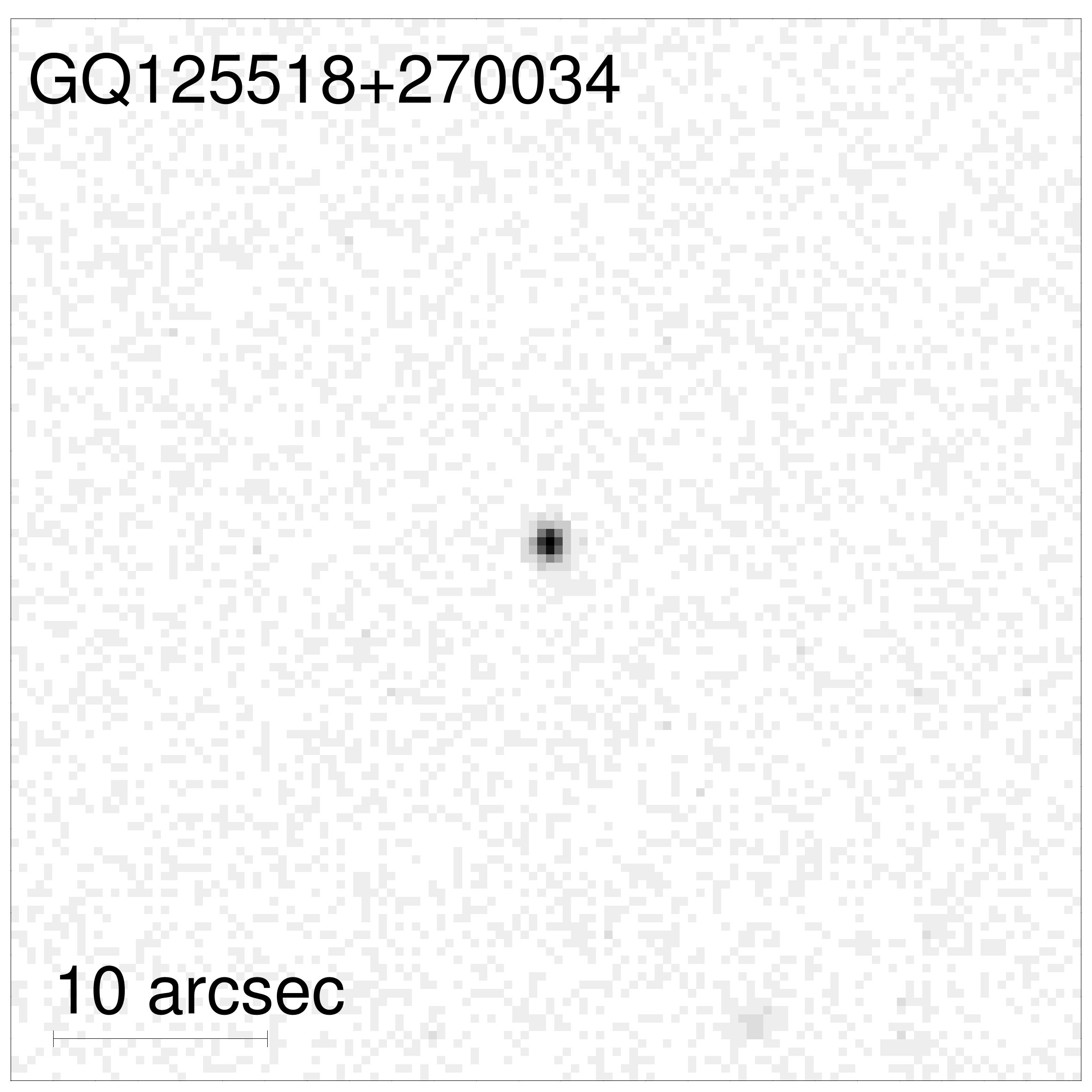,width=3.5cm} 
\epsfig{file=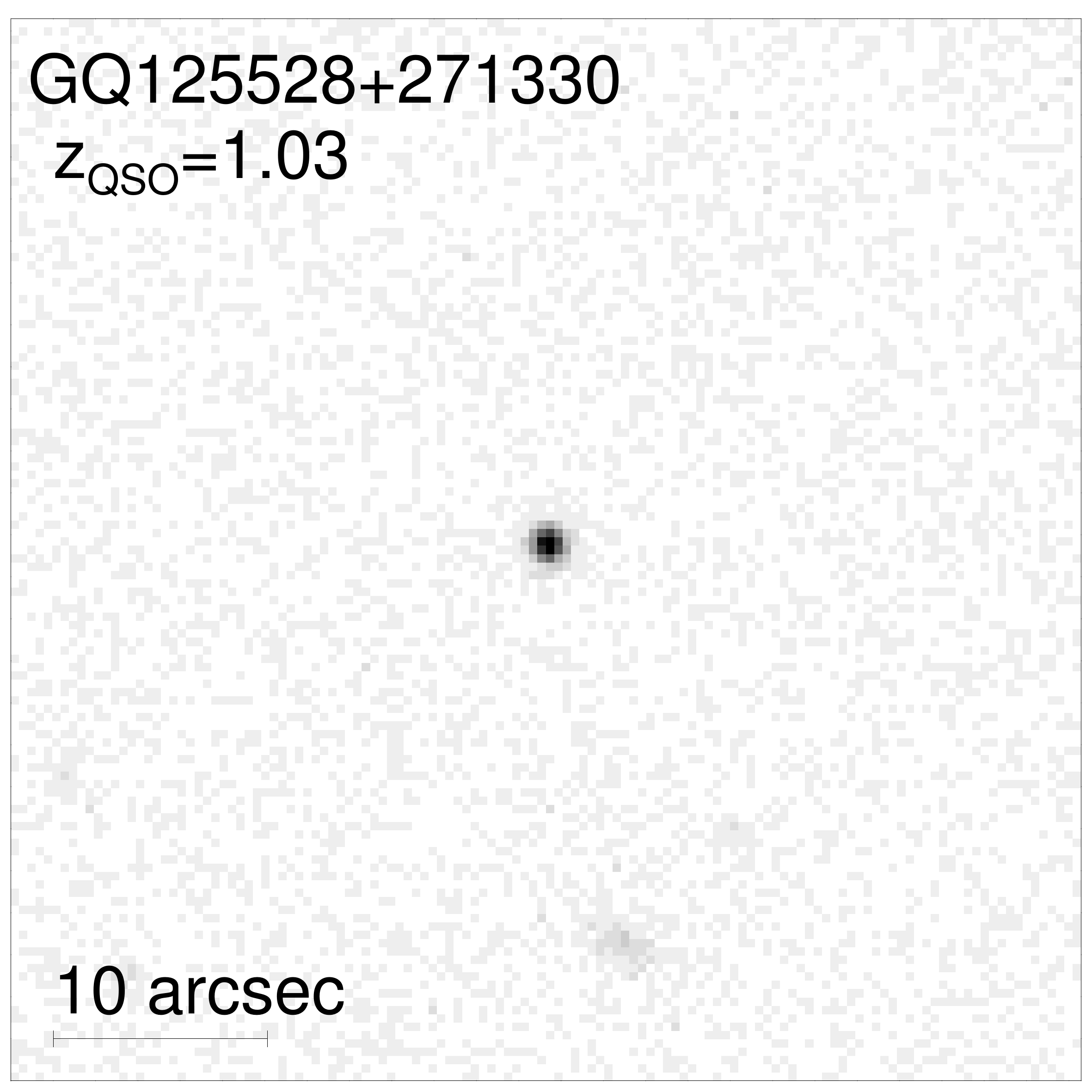,width=3.5cm} 
\epsfig{file=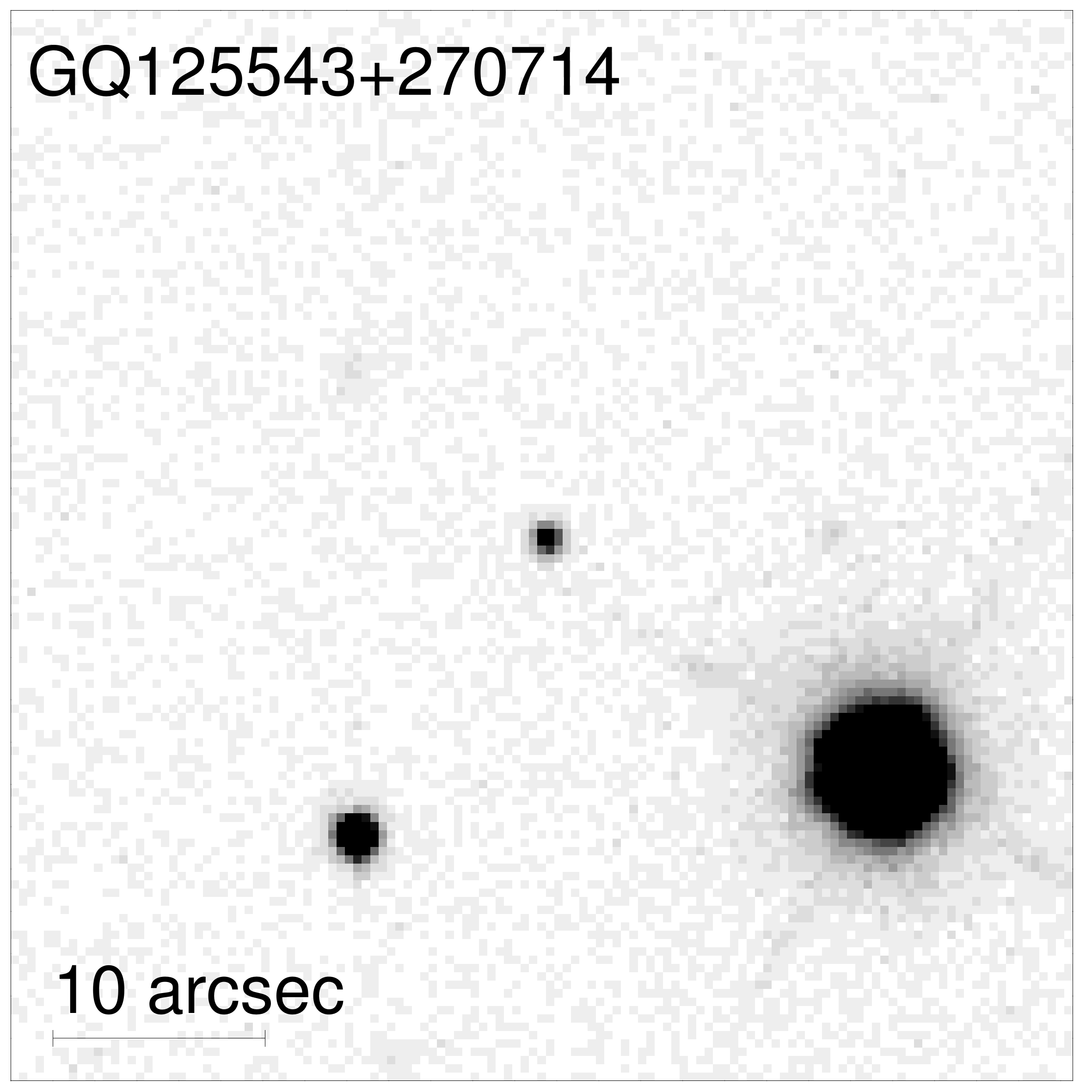,width=3.5cm} 
    \caption{50$\times$50 arcsec$^2$ thumbnails around each stationary source.}
	\label{fig:mosaic3}
\end{figure*}

\end{appendix}

\end{document}